%% file: Yang_Zhang_Thesis_arxiv.tex
\documentclass[a4paper,twoside,12pt,titlepage]{mybook}
\usepackage{a4}

\usepackage{amsmath,amsfonts,amssymb,amsthm}
\usepackage{url}
\usepackage{float}
\usepackage{graphicx}
\usepackage{subcaption}
\usepackage[sectionbib]{bibunits}
\usepackage{algorithmic}
\usepackage{setspace}
\usepackage{fancyhdr}
  \renewcommand{\chaptermark}[1]{\markboth{\chaptername \ \thechapter \ \ #1}{}}
  
\usepackage{adfatitlepage}
\usepackage{rotating}
\usepackage[none]{hyphenat}
\usepackage[intoc]{nomen}
\usepackage{pstricks}
\usepackage{array}
\usepackage{substr}
\usepackage[chapter]{algorithm}
\usepackage{url}
\usepackage{verbatim}
\usepackage{cite}
\usepackage{mathrsfs}
\usepackage{multirow}
\usepackage{bm}
\usepackage{lineno}
\usepackage{array}
\usepackage{color}
\usepackage{latexsym}
\usepackage{amsxtra}
\usepackage{acronym}
\usepackage{longtable}
\usepackage{nomencl}
\usepackage{titletoc}
\usepackage{stfloats}
\usepackage{hhline}
\usepackage{tabularx}
\usepackage{makecell}
\usepackage{enumerate}
\usepackage{epstopdf}
\usepackage{epsfig}
\usepackage{hyperref}
\usepackage{pifont}
\usepackage{pdfpages}
\newcommand{\iid}{i.\@i.\@d.\ }

\DeclareUnicodeCharacter{2212}{-}

\makeatletter
\usepackage{etoolbox}
\pretocmd{\tableofcontents}{%
  \if@openright\cleardoublepage\else\clearpage\fi
  \pdfbookmark[0]{\contentsname}{toc}%
}{}{}%
\makeatother

\makenomenclature

\usepackage{geometry}
\begin{document}
\begin{titlepage}
\begin{center}
\pagenumbering{Roman}
\vspace*{1cm}
\Huge \hspace{-15mm}\textbf{Algorithms of Real-Time Navigation and Control of Autonomous Unmanned Vehicles}\\
\vspace{4cm}
\hspace{-15mm}\large\textbf{Yang Zhang}
\\
\vspace{2cm}
\normalsize
\vspace{2cm}
\hspace{-15mm}\includegraphics[width=0.4\textwidth]{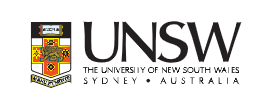}\\
\vspace{2cm}
\textbf{\hspace{-15mm}School of Electrical Engineering and Telecommunications\\
\hspace{-15mm}Faculty of Engineering\\
\hspace{-15mm}The University of New South Wales}\\
\vspace{2cm}
\end{center}
\end{titlepage}

\frontmatter

\pagenumbering{roman}
\pagestyle{fancy}
\fancyhf{}
\setlength{\headheight}{15pt}
\fancyhead[LE,RO]{\footnotesize \textbf \thepage}
\doublespacing
\include{misc/abstract}
\doublespacing

\doublespacing

\fancyhead[CE]{\footnotesize \leftmark}
\fancyhead[CO]{\footnotesize \rightmark}

\renewcommand{\chaptermark}[1]{%
\markboth{\MakeUppercase{%
\thechapter.%
\ #1}}{}}







\mainmatter
\doublespacing

\include{chapter/Chapter1}
\include{chapter/Chapter3}

\include{chapter/Chapter4}

\include{chapter/Chapter5}

\include{chapter/Chapter6}

\include{chapter/Chapter71}
\include{chapter/Chapter72}

\include{chapter/Chapter8}



\clearpage{\pagestyle{empty}\cleardoublepage}

\renewcommand{\bibname}{Bibliography}
\addcontentsline{toc}{chapter}{\protect\numberline{}{Bibliography}}
\singlespacing
{\bibliographystyle{IEEEtran}
\bibliography{bibs/Yang_Zhang_Thesis_arxiv}
}

\end{document}

%% file: misc/abstract.tex
\chapter*{Abstract}
\addcontentsline{toc}{chapter}{\protect\numberline{}{Abstract}}
The rapid development of robotics has benefited by more and more people putting their attention to it.
In the 1920s, ‘Robota', a similar concept, was first known to the world. It is proposed in Karel Capek' s drama, Rossum' s Universal Robots (RUR).
From then on, numbers of automatic machines were created all over the world, which are known as the robots of the early periods.
Gradually, the demand for robots is growing for the purpose of fulfilling tasks instead of humans.
From industrial uses, to the military, to education and entertainment, different kinds of robots began to serve humans in various scenarios.
Based on this, how to control the robot better is becoming a hot topic.
\par
For the topic of navigating and controlling mobile robots, number of related problems have been carried out.
Obstacle avoidance, path planning, cooperative work of multi-robots.
In this report, we focus on the first two problems, and mention the last one as a future direction in the last part.
\par
For obstacle avoidance, we proposed algorithms for both 2D planar environments and 3D space environments.
The example cases we raise are those that need to be addressed but have always been ignored.
To be specific, the motion of the obstacles are not fixed, the shape of the obstacles are changeable, and the sensors that could be deployed for underwater environments are limited.
We even put those problems together to solve them.
The methods we proposed are based on the biologically inspired algorithm and Back Propagation Neural network (BPNN).
In addition, we put efforts into trajectory planning for robots. The two scenarios we set are self-driving cars on the road and reconnaissance and surveillance of drones.
The methods we deployed are the Convolutional Neural Network (CNN) method and the two-phase strategy, respectively.
When we proposed the strategies, we gave a detailed description of the robot systems, the proposed algorithms. We showed the performance with simulation results to demonstrate the solutions proposed are feasible.
\par
For future expectations, there are some possible directions.
When applying traditional navigation algorithms, for example, biologically inspired algorithms, we have to pay attention to the limitations of the environment.
However, high-tech algorithms sometimes are not computationally friendly.
How to combine them together so as to fulfill the tasks perfectly while the computational efficiency is not too high is a worthy topic.
In addition, extending the obstacle avoidance algorithms to more competitive situations, such as applying to autonomous UAVs, is also being considered.
Moreover, for cooperation among multi robots, which could be regarded as Network Control System (NCS), the issues, such as how to complete their respective tasks, how to choose the optimal routes for them are worth attention by researchers.
\par
All in all, there is still a long way to go for the development of navigation and control of mobile robots. Despite this, we believe we do not need to wait for too long time to see the revolution of robots.
\par

%% file: chapter/Chapter1.tex
  \chapter{Introduction}
\label{C1:chapter1}

%
%
%
%
%
%
%
%
\section{Background}
In 1921, Karel Capek, a famous writer from Czechoslovakia, brought his drama called Rossum' s Universal Robots (RUR) to the world \cite{Moran2008RUR}.
In the play, the hero Rossum developed a kind of automatic worker and named them ‘Robota'.
These automatic workers, appear as labor force, they look like humans and they can think for themselves like humans\cite{Capek2004RUR}.
%
It was the first time that ‘Robot' was known by the world.
About fifty years later, in 1979, the Robotic Institute of America gave the definition of robot.
A robot is “a re-programmable, multi-functional manipulator designed to move materials, parts, tools, or specialized devices through various programmed motions for the performance of a variety of tasks” \cite{Reilly2011FATA}.
\par
%
Since then on, researchers from all over the world have put their efforts into robot control.
Numbers of automatic machineries have been invented.
In the 1960s, the Johns Hopkins University Applied Physics Laboratory designed the Johns Hopkins Beast. It was an early mobile transport robot. This machine has basic intelligence and the ability to survive on its own \cite{1960}.
In the 1970s, The Stanford Research Institute designed the robot Sarky, which was a landmark event in robotics because it combined hardware and software to sense its environment.
By the beginning of the 20th century, the Danish company Lego launched the Mind-Storms set, Honda brought their humanoid robot ASIMO, and Microsoft introduced the Microsoft Robotics Studio to the public \cite{20thcentury}.
Every time a new product appears in the sight of the public, people are shocked by the huge changes in the world.
\par
%
Nowadays, with the control of robots becoming a more and more popular issue being discussed all around the world, the application of robots has made great achievements in many fields.
According to the profession of robots, it can be divided into industrial fields, special fields and service fields \cite{3.1}.
For industrial machinery, robots could help humans with handling, welding, assembly, painting, inspection and others, which are mainly used in modern factories and flexible processing systems \cite{indust.app}.
For the extreme operation robot, they are becoming the necessary tools to implement activities from construction, to agriculture, to search and rescue \cite{6.2} \cite{rescue.app}. They can operate in extremely hostile circumstances that are not friendly to humans, like nuclear power plants, underseas, space \cite{extreme.app}.
For entertainment robots, they can sing, dance, and even play musical instruments according to the instructions given by their hosts. They are becoming an indispensable part of the human world.
\par
%
\section{Literature Review}
With more and more researchers putting their efforts into robotics, the application of robots is becoming increasingly extensive.
In order to make the robots implement tasks perfectly, precise control of robots has becoming a popular and necessary topic.
For intelligent autonomous robots, one of those important tasks in robot control is navigating them safely in cluttered environments.
To be specific, guide the robot to its destination while avoiding obstacles that are in its way \cite{yang.1}.
\par
%
%
%
%
%
%
%
\par
\subsection{Classification of Navigation Strategies}
In order to apply to various scenarios, several navigation strategies for moving robots have been carried out by researchers all over the world \cite{3.1} \cite{3.2}. Those algorithms are quite different from system modeling to algorithm description to the application scenarios. When we do research on an existing navigation strategy, what we focus on are the model of the robot, the environment it applied to, the status of the obstacles, and the algorithm itself. To comprehend and evaluate these strategies, classification is necessary. According to different standards, a number of classification methods are proposed. Here, we will highlight some of them.
\par
%
%
In general, all the algorithms could be divided into global path planning and local path planning based on the knowledge of the environment. Global path planning acquires the robot to master prior complete information about the environment, while local path planning is based on real-time environmental information collected by sensors \cite{3.3} \cite{3.4}.
Further, they could be divided into offline path planning and online path planning. For offline path planning, the robot could build a map without the help of the sensors as they get the information of the environment before they depart. For online path planning, the robot needs to obtain information of the environment by the sensors placed on them \cite{plus.1}.
Moreover, static path planning and dynamic path planning distinguish the status of the obstacles. To be specific, a navigating strategy could be regarded as a static path planning strategy when the information of obstacles obtained by robot is static, otherwise it is a dynamic path planning strategy. 
%
%
Here, some algorithms which are well-known by people in the past two decades will be mentioned to illustrate the classification methods.
\par
A well-known obstacle avoidance algorithm is the dynamic windows approach, which was first proposed in 1997 \cite{3.6}. It is an approach based on the motion of robot. The dynamic window restricts the range of both angular velocity and linear velocity. Sampling is carried out around the current state to obtain the reasonable movement change at the next moment, which includes angular velocity and linear velocity. 
The curvature velocity method and the lane curvature method were derived from the dynamic windows approach \cite{3.7} \cite{3.8}. 
These three obstacle avoidance strategies are local path planning, and they are usually applied to static obstacles. 
For dynamic obstacles, velocity obstacles avoidance method and collision cone approach were introduced to the world in 1998 \cite{3.9} \cite{3.11}. Reciprocal velocity obstacles avoidance method for multi-agent safely navigating, which was inspired by the velocity obstacles avoidance method, is also for dynamic obstacles \cite{3.10}. Another well known method is the inevitable collision states strategy, in which both the dynamics of the robotic system and the obstacle is considered. Obviously, it is a dynamic path planning strategy \cite{3.12} \cite{3.13}.
\par
Although it is possible to find the optimal solutions for local path planning method, it is common to sacrifice the optimal solution to ensure the computation speed is not too slow in practice \cite{plus.2}. 
In addition, global path planning and local path planning could be combined to perform better in different application scenarios. A reactive probabilistic path planning strategy for wheeled mobile robot was successfully employed by Savkin and Hoy \cite{3.5}. The global path planning strategy is used to reach the target while the local path planning strategy is for obstacle avoidance. The whole control law consist of some separate modes, switching among these modes until the robot reaches the goal. The most remarkable feature of it is that the trajectory is the shortest one for a known environment while safe navigation also be given for unknown environment.
\par
%
Based on the navigation objective, the existing obstacle avoidance methods could be divided into three categories, the shortest path planning, the fastest path planning, and the balance planning. The shortest path planning, for which only distance is considered, no time or other factors. The fastest path planning, for which only time is considered, no distance or other factors. Balance planning, for which both time and distance factors must be considered. Obviously, balance path planning is the most difficult one to achieve. For researchers, how to balance the distance and the time-consuming, improving one of them, while not sacrificing the other is always a historically arduous task.  
\par
%
There is another classification method that needs to be mentioned. According to this classification method, all the algorithms are divided into three categories. To be specific, traditional algorithms, high-tech algorithms, and combination algorithms. Compared with the high-tech algorithms, the history of traditional algorithms is longer, and the development is steadier, it can solve some deterministic problems perfectly, as their computational efficiency is always higher. The Dijkstra algorithm, A* algorithm, and D* algorithm are typical traditional algorithms \cite{Dijkstra1} \cite{Dijkstra2} \cite{A*1} \cite{A*2}.
High-tech algorithms are more intelligent when dealing with those more challenging problems. For those situations, the internal structure is not clear, the problem cannot be described by standard models, high-tech algorithms, such as PSO algorithm, genetic algorithm, and reinforcement learning-based algorithms, perform better \cite{reinforcement}. 
For some application scenarios, combining traditional algorithms with high-tech algorithms could get a better result, as they could play their respective strengths. 
Anyway, it is all about getting people to better understand and apply these algorithms.
\par
\subsection{Typical Algorithms}
Over the past few decades, numbers of studies have been carried out all over the world. The Dijkstra algorithm, as one of the traditional algorithms mentioned in the last part, could be regarded as a representative algorithm, as many algorithms are evolved from it. 
The Dijkstra algorithm was proposed by E. W. Dijkstra in 1959. The aim of the algorithm was to solve the shortest path problem from one node to another. The most outstanding point of the Dijkstra algorithm is that the next node selected in each iteration process is the nearest child node of the current one. The whole process could be described as initialization, searching for the minimum point, optimizing the path selection through relaxation, and iterating over and over again \cite{Dijkstra2}. In order to ensure the path it chooses is the shortest one, the smallest nodes are visited one by one according to the greedy principle during each iteration process, so as to solve the optimal path planning problem.
\par
The A* algorithm is the most efficient direct search method for solving the shortest path planning problem in static environments. In the search process, heuristic search rules are established to measure the distance relationship between the real-time search location and the target location, so that the search direction is oriented towards the target location, and finally, the search efficiency is improved.
The estimation function of the current node x is defined as:
\begin{equation}
f(x) = g(x) + h(x),
\end{equation}
where g(x) is the actual distance from the starting point to the current node x, and h(x) is the minimum distance estimated from node x to the ending point. The form of h(x) could be from Euclidean distance or Manhattan distance.
The basic realization process of the algorithm is described as follows: calculating the f value of each child node from the starting point, selecting the child node with the lowest f value as the next point of searching, and iterating until the next child node is the target point. The closer the distance between the estimated value and the actual value is, the faster the final search speed is \cite{A*1} \cite{A*2}.
\par
%
%
In addition, a tangent graph-based reactive algorithm needs to be mentioned, which is proposed by Savkin et al. \cite{3.5}.
The authors gave four distinct types of segments, (OT), (CT), (OO) and (CO) segments. To be specific,  
\begin{itemize}
\item (OT)-segments: the line \emph{L} is tangent to a boundary and crosses
the target \emph{T}.
\item (CT)-segments: the line \emph{L} is tangent to an initial circle and crosses
the target \emph{T}.
\item(OO)-segments: the line \emph{L} is tangent to a boundary on two different points or simultaneously tangent to two boundaries.
\item (CO)-segments: the line \emph{L} is simultaneously tangent to a boundary and an initial circle.
\end{itemize}
Fig. 1.1 is an example of a tangent graph for the given scenario. The initial position is marked with an arrow between the two initial circles, and the target position is marked with \emph{T}. In addition, all the tangent points and tangent lines to the boundaries corresponding to the given four types of segments have been marked \cite{3.5}.
With the probability value \emph{p}, where $0<p<1$, the trajectory is determined. 
\begin{figure}[h!]
\centering
\includegraphics[width=4.2in]{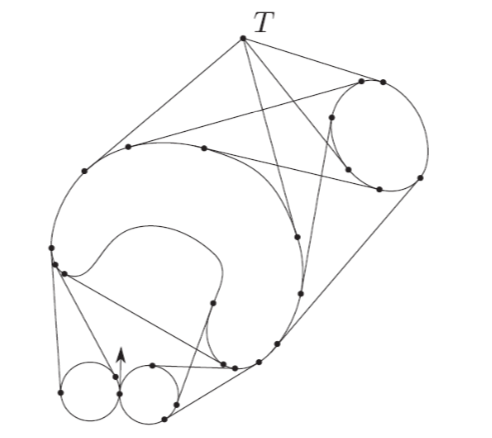}
\caption{An example of tangent graph, with a target point \emph{T}.\cite{3.5}}
\end{figure}
\begin{figure}[h!]
\centering
\includegraphics[width=4.2in]{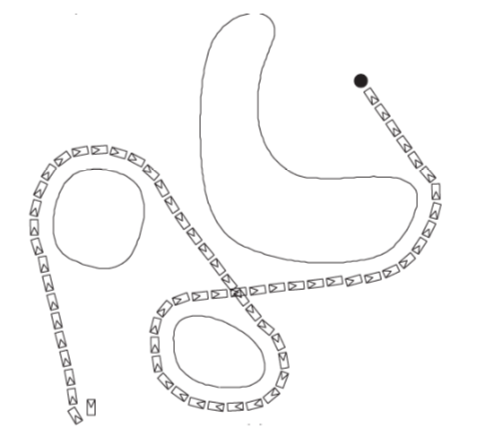}
\caption{A simulation result for the tangent graph-based reactive algorithm. \cite{3.5}}
\end{figure}
Fig. 1.2 is the simulation result of a simple scenario. It is obvious that the robot could depart from the initial point and arrive at the target point successfully without any collisions with the three obstacles. 
For the same scenario, when the value chosen for \emph{p} changes, the path for the robot will be different.
Some more advanced algorithms for robot navigation in environments with obstacles were proposed in \cite{plus.3,3.16,plus.4,3.15}.
In 2016, a binary map building method for planar environment, which is based on the tangent graph-based reactive algorithm was proposed \cite{plus.3}. After this, the map building method has been extended to 3D environment in 2017, in which the sensors employed are 2D range finder sensors \cite{plus.4}. 
\par
Biologically inspired algorithm is an algorithm learning from Nature. Here, we introduce one, which is learned from squirrels, imitating a squirrel running around a tree.
At the beginning, an Equiangular Navigation Guidance (ENG) laws was introduced by Teimoori and Savkin in 2008 \cite{3.16}. Then, in 2010, they brought a static obstacle avoidance strategy to the world \cite{3.15}. After that, the biologically inspired reactive algorithm for dynamic environments with moving obstacles was proposed in 2013 by Savkin and Wang \cite{3.14}. 
They described the mobile robot with a non-holonomic model and gave strict constraints to the movement of obstacles. Then the authors gave the navigation algorithm, switching between two modes, eluding the obstacles, and moving straightly to the target, with some limiting conditions. 
\begin{figure}[h!]
\centering
\includegraphics[width=4.2in]{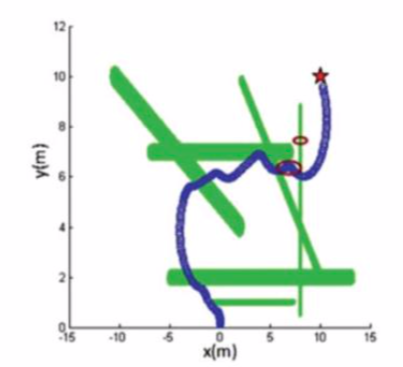}
\caption{A simulation result for the biologically inspired reactive algorithm. \cite{3.14}}
\end{figure}
A simulation result is given in Fig. 1.3. The robot starts from the initial point, the coordinate position of which is $(0, 0)$, and the target position is $(10, 10)$, which is marked with a red star. The trajectory of the robot is shown in blue, and the movements of obstacles are shown with green one. We can get from the figure that the robot could arrive at the target successfully without collisions with the obstacles. It is a viable navigation strategy for the given scenario. 
\par
\subsection{Conclusion}
As we focus on the topic of navigation and control of mobile robots, we started from researching existing algorithms. For the representative path planning algorithms, the Dijkstra algorithm, A* algorithm, and D* algorithm, their performance is acceptable for static environment. The shortest path could be found when the environment is not too complicated. However, the logic of those algorithms is too simple and basic to deal with some challenging situations. 
For the tangent graph-based reactive algorithm, the value chosen for \emph{p} is still a tough task to solve. As mentioned before, the trajectory for the robot will not be the shortest one if the value chosen for \emph{p} is not a suitable one. In addition, the obstacles in the given scenarios are always stationary ones.
The biologically inspired algorithm, which is based on the Equiangular Navigation Guidance (ENG) laws, focuses on dynamic environments, where the movement of obstacles are taken into consideration. However, the limitations of the obstacles are too restrictive. To be specific, the obstacles could only move in a straight line with constant speed without any rotations. In practical situations, the obstacles may be speed changeable with rotation. In addition, whether the path is the most optimal one is also not been taken into consideration. 
After careful study and analysis, we divided the problem into two parts, obstacle avoidance and path planning.
\par
For the obstacle avoidance part, when we did the literature review of existing algorithms, we found that each single algorithm has its limitations \cite{1.112} \cite{1.113}.
For the motion of the obstacles, the speed of the robot is always constant, and the angular velocity is always zero in some prior algorithms.
However, in a practical scene, changeable motion is more common.
For the shape of the obstacles, they were always approximated as circles in planar environments and spheres in three-dimension environments.
But shape changeable obstacles are also common in real life. For example, birds in the sky, humans on the ground, and fish in the sea. 
%
%
Thus, we set the main goal of our research to reduce the limitations of the obstacles.
To be specific, reducing the limitation of motion and the shape of obstacles.
%
%
Moreover, to make the application scenarios of our proposed algorithms more feasible, we also considered the problem of dimensions.
The complexity of the model would become higher when considering the 3D environments compared with 2D environments.
In addition, for 3D environments, implementing in air conditions and in underwater conditions cannot be considered as a same task.
%
%
%
This is because the sensors that could be deployed are limited.
As a result, the information about the environment that could be acquired is not sufficient as in other situations.
The difficulty of designing algorithms is increasing.
\par
When considering path planning problems, there are several problems that need to be addressed \cite{1.111}.
As we mentioned in the prior section, robots are widely used in many fields. The diversity of the application scenarios causes the focus of each algorithm to be different. 
We focus on two typical scenarios, self-driving cars on the road scenario and Unmanned Aerial Vehicle (UAV) reconnaissance and surveillance scenario.
Self-driving cars develop rapidly these years with the satisfactory progress of robotic technology. Several approaches to machine learning are used to solve trajectory planning problem. However, a significant amount of data is required in advance for learning. What we want to achieve is predicting the motion of moving vehicles based on the insufficient amount of data. 
Furthermore, flying drones are employed for military and rescue purpose. For reconnaissance and surveillance drones, a lower attitude of the travelling path is preferred, in order to observe the terrain better. In this case, the length of the path is unsatisfactory if targets of interest on the ground are covered completely. Our goal is solving this challenging task, finding an optimal trajectory for the drones at a given attitude while the two requirements reach a compromise. 
\par

%% file: chapter/Chapter3.tex
\chapter{A Method for Collision-free Guidance of Mobile Robots in Unknown Dynamic Environments with Moving Obstacles}
\label{C3:chapter3}
From this chapter, we begin to propose navigation strategies for robots. 
In this chapter, we focus on the problem of avoiding moving obstacles in planar environments.
Although some strategies have been proposed by other researchers before, the constraints of the obstacles' motion are too restrictive.
Our aim is to relax the limitation of both speed and angular velocity of the obstacles.
Moreover, as we are inspired by the on-line reactive approach proposed by Savkin et al. \cite{3.14}, we will try to propose a better solution while achieving the same goal.
\par
%
The results of this chapter were originally published in the conference papers: 
%
\textbf{Y.~{Zhang}}, ``A Method for Collision-free Guidance of Mobile Robots in Unknown Dynamic Environments with Moving Obstacles,'' in {\em 2019 IEEE International Conference on Robotics and Biomimetics (ROBIO)}, Dali, China, 6-8 Dec. 2019. \emph{Accepted for publication.}
%
\section{Introduction}
Planning path for mobile robots while avoiding obstacles effectively is an important problem of robotics. A lot of researches have been carried out to solve the problem, e.g. \cite{3.1}, \cite{3.2}. Although these approaches are all giving solutions to the problem of finding trajectory for the robots, they use quite different assumptions on obstacles, very different robot's motion models, and different sensors. 
Although several strategies have been proposed, they are not effective for rapidly changing environments.
\par
In order to solve these problems, a biologically inspired reactive algorithm for dynamic environments with moving obstacles was proposed in 2013 \cite{3.14}. It is an approach based on Equiangular Navigation Guidance (ENG) laws \cite{3.15} \cite{3.16}, where they described the mobile robot with non-holonomic model and gave a strict constraint to the movement of obstacles. The proposed algorithm can successfully guide the robot to the target point with a number of assumptions. Although such algorithm is realizable in some situations, it is not always feasible in practice. It is because that on one hand the shapes of obstacles are limited strictly and on other hand obstacles can only move along a straight line with a constant speed.
\par
Therefore, we propose a navigation approach in this chapter, which is based on the biologically inspired algorithm \cite{3.14}. It is a speed and angular velocity based on-line reactive approach. The method can be applied to dynamic environments with moving obstacles. Without loss of generality, speed and angular velocity are changeable within a certain range, we do not assume that all the obstacles are convex. Our navigation approach can be described as a sliding mode control law, which switches between two modes, i.e. target approaching mode and obstacle avoidance mode, respectively. Compared with the previous works, constraints to the obstacles are less. 
With successfully achieving those tasks, our proposed algorithm does not need such high computational complexity, unlike those high-level decision-making algorithm, such as \cite{3.17}. Hence, the significance for practice of our proposed method is obvious.
\par
The remainder of the chapter is organized as follows. Section 2.2 presents problem statements including system description and some assumptions. Section 2.3 introduces our navigation algorithm. Computer simulations and a comparison with the biologically inspired algorithm proposed by Savkin et al. are given in section 2.4. Finally, section 2.5 presents brief conclusions.
\par

\section{Problem Statement}
In this section, we consider a planar vehicle or wheeled mobile robot modeled as a unicycle. For the controlled object, the control inputs are angular velocity and speed, both limited by given constants. The kinematics of the considered vehicle is described as follows:
\begin{equation}
\dot{x}{(t)}=V_R{(t)}\cos\theta,x{(0)}=x_0,
\end{equation}
\begin{equation}
\dot{y}{(t)}=V_R{(t)}\sin\theta,y{(0)}=y_0,
\end{equation}
\begin{equation}
\dot{\theta}{(t)}=u_R{(t)}\cos\theta,\theta{(0)}=\theta_0,
\end{equation}
where\\
\begin{equation}
V_R{(t)}\in\left[0, V_{max}\right], u_R{(t)}\in\left[ -U_{max}, U_{max}\right]
\end{equation}
Here, $\left[x(t), y(t)\right]$ is the vector of the vehicle's Cartesian coordinates and $\theta{(t)}$ gives its orientation. $V_R (t)$ and $u_R (t)$ are the speed and angular velocity, respectively. The angle $\theta{(t)}$ is measured in the counter-clockwise direction. The maximum speed $V_{max}$ and the maximum angular velocity $U_{max}$ are given.
\par
Now, describe the vector of the robot's coordinates with
\begin{equation}
c_R{(t)}:=\left[x(t), y(t)\right]
\end{equation}
In order to introduce our proposed navigation approach, we set a scene example with essential elements. We definite a initial point \emph{\textbf{I}} and a target point \emph{\textbf{T}}, which are always stationary. There are also several random moving obstacles $\Diamond_1{(t)}, \Diamond_2{(t)}, $...$, \Diamond_k{(t)}$ in the plane. Each obstacle is closed bounded planar set. We do not assume that all the obstacles are convex. An illustration is (a) given in Fig. 2.1. All points on the boundaries of obstacles should be seen by the robot, or it will not be considered, e.g. (b) in Fig. 2.1. There is some inner space in the obstacle, the red part could be seen by the robot only when the robot goes into the obstacle. Narrow U-trap situation is not considered because it is very difficult to avoid collisions in narrow space and the robot may be trapped, e.g. (c) in Fig. 2.1. Moreover, collisions between obstacles are not considered.
\par
\begin{figure}[h]
	\centering
	\includegraphics[width=4.2in]{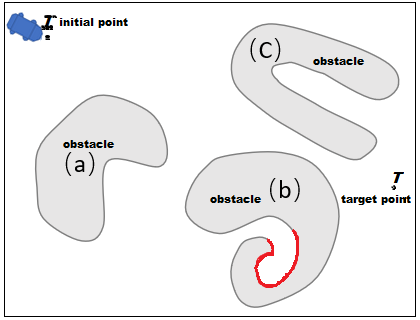}
	\caption{Illustration of obstacles.}
\end{figure}
The objective of our algorithm is to drive the vehicle to the target point \emph{\textbf{T}} while avoiding collisions with all moving obstacles in the plane. The current distance between the vehicle and the border of the obstacle $\Diamond_i$ defined as
\begin{equation}
d_i{(t)}:=\min_{r{(t)}\in{\Diamond_i{(t)}}}\lVert{r{(t)}-c_R{(t-1)}}\rVert.
\end{equation}
Here, $\lVert\cdot\rVert$ denotes the standard Euclidean vector norm, r(t) is the point sets of the obstacle. We suppose that the point  $r_{min}{(t)}=\left[x_{min}{(t)}, y_{min}{(t)}\right]$ makes the minimum value achieved.
\par
We define a minimum distance between the robot current position and the nearest obstacle $\Diamond_i$ as
\begin{equation}
d_{\rm{min}}{(t)}:=\min_{r{(t)} \in{\Diamond_i{(t)}}}\lVert{r{(t)}-c_R{(t-1)}}\rVert.
\end{equation}
\par
The obstacle $\Diamond_i$ is moving with the velocity $v_i{(t)}$ and revolving around the mass center with the angular velocity $u_i{(t)}$. For obstacle $\Diamond_i$, we suppose that the shape and mass center of each obstacle are known, the motion of the obstacles should be restricted rigorously, as $v_i{(t)}$ and $u_i{(t)}$ are contributing to the value of $d_{min} {(t)}$.
\par
We assume that there is a point on the border of the obstacle, and the distance between the mass center and that point is $B_{initial}$. With the effect of angular velocity $u_i(t)$, the point will stop at the end one, and the distance between the mass center and that end point is $B_{end}$. The maximum distance achieved in the point section [initial, end], which is shown as $B_{max}$. We define $\Delta distance=B_{max}-B_{initial}$, see Fig. 2.2.
%
\par
\begin{figure}[h]
	\centering
	\includegraphics[width=4.2in]{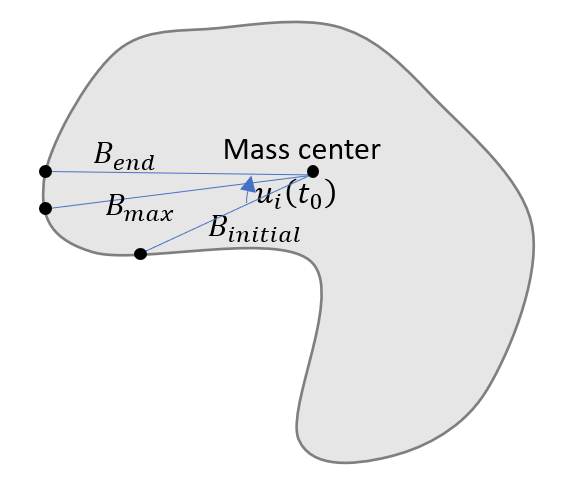}
	\caption{The illustration of the constraints of motion of the obstacles to avoid collision.}
\end{figure}
The constraints of velocities $v_i$ and angular velocity $u_i$ are:
\begin{equation}
0<\lVert v_i{(t)}\rVert<\lVert v_R{(t)}\rVert<V_{max},
\end{equation}
\begin{equation}
-U_{max}<u_i(t)<U_{max},
\end{equation}
\begin{equation}
\Delta distance<\lVert v_R{(t)}\rVert-\lVert v_i{(t)}\rVert.
\end{equation}
For some $V_R > 0$ and all $\Diamond_i$, t. the collision avoidance may be impossible if (6)-(8) does not hold.
\par
\section{Navigation Algorithm}
In this section, we give a detailed description of our proposed collision-free navigation algorithm. We suppose that there are some moving obstacles on the plane, the robot starting moving from initial point \emph{\textbf{I}} and aiming to get the target point \emph{\textbf{T}}. Fig. 2.3 is a block diagram which shows how the algorithm works.\par
There are three steps for the robot to get the coordinate of every single second when the robot is avoiding an obstacle, see the gray dotted box in Fig. 2.3. Then we will describe these three steps separately.
\begin{figure}[h]
	\centering
	\includegraphics[width=4.2in]{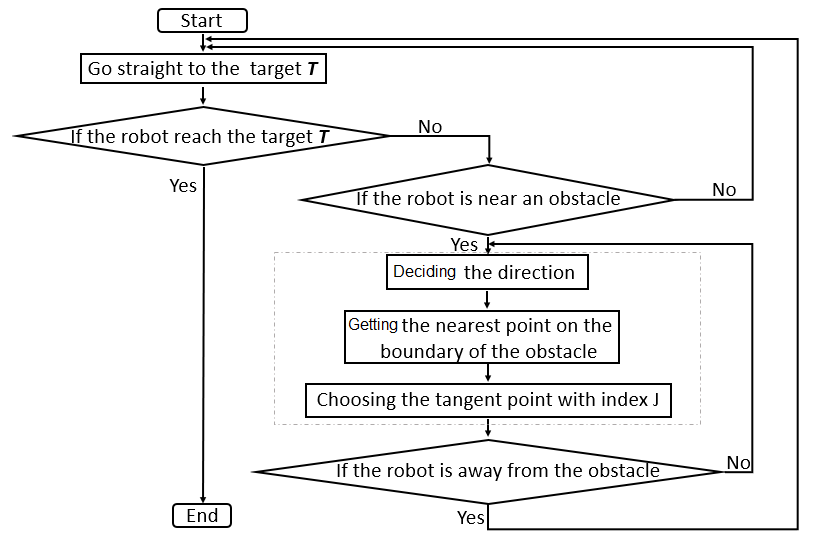}
	\caption{Block diagram of the algorithm.}
\end{figure}

\subsection{Deciding the direction.}
We define a line \emph{\textbf{$L_T{(t_0)}$}} and a line \emph{\textbf{$L_M{(t_0)}$}}, which are the segment between the robot and target \emph{\textbf{T}} and segment between the robot and the mass center of obstacle $\Diamond_i$, respectively. Then, measure the angle of line \emph{\textbf{$L_M{(t_0)}$}} from line \emph{\textbf{$L_T{(t_0)}$}}, which is denoted as $\alpha(t_0)$.\\
\begin{equation}
\alpha{(t_0)}\in{(-\frac{\pi}{2}, \frac{\pi}{2})}
\end{equation}
Moreover, we assume that the counter-clockwise direction is the positive direction. If $\alpha{(t_0)}\in{\left[0, {\frac{\pi}{2}}\right)}$, the robot will avoid the obstacle in the positive direction, denoted as '+', e.g. Fig. 2.4(a). If $\alpha(t_0)\in{\left[-\frac{\pi}{2}, 0\right)}$, the robot will avoid the obstacle in the negative direction, denoted as '-', e.g. Fig. 2.4(b).
\begin{figure}[h!]
	\centering 
	\begin{subfigure}[b]{\textwidth}
		\centering
		\includegraphics[width=4.2in]{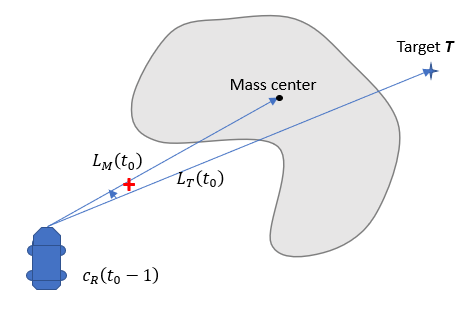}
		\caption*{(a)}
	\end{subfigure}
	\begin{subfigure}[b]{\textwidth}
		\centering
		\includegraphics[width=4.2in]{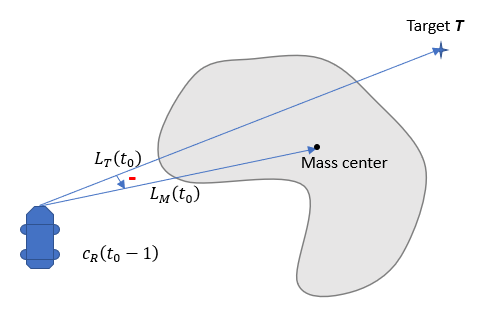}
		\caption*{(b)}
	\end{subfigure}	
	\caption{The position relationship between $L_M(t_0)$ and $L_T(t_0)$.}
\end{figure}

\subsection{Getting the nearest point on the boundary of the obstacle.}
The point on the boundary of the obstacle $\Diamond_i$ is defined as $r_{min}{(t_0)}= [x_{min}{(t_0)}, y_{min}{(t_0)}]$, which satisfies: 
\begin{equation}
\lVert r_{min}{(t_0)}-c_R{(t_0-1)}\rVert=d_{min}{(t_0)}.
\end{equation}
We modify the $r_{min}{(t_0)}$ to $r_{min}^* {(t_0)}$. Let $\alpha_0$ be a given angle. The two intersections with the obstacle $\Diamond_i$ of two rays from the robot are $r_{min}^+ (t_0)$ and $r_{min}^- {(t_0)}$, $\alpha_0^+$ and $\alpha_0^-$ are the positive intersection angle and negative intersection angle, respectively, e.g. Fig. 2.5. With the direction decided in the last step, we can get $r_{min}^* {(t_0)}$ equal to $r_{min}^+ {(t_0)}$ or $r_{min}^- {(t_0)}$, given as $r_{min}^* {(t_0)}=\left[x_{min}^* {(t_0)}, y_{min}^* {(t_0)}\right]$. In order to follow the boundaries of obstacles precisely, $\alpha_0$ should be a small one. It is a constant value which we determined by doing a number of simulation.
\begin{figure}[h]
	\centering
	\includegraphics[width=4.2in]{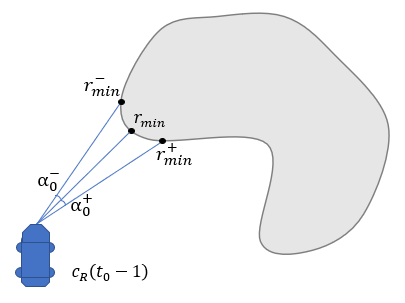}
	\caption{The illustration of intersection angles.}
\end{figure}

\subsection{Choosing the tangent point with index.}
We define a forecast circle $O_{t_0}$,
\begin{equation}
{(x-x^*_{min}{(t_0)})}^2+{(y-y^*_{min}{(t_0)})}^2={r_i{(t_0)}}^2\\
\end{equation}
The center of $O_{t_0}$ is $r_{min}^*{(t_0)}$ and the radius $r_i{(t_0)}$ is $\lVert v_i{(t_0)}\rVert$. There are two tangent lines for the forecast circle $O_{t_0}$ from $c_R {(t_0-1)}$, the angles $\alpha_i^1{(t_0)}$ and $\alpha_i^2{(t_0)}$ are angles of the tangent lines measured from horizontal line in counter-clockwise direction, e.g. Fig. 2.6. The two intersections are called $p_1{(t_0)}$ and $p_2{(t_0)}$, respectively.\par 
\begin{figure}[h]
	\centering
	\includegraphics[width=4.2in]{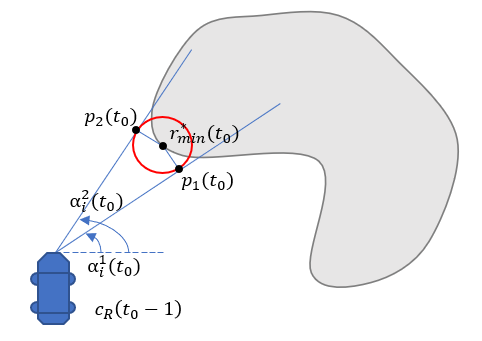}
	\caption{The coordinate of the robot when time $t_0$ equals to $p_J(t_0)$.}
\end{figure}
Here, we define $\gamma$,
\begin{equation}
\gamma= \sin^{-1}\frac{\lVert v_i{(t_0)} \rVert}{d_{min}{(t_0)}}.
\end{equation}
Then, we assume a pair of forecast vectors $l^1_i{(t_0)}$ and $l^2_i{(t_0)}$, given by
\begin{equation}
l^1_i(t_0):=\frac{\lVert v_i{(t_0)}\rVert}{\tan\gamma}\left[\cos{\left(\alpha_i^1{(t_0)}\right)}, \sin{(\alpha_i^1{(t_0)})}\right],
\end{equation}
\begin{equation}
l^2_i{(t_0)}:=\frac{\lVert v_i{(t_0)}\rVert}{\tan\gamma}\left[\cos{\left(\alpha_i^2{(t_0)}\right)}, \sin{(\alpha_i^2{(t_0)})}\right].
\end{equation}
For moving obstacle $\Diamond_i$, $\beta_i^1{(t_0)}$ is the angle between $l_i^1{(t_0)}$ and $v_R{(t_0-1)}$ measured from $v_R{(t_0-1)}$ and $\beta_i^2{(t_0)}$ is the angle between $l_i^2{(t_0)}$ and $v_R{(t_0-1)}$ measured from $v_R{(t_0-1)}$. \par

We denote an index $J{(t_0)}$ by comparing the absolute value of $\beta_i^1 {(t_0)}$ and $\beta_i^2{(t_0)}$, such that $J{(t_0)}=1$ when $\left|l_i^1 {(t_0)}\right|  \leq \left|l_i^2 (t_0)\right|$, $J{(t_0)}=2$ when $\left|l_i^1 {(t_0)}\right| > \left|l_i^2 {(t_0)}\right|$. \par
Now, we introduce the following function:
\begin{equation}
f\left(l_J{(t_0)}, v_R{(t_0-1)})\right)=
\begin{cases}
0, & \beta_J{(t_0)}=0 \\
1, & 0 < \beta_J{(t_0)}\leq \pi \\
-1,& -\pi < \beta_J{(t_0)}<0 \\
\end{cases}
\end{equation}
\par
The navigation law is proposed as:
\begin{equation}
\begin{aligned}
& u_R(t) = -U_{max}f\left(l_J(t), v_R{(t-1)} \right);\\
& V_R(t) = \lVert l_J(t) \rVert.
\end{aligned}
\end{equation}
Obviously, $d_{min}^2 (t)={\lVert v_i (t)\rVert}^2+{\lvert l_J (t) \rVert}^2$. As $d_{min}(t)<V_{max}, V_R (t)=\lVert l_J (t) \rVert <V_{max}$, satisfy constraints (2).\\
Specifically, \emph{\textbf{M1}} is target approaching mode. The robot goes straight to the target \emph{\textbf{T}} with maximum speed:
\begin{equation}
u_R (t)=0;  V_R (t)=V_{max}.
\end{equation}
\emph{\textbf{M2}} is obstacle avoidance mode. The motion of robot follows law (15). The robot starts to move with \emph{\textbf{M1}} and switches to \emph{\textbf{M2}} when the distance between the robot and the nearest obstacle reduces to $d_{min}$. When the robot is away from the obstacle and orientates towards the target, it will switch to \emph{\textbf{M1}}. Switching between these two modes until the robot reach the target \emph{\textbf{T}}. Furthermore, the robot is away from the obstacle means that the distance between the robot and the obstacle is more than $d_{min}$, line \emph{\textbf{$L_T{(t)}$}} does not across the obstacle.\par
So far, we have introduced our collision-free navigation algorithm for mobile robots which can be applied to dynamic environment with a number of moving obstacles.
\section{Computer Simulation}
In this section, we present computer simulation results for a wheeled mobile robot
navigating in dynamic environments with moving obstacles. The simulations are performed with Matlab. Our aim is to simulate  our proposed algorithm and compare it with other algorithms.\par
First, we examine our approach for a simplest case: the environment with only one moving obstacle. 
The coordinate position of the initial point \emph{\textbf{I}} and the target point \emph{\textbf{T}} are [0, 0] and [10, 10]. We set the maximum speed $V_{max}$ and maximum angular velocity $U_{max}$ of the robot are 0.707 m/s and 1.414 rad/s, respectively. 
\par
\begin{figure}[h!]
	\centering 
	\begin{subfigure}[h]{\textwidth}
		\centering
		\includegraphics[width=4.2in]{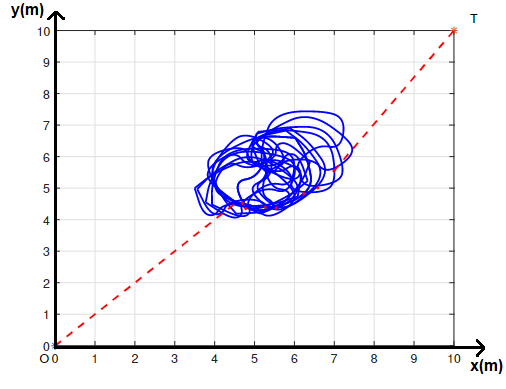}
		\caption*{(a)}
	\end{subfigure}
	\begin{subfigure}[h]{\textwidth}
		\centering
		\includegraphics[width=4.2in]{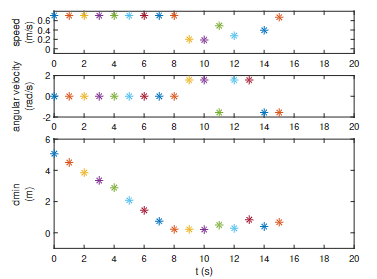}
		\caption*{(b)}
	\end{subfigure}	
	\caption{Simulation results with one moving obstacle.} 
\end{figure}
The result is shown in Fig. 2.7. Fig. 2.7(a) shows the path of the moving obstacle and the path of the robot from the initial point \emph{\textbf{I}} to the target point \emph{\textbf{T}}. The trajectory of the robot is shown in red dotted line. The movement of the obstacles is randomly, which is shown in blue closed curve. The speed and angular velocity of the robot and the minimum distance between the robot and the nearest obstacle are shown in Fig. 2.7(b). The speed and angular velocity are always within the range, and the minimum distance never reach 0. We can see that the robot is moving towards the target point \emph{\textbf{T}} until it is near the moving obstacle, and the robot is close to the moving obstacle when it eludes the moving obstacle. It means that the robot could spend less time reaching the destination.
\par
\begin{figure}[h!]
	\centering
	\begin{subfigure}[h]{\textwidth}
		\centering
		\includegraphics[width=4.2in]{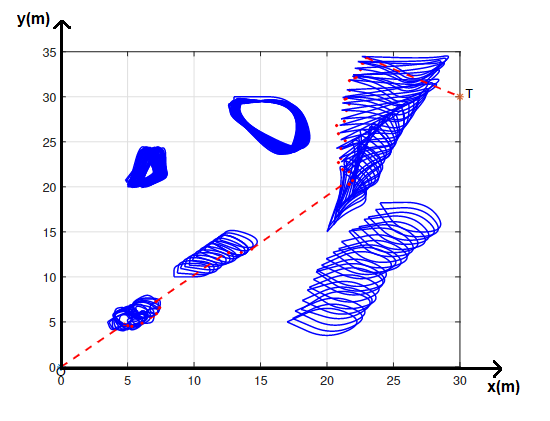}
		\caption*{(a)}
	\end{subfigure}
	\begin{subfigure}[h]{0.5\textwidth}
		\centering
		\includegraphics[width=2.95in]{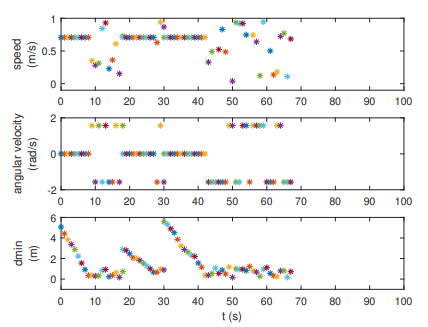}
		\caption*{(b)}
	\end{subfigure}	
	\begin{subfigure}[h]{0.4\textwidth}
		\centering
		\includegraphics[width=2.95in]{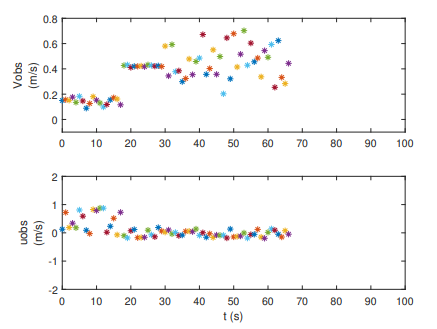}
		\caption*{(c)}
	\end{subfigure}	
	\caption{Simulation results with six moving obstacles.} 
\end{figure}
Fig. 2.8 shows a more challenging situation with six obstacles. As we set, three of the obstacles are in the robot's way, and others are not. The robot would only pay attention to those obstacles which are in its way and take action to the nearest one, although the robot can see all the moving obstacles appeared in its sight. We can confirm above information from the simulation results. Fig. 2.8(a) shows the paths of the robot and the obstacles. Fig. 2.8(b) shows the speed of the robot $v_R{(t)}$, angular velocity of the robot $u_R{(t)}$ and the shortest distance between the robot and the nearest obstacle $d_{min}{(t)}$. Fig. 2.8(c) shows the speed and angular velocity of the nearest obstacle to the robot, $v_i{(t)}$ and $u_i{(t)}$, respectively. Obviously, the robot can reach the target successfully without any collision with obstacles as implementing our proposed algorithm.\par
\begin{figure}[h]
	\centering
	\includegraphics[width=4.2in]{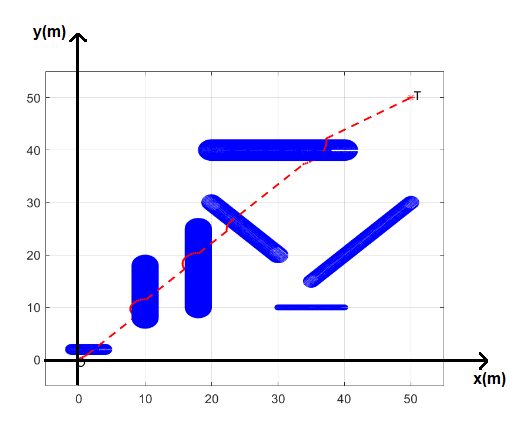}
	\caption{Example of implementing the biologically inspired algorithm.}
\end{figure}
Moreover, we compare our proposed algorithm with the biologically inspired algorithm \cite{3.14}. For the algorithm of Savkin et al, the shapes of obstacles must be discs and obstacles can only move with a constant speed without changing their directions. We tried a scenario with some obstacles moving with changing their directions, the simulation result of the biologically inspired algorithm proposed by Savkin. et al. is shown in Fig. 2.9. As mentioned, our proposed algorithm can be applied to more complex environment, for both shape and motion of obstacles. 
\par
\begin{figure}[h]
	\centering 
	\begin{subfigure}[h]{0.5\textwidth}
		\centering
		\includegraphics[width=2.95in]{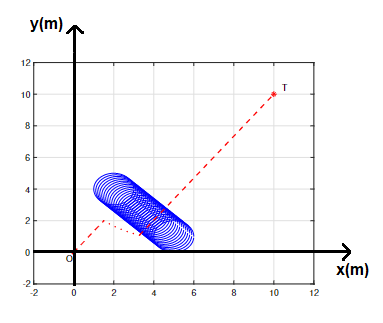}
		\caption*{(a)Robot's path for the algorithm of Savkin et al.}
	\end{subfigure}
	\begin{subfigure}[h]{0.4\textwidth}
		\centering
		\includegraphics[width=2.95in]{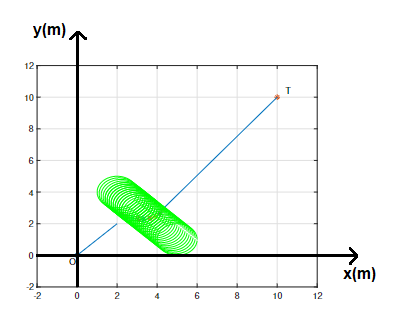} 
		\caption*{(b)Robot's path for our algorithm.}
	\end{subfigure}	
	\begin{subfigure}[h]{0.5\textwidth}
		\centering
		\includegraphics[width=2.95in]{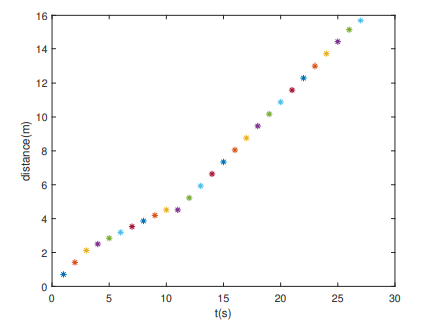} 
		\caption*{(c)Length of robot's trajectory for the algorithm of Savkin et al.}
	\end{subfigure}	
	\begin{subfigure}[h]{0.4\textwidth}
		\centering
		\includegraphics[width=2.95in]{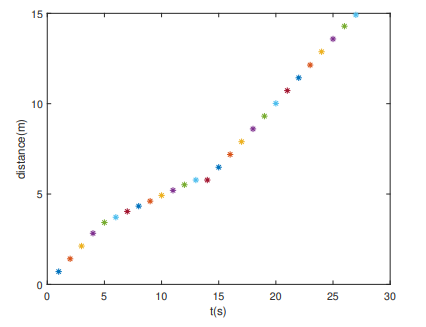}
		\caption*{(d)Length of robot's trajectory for our algorithm.}
	\end{subfigure}	
	\caption{Comparison of the two algorithms.}
\end{figure}
Here, we try a simple situation for comparison these two algorithms, see Fig. 2.10. Fig. 2.10(a) and Fig. 2.10(b) show the paths of the robot navigated by the algorithm proposed by Savkin et al. and our proposed algorithm, respectively. With the limitation of the previous algorithm, the obstacle in the example scene could only move with a constant speed, and angular velocity always be 0. It is clear that with our proposed algorithm, the robot performs better on following the boundaries of obstacles. As the robot is going directly to the target before obstacles appear in its sight, the less the robot spent on avoiding collisions with the obstacles, the shorter the path length is Fig. 2.10(c) and Fig. 2.10(d) illustrate that clearly with the lengths of paths under the two compared algorithms, the experimental results are 15.7(m) and 14.9(m), respectively. Obviously, the path of the robot navigated by our proposed algorithm is shorter, however, the margin is just about 5\%, so with taking into account inaccuracies of the computational algorithms (impact of different integration steps, etc.), we can conclude that the both algorithms produce comparable results for simplest scenarios with disc-shaped obstacles and obstacle movements along straight lines without angular acceleration. In other words, the proposed algorithm has a significant advantage over the algorithm of \cite{3.14} as the proposed algorithm unlike the method of \cite{3.14} works in scenarios with obstacles with complex shapes and complex motions, and in simplest scenarios for which the algorithm of \cite{3.14} was designed, the proposed algorithm demonstrates similar effectiveness.

\section{Conclusion}
We proposed a collision-free navigation approach for 2D environment with moving obstacles. The speed and angular velocity of obstacles are changeable with some constraints.
The robot moves from the initial point with the maximum speed without any rotations, and shifts to obstacle avoidance mode when an obstacle is detected. The algorithm switches between the two modes until the robot successfully reaches the target point. 
In our further research, we will focus on reactive navigation approaches that are based on local sensor measurements.
%
%

%% file: chapter/Chapter4.tex
\chapter{Collision-free Navigation of Flying Robots in Outdoor Environments with Dynamic Obstacles}
\label{C4:chapter4}
In the previous chapter, we proposed a navigation strategy for motion changeable obstacles in planar environments. In this chapter, we propose a solution to the problem of obstacle avoidance in 3D environments. Notice that 3D navigation problems are much harder to solve than 2D navigation problems.
%
%
%
%
For obstacles with time-varying motion, it is important to predict future motion. A local obstacle avoidance strategy combined with an effective path planning method could achieve a shorter trajectory. Moreover, we wish to achieve computational efficiency, which often makes
high-level decision-making methods not applicable.
%
%
%
%
\par	
The results of this chapter were originally published in the conference papers: 
\textbf{Y.~{Zhang}}, ``Collision-free Navigation of Flying Robots in Outdoor Environments with Dynamic Obstacles,'' in {\em 2020 6th International Conference on Control, Automation and Robotics (ICCAR)}, Singapore, 20-23 Apr. 2020.

\section{Introduction}
From industrial perspective to agricultural perspective, even to military perspective, robots are widely used \cite{4.8}, \cite{4.9}. With more and more applications of robots, the requirements of robots are becoming higher and higher. Many approaches to robots safe driving have been proposed, however, they are designed for planar environment. An approach of aircrafts collision-free navigation was addressed in 2008 \cite{4.1}. From then on, more and more researchers focus on solving the problems in three-dimensional environments. Although some previous works have been done to solve the problem of implementing in three-dimensional environments, these approaches are not realizable in some situations as assumptions are too many. For example, \cite{4.10} focuses on the navigation part in a tunnel like environments, obstacles avoiding part has not been addressed. An obstacle avoidance strategy was proposed in 2017, but there are too many constraints to the shape and the motion of the obstacles \cite{4.11}.
\par

%
It is a more difficult task as much more factors need to be considered. Take sensors as an example. For indoor environments, we can fix some sensors at the corner of the closed space, e.g. \cite{4.5}. However, for outdoor situations, we can only rely on the sensors equipped on the robots, e.g. \cite{3.5}, \cite{4.7}. In order to solve the problem perfectly, a proper model to describe the controlled object is needed. The kinematics equation should be given with some necessary constraints. Compared with holonomic one, a nonholonomic system is more appropriate for the practical situations as the more irresistible factors are considered with nonholonomic constraints \cite{3.14}.
\par
We propose an on-line reactive approach, which can be applied to three-dimensional environments with number of moving obstacles in the space. Motion is variable, and we do not assume that all the obstacles are convex. 
We drew inspiration from \cite{4.13}, \cite{4.14} and \cite{4.15}, as they introduced a statistic model for motion forecast, combined two coordinates together and gave the concept of vision cone under three-dimension conditions, respectively. The robot starts moving from the starting point and switches between these two modes while observing the surroundings until reaching the ending point.
\par
The remainder of the chapter is organized as follows. Section 3.2 gives the mathematical model of the system. Section 3.3 elaborates the navigation algorithm with mathematical analysis and also gives some illustrations. Computer simulations of our proposed navigation algorithm is shown in section 3.4. Finally, section 3.5 presents a brief conclusion.
\par
	
\section{System Description}
In this section, we consider a three-dimensional nonholonomic vehicle or flying mobile robot. The mathematical model of the 3D environment is described as follows:
	\begin{equation}
	c_R{(t)}:=\left[x(t),y(t),z(t)\right].
	\end{equation}
	\begin{equation}
	\dot{c_R}{(t)}=V_R{(t)}\cdot a_R{(t)},
	\end{equation}
	\begin{equation}
	\dot{a_R}{(t)}=u_R{(t)}.
	\end{equation}
where $V_R{(t)} \in R$, $\lVert a_R{(t)} \rVert = 1 and u_R{(t)}\in R^{3}$ for all \emph{t}. Here, $(3.1)$ is the three-dimensional vector of the robot's Cartesian coordinates, $(3.2)$ and $(3.3)$ give the motion of the robot. $V_R{(t)}$ is the linear velocity of the robot and $u_R{(t)}$ shows its orientation. Moreover, the following constraints hold:\\
	\begin{equation}
	V_R{(t)}\in\left[0, V_{max}\right], 
	\end{equation}
	\begin{equation}
	\lVert{u_R{(t)}}\rVert \leq {U_{max}}.
	\end{equation}
	\begin{equation}
	(a_R{(t)},u_R{(t)})=0,
	\end{equation}
for all \emph{t}. The constants $V_{max}$ and $U_{max}$ are given. In addition, $\lVert\cdot\rVert$ denotes the standard Euclidean vector norm and $(\cdot,\cdot)$ denotes the scalar product of two vectors. Equation (3.6) can guarantee that $a_R{(t)}$ and $u_R{(t)}$ are always orthogonal. Furthermore, let $v_R(t):=\dot{c}_R{(t)}$ denote the velocity vector of the robot. Obviously,
	\begin{equation}
	a_R{(t)}=\frac{1}{\lVert v_R{(t)}\rVert} v_R{(t)}.
	\end{equation}
\par
	
This model is a kinematic model, in real life applications this model is often supplemented by a dynamic model of a ground mobile vehicle with advanced controllers and state estimators such as H-infinity controllers \cite{R1} \cite{Hinfinitycontroller1} \cite{Hinfinitycontroller2} and robust state estimators \cite{R2} \cite{R3} \cite{R4} \cite{R5} \cite{robuststateestimators1} \cite{robuststateestimators2} \cite{tuft1}.
\par
In order to describe our proposed navigation approach clearly, we set a general three-dimensional scene with some essential elements as an example. We define a starting point \emph{\textbf{S}} and an ending point \emph{\textbf{E}}, which are always stationary. There are also several moving obstacles $\Diamond_1{(t)}, \Diamond_2{(t)}, $...$, \Diamond_k{(t)}$ in the space. Obstacles are solid sets. We do not assume that all the obstacles are convex. The motion of them is not known in advance. Here, we give some constraints to the shapes of the obstacles. In order to illustrate the constraints clearly, we describe it with the obstacles cross sections. Fig. 3.1(a) is an advisable illustration of the cross section of the obstacles. The boundary should be closed curve. Some conditions do not be taken into consideration. Firstly, there is some inner space can not be seen by the robot, as shown in Fig. 3.1(b). Secondly, the shape of the obstacle looks like a magnet, the cross section of it is like a narrow U-trap, e.g. Fig. 3.1(c). Moreover, collisions between obstacles are not considered.\par
	\begin{figure}[h!]
	\centering
	\begin{subfigure}[b]{\textwidth}
		\centering
		\includegraphics[width=4.2in]{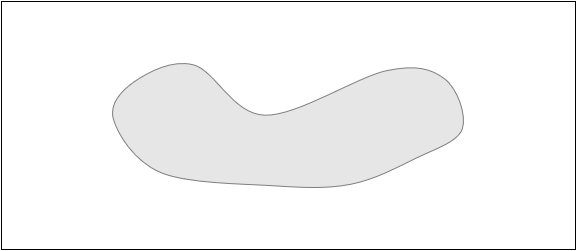}
		\captionsetup{}
		\caption*{(a)}
	\end{subfigure}
	\begin{subfigure}[b]{0.5\textwidth}
		\centering
		\includegraphics[width=2.9in]{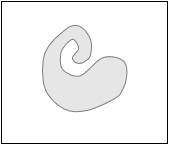}
		\caption*{(b)}
	\end{subfigure}
	\begin{subfigure}[b]{0.4\textwidth}
		\centering
		\includegraphics[width=2.9in]{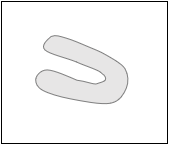}
		\caption*{(c)}
	\end{subfigure}
	\caption{Illustration of the obstacles shape.}
    \end{figure}
The objective of our algorithm is to guide a flying robot to move from the starting point \emph{\textbf{S}} and stop when reaching the ending point \emph{\textbf{E}} while avoiding moving obstacles on its way.
\par
Due to the constraints of the method for predicting the motion of the obstacle we adopted, discretization of continuous variables should be done. The sampling period $T_s$ is set to 1s.
\par
We define a $d{(t)}$ here, which shows the current distance between the robot and the surface of the obstacle $\Diamond_i$ and we suppose that $r_{min}{(t)}$ is the point on the surface of the obstacle $\Diamond_i$ which makes the minimum value $d_{min}{(t)}$ achieved. Then,
	\begin{equation}
	d_{min}{(t)}:=\min_{r_{i}\in{\Diamond_i{(t)}}}\lVert{r_{i}{(t)}-c_R{(t-1)}}\rVert.
	\end{equation}
\par
The motion of the obstacle $\Diamond_i$ is described as follows. We suppose that the mass center of the obstacle $\Diamond_i$ is known. It is moving with the velocity $v_i{(t)}=V_i{(t)}\cdot a_i{(t)}$ and revolving around the mass center with the rotating vector $u_i{(t)}$. Also, we need to give some constraints to the motion of the obstacles, as $v_i{(t)}$ and $u_i{(t)}$ have impacts on the value of $d_{min} {(t)}$. We assume that the velocities $v_i$ and angular velocity $u_i$ satisfy the constraints:
	\begin{equation}
	0<\lVert v_i{(t)}\rVert<\lVert v_R{(t)}\rVert<V_{max},
	\end{equation}
	\begin{equation}
	\lVert u_i(t)\rVert <\lVert u_R{(t)}\rVert <U_{max}.
	\end{equation}
%
\section{Navigation Algorithm}
In this section, we describe our proposed algorithm with some analysis and illustrations. We suppose that there are some moving obstacles in the space, the robot starts moving from starting point \emph{\textbf{S}} and its aim is to reach the ending point \emph{\textbf{E}} while avoiding collisions with the obstacles on its way. There is a block diagram which shows how the algorithm works in Fig. 3.2. The algorithm is a sliding mode control law, switching between two modes, namely, the target approaching mode (\emph{\textbf{M1}}) and the obstacle avoidance mode (\emph{\textbf{M2}}).\par
	\begin{figure}[h!]
	\centering
	\begin{subfigure}[b]{\textwidth}
		\centering
		\includegraphics[width=4.2in]{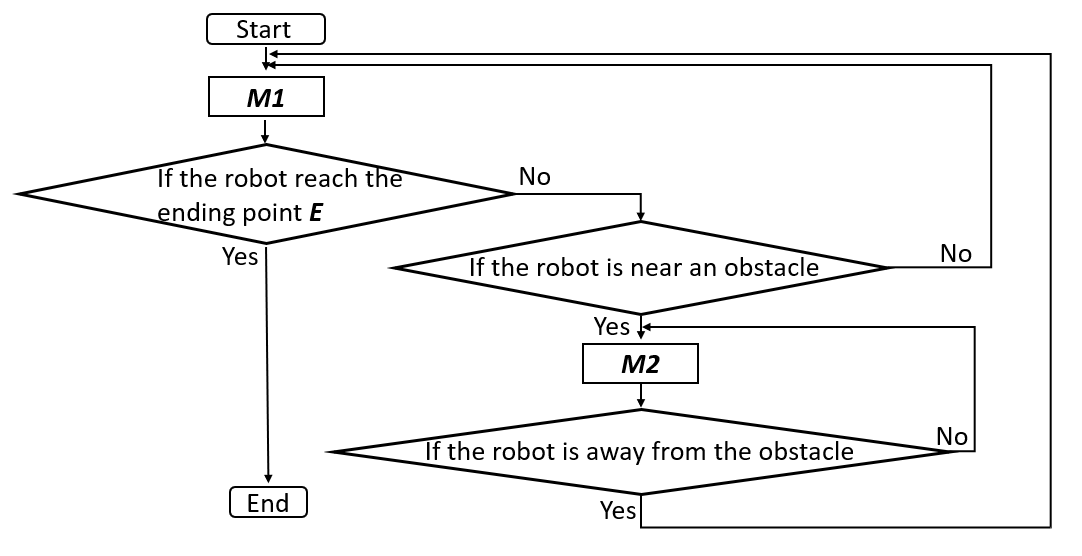}
		\caption*{(a)}
	\end{subfigure}
	\begin{subfigure}[b]{\textwidth}
		\centering
		\includegraphics[width=4.2in]{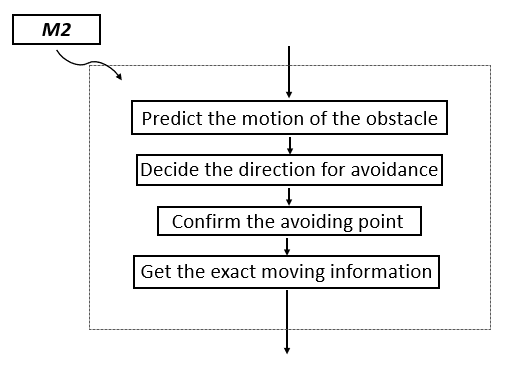}
		\caption*{(b)}
	\end{subfigure}
	\caption{Block diagram of the algorithm.}
    \end{figure}
For the target approaching mode (\emph{\textbf{M1}}), the robot goes straight to the ending point \emph{\textbf{E}} with maximum speed without any rotation. The motion equation is as follows:
	\begin{equation}
	V_R (t)=V_{max},
	\end{equation}
		\begin{equation}
	u_R (t)=0.
	\end{equation}
\par
For the obstacle avoidance mode (\emph{\textbf{M2}}). There are four steps for the robot to get a new motion information each time point: predict the motion of the obstacle, decide the direction for avoidance, confirm the avoiding point and get the exact motion information. Then, we will explain each step carefully.
\subsection{Predict the motion of the obstacle.}
We introduce an Auto-Regressive Model (AR Model) here, which is a statistical model and given as follows:
	\begin{equation}
	X_t=c+\sum^p_{i=1}\varphi_{i}X_{t-i}+\varepsilon_t,
	\end{equation}
where $c$ is a constant, $p$ denotes the order of the model, $\varphi_1, \varphi_2,...,\varphi_p$ are the parameters of the model and $\varepsilon_{t}$ is the predicting error.
\par
We suppose that the motion of the obstacles is changing slowly. There are two independent model for the motion of an obstacle, velocity $v_i{(t)}$ and rotation $u_i{(t)}$. Here, we give detailed description on how to predict the velocity of obstacle $\Diamond_i$ when time is $t_0$ as an example, denotes as $v_i{(t_0)}$.
\par
As we got some previous data $v_i{(1)}, v_i{(2)},...v_i{(t_0-1)}$, we can get from (3.13),
	\begin{equation}
	v_i{(t_0)}=a_{1}v{(t_0-1)}+a_{2}v{(t_0-2)}+...+a_{p}v{(t_0-p)}+\varepsilon_{t_0}.
	\end{equation}
We determine the value of the order $p$ for the AR model using the Ordinary Least Squares (OLS) Method. Moreover, an inspection of whether the model is reasonable is needed. Furthermore, the same operation should be applied to the obstacle's rotation $u_i{(t)}$.
\par
\subsection{Decide the direction for avoidance.}
We have built a three-dimensional coordinate $(X,Y,Z)$ and denoted it as geodetic coordinate when we described the system. Now, we give a two-dimensional coordinate which is an on-line coordinate and denote it as $(X^{'},Y^{'})$. In the beginning, the two-dimensional coordinate is vertical to the $Z$-axis as a original status. For time $t_0$, the original point is the real-time position of the robot, the positive direction of $Y^{'}$-axis is from original point to the ending point \emph{\textbf{E}} and then we can get the $X^{'}$-axis.
\par
We define a line \emph{\textbf{$L_{center}{(t_0)}$}}, which is the line segment between the robot and the mass center of the obstacle. Then, measure the angle of the line \emph{\textbf{$L_{center}{(t_0)}$}} from $Y^{'}$-axis and denote it as $\alpha(t_0)$,
	\begin{equation}
	\alpha{(t_0)}\in{(-\frac{\pi}{2}, \frac{\pi}{2})}.
	\end{equation}
\par
Now, we need to assume a positive direction for the robot. Thus, we suppose that the direction which is the same direction with the $X$-axis is the positive direction. If $\alpha{(t_0)}\in{\left[0, {\frac{\pi}{2}}\right)}$, the robot will avoid the obstacle from the positive direction and denote it as '+'. If $\alpha(t_0)\in{\left[-\frac{\pi}{2}, 0\right)}$, the robot will avoid the obstacle from the negative direction and denote it as '-'.
\par
\subsection{Confirm the avoiding point.}
We denote the point on the surface of the obstacle $\Diamond_i$ which makes the shortest distance between the robot and the obstacle $\Diamond_i$ happened as $r_{min}{(t_0)}= [x_{min}{(t_0)}, y_{min}{(t_0)}, z_{min}{(t_0)}]$, which satisfies: 
	\begin{equation}
	\lVert r_{min}{(t_0)}-c_R{(t_0-1)}\rVert=d_{min}{(t_0)}.
	\end{equation}
Here, give a $r_{min}^* {(t_0)}$ as the point on the surface of the obstacle $\Diamond_i$ we will operate on. We set a given angle $\alpha_0$ and it brings the robot an enlarged vision cone. Fig. 3.3(a) is an illustration, Fig. 3.3(b) is shown from the tangent plane perspective.
\par
	\begin{figure}[h!]
	\centering
	\begin{subfigure}[b]{0.5\textwidth}
		\centering
		\includegraphics[width=2.9in]{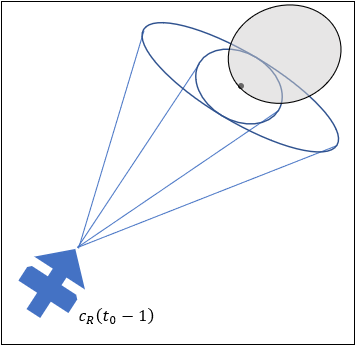}
		\captionsetup{}
		\caption*{(a)}
	\end{subfigure}
	\begin{subfigure}[b]{0.4\textwidth}
		\centering
		\includegraphics[width=2.9in]{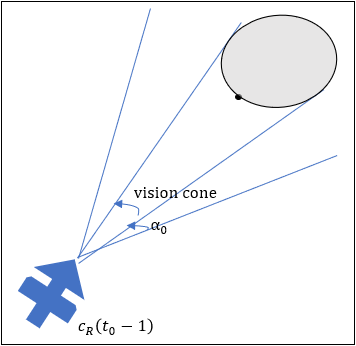}
		\caption*{(b)}
	\end{subfigure}
	\caption{The illustration of the vision cone.}
    \end{figure}
On the two-dimensional on-line coordinate $(X^{'},Y^{'})$, the two intersection between the enlarged vision cone and the obstacle are $r_{min}^+ (t_0)$ and $r_{min}^- {(t_0)}$. For the corresponding, we have $\alpha_0^+$ and $\alpha_0^-$, which denote the positive intersection angle and negative intersection angle, respectively. Then, we can get that $r_{min}^* {(t_0)}$ equals to $r_{min}^+ {(t_0)}$ or $r_{min}^- {(t_0)}$, the superscripts '+' or '-' is obtained in the last step. The $r_{min}^* {(t_0)}$ is described as follows:
	\begin{equation}
	r_{min}^* {(t_0)}=[x_{min}^* {(t_0)}, y_{min}^* {(t_0)}].
	\end{equation}
Moreover, $\alpha_0$ should be a small one and it should be a constant which is determined by the model of the system.
\subsection{Get the exact motion information.}
We define a forecast sphere $O_{t_0}$,
	\begin{equation}
	{(x-x^*_{min}{(t_0)})}^2+{(y-y^*_{min}{(t_0)})}^2+{(z-z^*_{min}{(t_0)})}^2={r_i{(t_0)}}^2.\\
	\end{equation}
We consider the problem in the on-line coordinate $(X^{'},Y^{'})$, the mass center of the sphere $O_{t_0}$ in the two-dimensional on-line coordinate is $r_{min}^*{(t_0)}$ and the radius $r_i{(t_0)}$ is $\lVert v_i{(t_0)}\rVert$. There are only two tangent lines from $c_R {(t_0-1)}$, the angles $\alpha_i^1{(t_0)}$ and $\alpha_i^2{(t_0)}$ are angles of the tangent lines measured from $X^{'}$-axis in counter-clockwise direction. An illustration is given in Fig. 3.4.\par 
	\begin{figure}[h]
		\centering
		\includegraphics[width=4.2in]{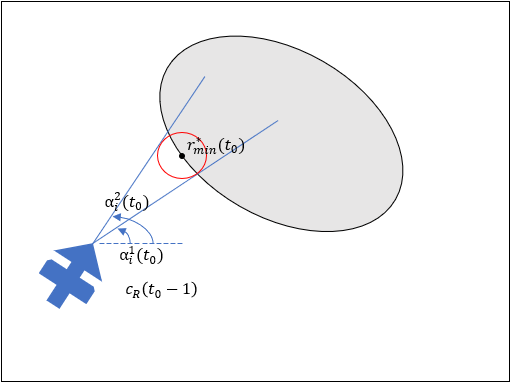}
		\caption{The illustration of how to choose $r_{min}^* {(t_0)}$.}
	\end{figure}
%
We define a $\gamma$,
	\begin{equation}
	\gamma= \sin^{-1}\frac{\lVert v_i^{*}{(t_0)} \rVert}{d_{min}^{*}{(t_0)}}.
	\end{equation}
Here, $v_i^{*}{(t_0)}$ is the transferred velocity in the on-line coordinate $(X^{'},Y^{'})$, $d_{min}^{*}{(t_0)}$ is the minimum distance between $c_R {(t_0-1)}$ and $r_{min}^*{(t_0)}$ in the on-line coordinate $(X^{'},Y^{'})$. Then, we assume a pair of vectors $l^1_i{(t_0)}$ and $l^2_i{(t_0)}$, given by
	\begin{equation}
	l^1_i(t_0):=\frac{\lVert v_i^{*}{(t_0)}\rVert}{\tan\gamma}\left[\cos{\left(\alpha_i^1{(t_0)}\right)}, \sin{(\alpha_i^1{(t_0)})}\right],
	\end{equation}
	\begin{equation}
	l^2_i{(t_0)}:=\frac{\lVert v_i^{*}{(t_0)}\rVert}{\tan\gamma}\left[\cos{\left(\alpha_i^2{(t_0)}\right)}, \sin{(\alpha_i^2{(t_0)})}\right].
	\end{equation}
	An index $J{(t_0)}$ is given by comparing the absolute value of $\beta_i^1 {(t_0)}$ and $\beta_i^2{(t_0)}$. Specifically, $J{(t_0)}=1$ when $\left|l_i^1 {(t_0)}\right|  \leq \left|l_i^2 (t_0)\right|$, $J{(t_0)}=2$ when $\left|l_i^1 {(t_0)}\right| > \left|l_i^2 {(t_0)}\right|$.
\par
Now, we introduce the following function:
	\begin{equation}
	f\left(l_J{(t_0)}, c_R^{*}{(t_0-1)})\right)=
	\begin{cases}
	0, & \beta_J{(t_0)}=0 \\
	1, & 0 < \beta_J{(t_0)}\leq \pi \\
	-1,& -\pi < \beta_J{(t_0)}<0 \\
	\end{cases}
	\end{equation}
$c_R^{*} {(t_0-1)}$ is the robot coordinate position for time point $t_0-1$ in the on-line coordinate $(X^{'},Y^{'})$.
\par
The motion of the robot in the on-line coordinate are expressed as following:
	\begin{equation}
	V_R^{*}(t) = \lVert l_J(t) \rVert,
	\end{equation}
	\begin{equation}
	u_R^{*}(t) = -U_{max}f\left(l_J(t), c_R^{*}{(t-1)} \right).
	\end{equation}
\par
We have to transfer it back to the three-dimensional geometric coordinate at the end of each circle.
In short, the robot starts to move with the target approaching mode and switches to the obstacle avoidance mode when the distance between the robot and the nearest obstacle is shorter than \emph{D}, which is a constant. When the robot is away from the obstacle (the distance between the robot and the nearest obstacle is shorter than \emph{D}), the robot would switch back to the target approaching mode. Switching between these two modes until the robot reaches the ending point \emph{\textbf{E}}. So far, we have introduced our collision-free navigation algorithm for flying robots which can be applied to outdoor environments with numbers of moving obstacles in the space.
\par
\section{Computer Simulation}
In this part, we will show the performance of our proposed algorithm with the simulation results implemented with the Matlab. The aim for this part is to show that our proposed algorithm is a viable one for navigating flying robots in outdoor environments while avoiding obstacles on the way.
\par
First, we examine the obstacle avoidance mode (\emph{\textbf{M2}}) in on-line coordinate. Fig. 3.5 shows a single circle for time point $t_0$ to avoid the moving obstacle. We can also get that the predicting motion of the obstacle is: $v_i^{*}{(t_0)}=[0.5, 0.5], u_i^{*}{(t_0)}=0.$
\par
	\begin{figure}[h!]
		\centering
		\includegraphics[width=4.2in]{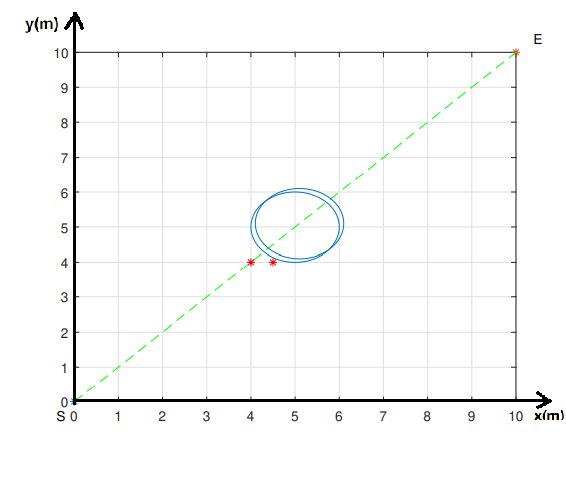}
		\caption{The planar part of obstacle avoidance mode.}
	\end{figure}
Moreover, we examine a whole process of navigating a robot to reach its aim while avoiding an obstacle on its way. The illustration is shown in Fig. 3.6. Fig. 3.6(a) and (b) show the trajectory from different angle of view. Obviously, the robot moving from the starting point \emph{\textbf{S}} to the ending point \emph{\textbf{E}}, and it avoids an moving obstacle on the way. Fig. 3.7(a) and (b) are the graphics of the obstacle motion $v_i{(t)}$ and $u_i{(t)}$. The $X$-axis of the graphics is \emph{time {t(s)}}, the $Y$-axis for Fig. 3.7(a) are three different axis components of $v_{obs}{(t)}$ and scalar $V_{obs}{(t)}$, also the $Y$-axis for Fig. 3.7(b) are component values of $u_{obs}{(t)}$ and scalar $U_{obs}{(t)}$. We can get from the graphics, (3.9) and (3.10) are satisfied. Obviously, our proposed algorithm can successfully reach the goal.
    \begin{figure}[h!]
	\centering
	\begin{subfigure}[b]{0.5\textwidth}
		\centering
		\includegraphics[width=2.9in]{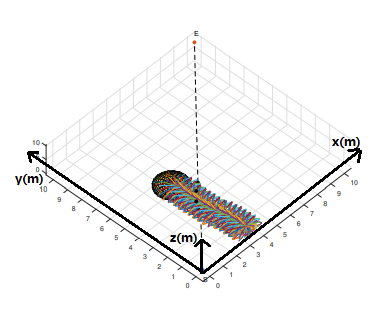}
		\caption*{(a)}
	\end{subfigure}
	\begin{subfigure}[b]{0.4\textwidth}
		\centering
		\includegraphics[width=2.9in]{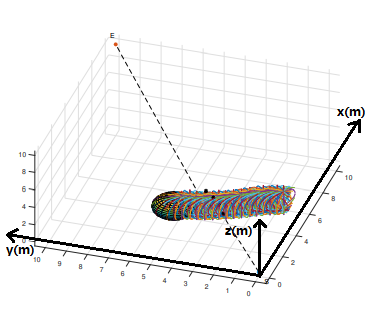} 
		\caption*{(b)}
	\end{subfigure}
	\caption{The simulation result of the trajectory.}
    \end{figure}
    \begin{figure}[h!]
	    \centering
	\begin{subfigure}[b]{0.5\textwidth}
		\centering
		\includegraphics[width=2.9in]{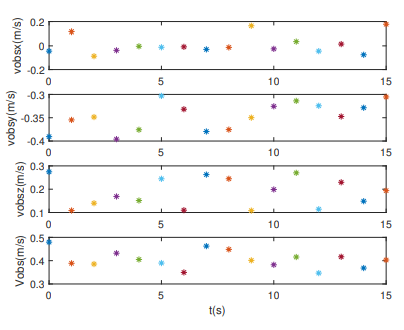}
		\captionsetup{}
		\caption*{(a)}
	\end{subfigure}
	\begin{subfigure}[b]{0.4\textwidth}
		\centering
		\includegraphics[width=2.9in]{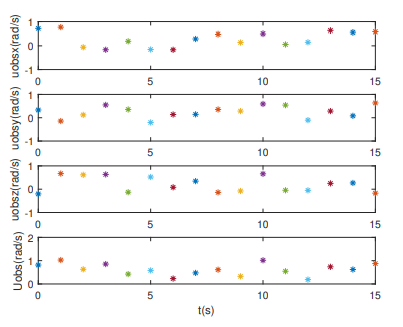} 
		\caption*{(b)}
	\end{subfigure}
	\caption{The motion of the obstacle.}
    \end{figure}
\section{Conclusion}
We proposed a collision-free navigation approach for flying robots in this chapter. It can be applied to three-dimensional environments with a number of dynamic obstacles. For the obstacles, we do not assume that all the obstacles are convex, and their motion are variable with some constraints. By means of a statistic model, we could forecast the motion of the obstacle. When proposing the obstacle avoidance strategy, a core is to combine earth coordinate system with an on-line coordinate system. Trajectory for UAVs could be planning successfully by our proposed algorithm.
\par

%% file: chapter/Chapter5.tex
\chapter{A Navigation Algorithm for UUVs to Avoid Shape Changeable Obstacles}
\label{C5:chapter5}
In the preceding two chapters, we have proposed obstacle avoidance strategies for motion changeable obstacles. The two strategies are aiming at 2D environments and 3D environments, respectively. 
From this chapter, we will try to solve the problem of avoiding shape changeable obstacles. 
In this chapter, we propose a robot navigating method in order to guide undersea unmanned vehicles (UUVs) avoiding shape changeable obstacles for the 2D undersea environments. 
One of the biggest challenge is to record and predict the deformation of the obstacles shape. 
We introduce an \emph{AMAPS} to solve the problem. 
Then, a switching mode control law will be deployed to navigate the UUV to its destination.
%
%
%
\par
The results of this chapter were originally published in the conference papers: 
\textbf{Y.~{Zhang}} and J.~{Zhang}, `` A Navigation Algorithm for UUVs to Avoid Shape Changeable Obstacles,'' in {\em 2022 the 14th international Conference on Computer and Automation Engineering (ICCAE)}, Brisbane, Australia, 25-27 Mar. 2022.
\section{Introduction}
With the development of navigating the moving robots on the ground, numbers of navigation strategy has been proposed \cite{3.1}. 
In recent years, researchers in the field of robots controlling put their effort into navigating robots in the underwater environment \cite{5.7}. Safe driving of underwater unmanned vehicle (UUV) is one of the most important task. Thus, numbers of algorithms are appeared, e.g. \cite{5.8}, \cite{5.9}. D. Li et al. proposed an obstacle avoidance algorithm for snake-like robots based on immersed boundary-lattice Boltzmann method (IB-LBM) and improved artificial potential field (APF), which can be applied to two-dimensional environment\cite{5.10}. From the classical one \cite{5.11} to the evolved methods\cite{5.12} \cite{5.13}, they are all based on artificial potential field (APF). Their mathematical description is simple, physical depiction is explicit, but, there are several limitations. To be specific, the application scenario always be an optimal one, some constraints in the practical environment cannot be satisfied. The computational efficiency of these algorithms is not ideal.
\par
Therefore, proposing an applicable solution to actual situations while the computational efficiency is acceptable is still a tough task. The navigation algorithm we propose is originated from observing the Nature, it is a biologically inspired algorithm. Similar researches have been conducted, however, they are focusing on avoiding moving obstacles on the ground \cite{3.14} \cite{yang.1}. For the undersea environment, one of the most common situation is the obstacles are shape changeable. For example, the fishes, the sea weeds, and others. Matveev et al. took the deformation of the obstacles into consideration \cite{5.15}. However, it could not be implemented in the underwater environments, because of the restriction of the sensor undersea. Hence, we choose sonar-based sensors to detect the obstacles, as the radar-based sensors and visual-based sensor are not applicable in underwater environments. The sonar-based ranging sensor could provide the shortest distance between the robot and the nearest obstacle, which is useful information for proposing the algorithm. For the deformation of obstacles, we introduce an \emph{AMAPS} to record and predict it. Thus, a safely navigating algorithm for UUVs could be implemented.
%
\par
The remainder of the chapter is organized as follows. Section 4.2 describes the underwater environment and the unmanned underwater vehicle (UUV) with a mathematical model, also gives some necessary constraints. Section 4.3 demonstrates how to plan trajectory for the UUV while avoiding shape changeable obstacles applied our proposed navigation algorithm. Section 4.4 illustrates the performance of our proposed algorithm in an example scene, which is achieved by Matlab. Section 4.5 draws a brief conclusion for the whole chapter.
\par
	
\section{System Description}
In order to describe the problem clearly, here we assume an unmanned underwater vehicle (UUV) in a planar environment. An illustration is given in Fig. 4.1. 
The mathematical model of the vehicle can be described as follows:
	\begin{equation}
	c_{vehicle}{(t)}:=\left[x(t),y(t)\right],
	\end{equation}
	\begin{equation}
	\dot{c_{vehicle}}{(t)}=V_{vehicle}{(t)}\cdot a_{vehicle}{(t)}.
	\end{equation}
Here, (1) is the two-dimensional vector of the UUV's Cartesian coordinates, (4.2) gives the motion of the UUV. $V_{vehicle}{(t)}$ is the linear velocity of the UUV and $a_{vehicle}{(t)}$ shows its orientation. 
	\begin{figure}[h!]
		\centering
    	\includegraphics[width=4.2in]{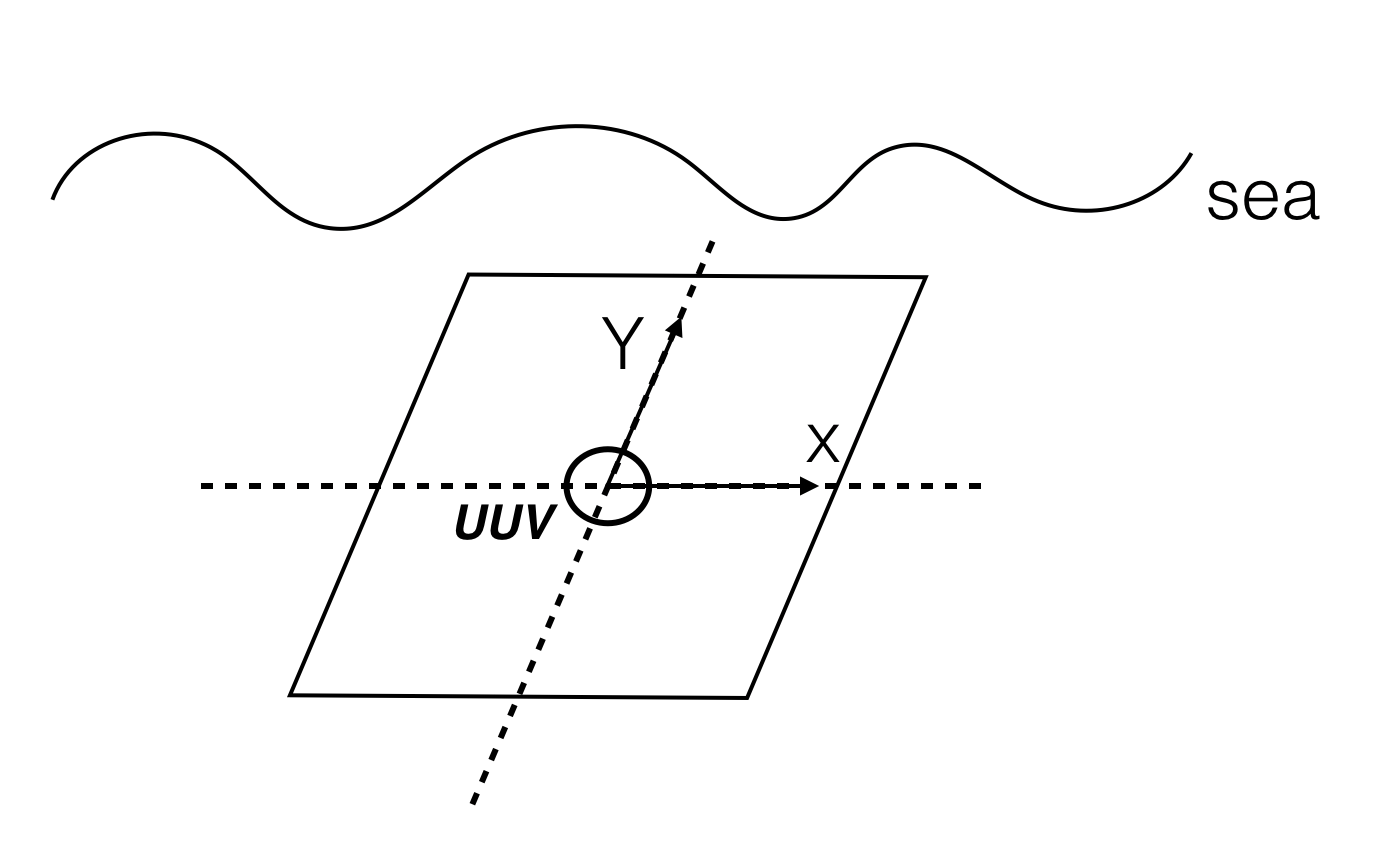}
    	\caption{The illustration of the UUV.}
	\end{figure}  
We can get:
    \begin{equation}
	{V_{vehicle}}{(t)}^2=V_{vehicleX}{(t)}^2+V_{vehicleY}{(t)}^2.
	\end{equation}
The following constraints hold:
	\begin{equation}
	V_{vehicleX}{(t)}\in\left[0, V_{max}\right], 
	\end{equation}
	\begin{equation}
	V_{vehicleY}{(t)}\in\left[0, V_{max}\right],
	\end{equation}
	\begin{equation}
	{a_{vehicle}{(t)}}\in\left[-U_{max}, U_{max}\right].
	\end{equation}
for all $\emph{t}$. Moreover, where $V_{vehicle}{(t)} \in R$, $a_{vehicle}{(t)} \in R^2$for all $\emph{t}$. The constants $V_{max}$ and $U_{max}$ are given, which depends on the designed parameter of the UUV. 
We can get:
	\begin{equation}
	V_{vehicle}{(t)}\in\left[0, \sqrt{2}V_{max}\right].
	\end{equation}
\par
For the underwater environment with shape changeable obstacles, we set an scene as an example. In the described scene, there are some necessary elements, initial point, end point and some obstacles. We named the initial point \emph{\textbf{I}} and the end point \emph{\textbf{E}}. They are stationary in the scene we set, also their location are known in advance.
	\begin{equation}
	I=\left[I_{x},I_{y}\right], 
	\end{equation}
	\begin{equation}
    E=\left[E_{x},E_{y}\right].
	\end{equation}
\par
The obstacles are named $\Diamond_1(t), \Diamond_2(t), ..., \Diamond_i(t)$. They are several random shapes obstacles in the plane. Here, we give some constraints to their shapes. Obstacles are solid sets and the boundary should be smooth closed curve. Fig. 4.2(a) is an satisfactory obstacle. However, if the minimum distance between the vehicle and the obstacle is much less than the maximum distance, our proposed algorithm may not solve the problem, because the inner space of the obstacle cannot be detected, and the vehicle may go into the obstacle and make itself get lost. A more extreme case is shown in Fig. 4.2(b), the minimum distance between the vehicle and the obstacle is denoted as \emph{\textbf{a}}, and the maximum distance between the vehicle and the obstacle is denoted as \emph{\textbf{b}}. Obviously, \emph{\textbf{a}} is much less than \emph{\textbf{b}}, it cannot satisfied the requirements. It is clearly that, the entry of the obstacle' s inner space is smaller than the width of the vehicle, the collision between the vehicle and the obstacle could not be avoided. Moreover, collisions between obstacles are not considered.
\par
    \begin{figure}[h!]
	\centering
	\begin{subfigure}[b]{0.5\textwidth}
		\centering
		\includegraphics[width=2.9in]{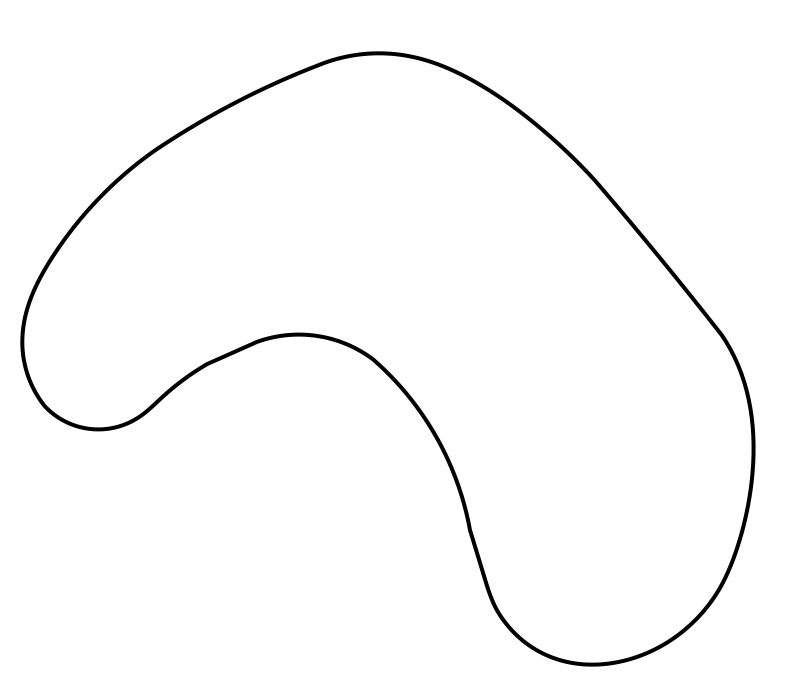}
		\caption*{(a)}
	\end{subfigure}
	\begin{subfigure}[b]{0.4\textwidth}
		\centering
		\includegraphics[width=2.9in]{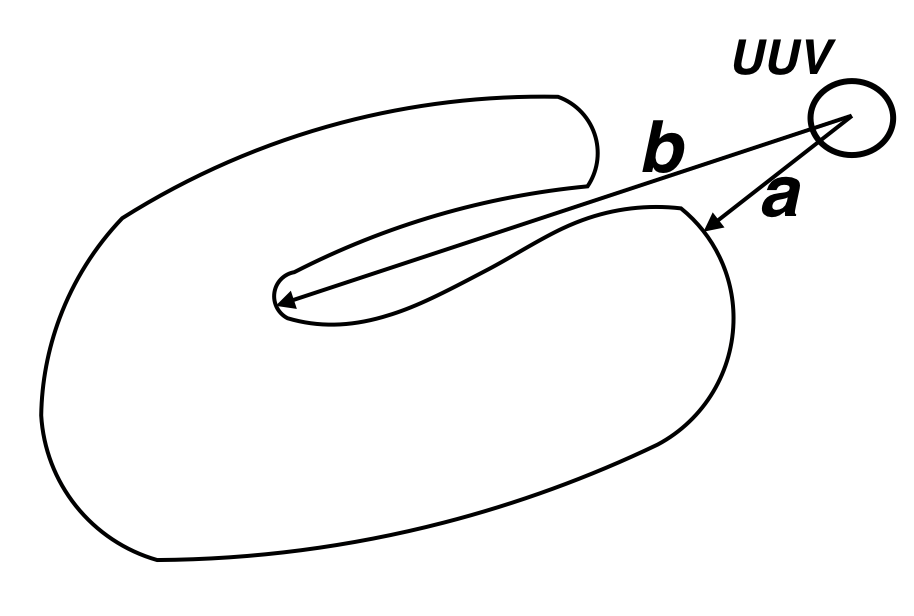} 
		\caption*{(b)}
	\end{subfigure}
	\caption{The illustration of the shape of the obstacles.}
    \end{figure}
In practice to deal with uncertainties and changes in shapes of obstacles, the proposed navigation laws may be supplemented by sliding-mode or switched controllers, see e.g. \cite{R6} \cite{R7} \cite{R8} \cite{R9}.
Our proposed algorithm is for the purpose of navigating the vehicle moving from the initial point \emph{\textbf{I}} and stop when it reaches the end point \emph{\textbf{E}}, and detours when meeting obstacles. We use a matrix to express the distance between the vehicle and the obstacle $\Diamond_i$ in its viewing angle, which is denoted as $d_i(t)$ for time t. The minimum distance and the maximum distance are $d_{imin}(t)$ and $d_{imax}(t)$. To be clear, only the part without covering could be detected. Take Fig. 4.3 as an example, part marked with heavy line can be detected, part marked with imaginary one is excluded. We suppose that $r_{imin}{(t)}$ and $r_{imax}{(t)}$ are the point on the surface of the obstacle $\Diamond_i$ which make the minimum value $d_{imin}{(t)}$ and maximum value $d_{imax}{(t)}$ achieved. \par
	\begin{figure}[h!]
		\centering
    	\includegraphics[width=4.2in]{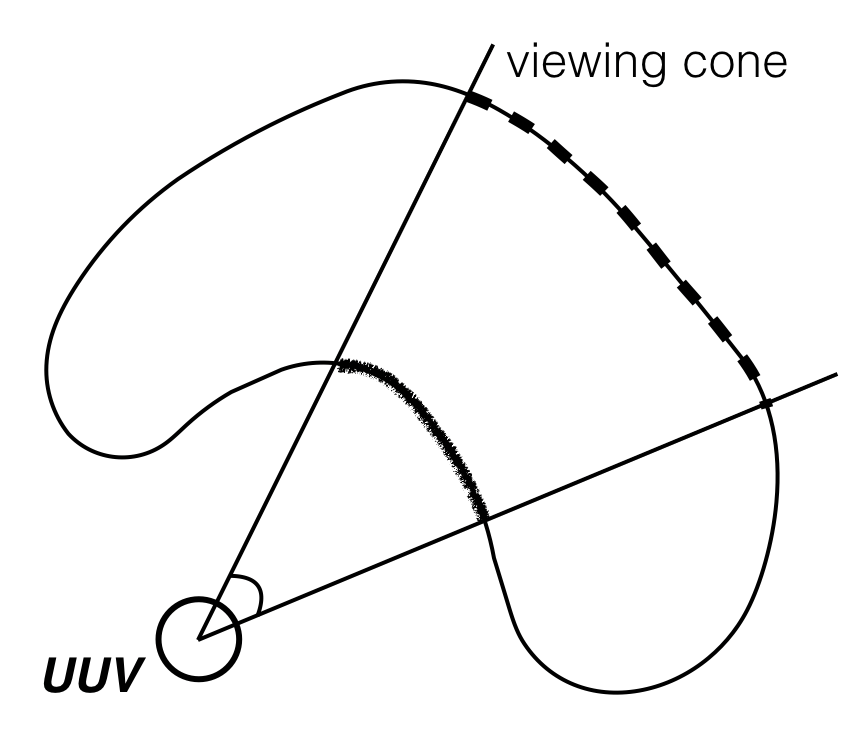}
    	\caption{The illustration of the robot viewing cone.}
	\end{figure}
Here, we describe the deformation of the obstacles. We introduce an \emph{As Minimum As Possible Square(AMAPS)}  here, which is similar with the \emph{pixel}. The \emph{AMAPS} is direction consistent with the coordinate axis. Fig. 4.4(a) is an example. Part of an obstacle is shown with amplifier, we cut the obstacle with the \emph{AMAPS}. We suppose that the mass centre of the obstacle $\Diamond_i(t)$ is known and never change, the area of it is a fixed one. The changing degree of the obstacle should comply the general rules, should not change suddenly. To be specific, anything cannot change from square to round in one second in natural world. With times of attempts, we get the obstacle deformation with probability distribution. Fig. 4.4(b) is part of the illustration. Within the limited times, the darker the color, the more times the \emph{AMAPS} was covered by the obstacle.\par
    \begin{figure}[h!]
	\centering
	\begin{subfigure}[b]{0.5\textwidth}
		\centering
		\includegraphics[width=2.9in]{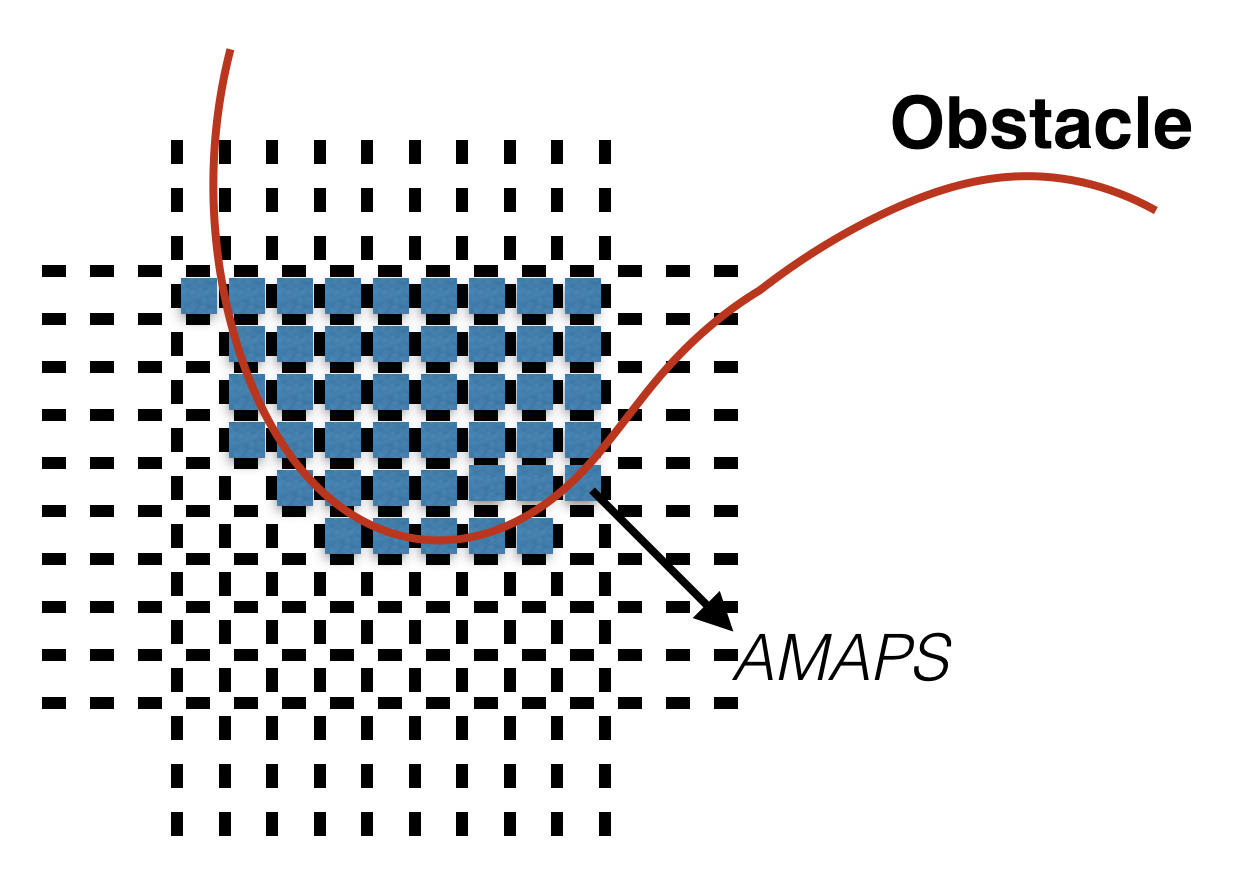}
		\caption*{(a)}
	\end{subfigure}
	\begin{subfigure}[b]{0.4\textwidth}
		\centering
		\includegraphics[width=2.9in]{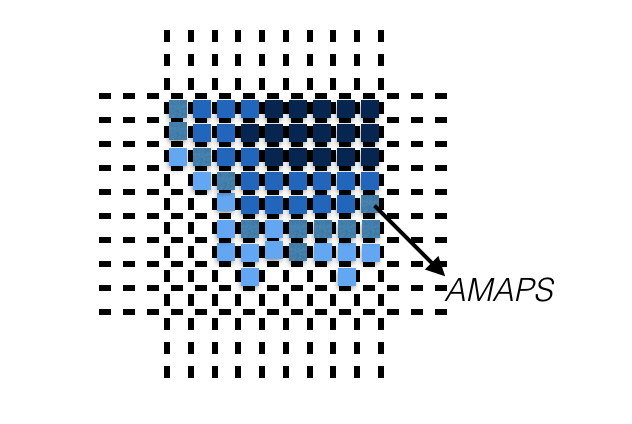} 
		\caption*{(b)}
	\end{subfigure}
	\caption{Cutting the obstacle with \emph{AMAPS}}
    \end{figure}
\section{Navigation Algorithm}
We give the detailed description of our proposed algorithm based on the model built in last section. In the scene, there is an initial point \emph{\textbf{I}} and an end point \emph{\textbf{E}}, also some shape changeable obstacles. The mission of the Undersea Unmanned Vehicle (UUV) is departing from the initial point \emph{\textbf{I}} to the end point \emph{\textbf{E}} while avoiding those shape changeable obstacles on its way. 
\par
Our proposed algorithm is a switching between two strategies, target approaching strategy and obstacle avoiding strategy, respectively.
For the target approaching strategy, the UUV goes straight to the end point \emph{\textbf{E}} with maximum speed without any rotation. 
For the obstacle avoiding strategy, there are a series of decisions need to be made. 
\par
First of all, the UUV need to confirm the nearest point on the surface of the obstacle in its viewing cone. As we introduced \emph{As Minimum As Possible Square (AMAPS)} before, which is a smaller enough square for the obstacles, we approximately hold that the centre of the square is the point on the surface of obstacle, which denoted as $r_{imin}(t)$. 
\par
Then, we would focus on the $r_{imin}(t)$, as it is the nearest point need to be avoided. The UUV will avoid collision with the obstacle if it can bypass the $r_{imin}(t)$.
We suppose that there is a line between the vehicle and the end point \emph{\textbf{E}}, and it can rotate around the vehicle counter-clockwise. If the line rotates ${\left[0, {\frac{\pi}{2}}\right)}$ when meets the mass centre of the vehicle, we denotes it as '+'. If the line rotates ${\left( {-\frac{\pi}{2}}, 0\right)}$ when meets the mass centre of the vehicle, we denotes it as '-'. It is a tag for avoiding direction. 
\par
Next, we define a forecast circle $O_{t}$, the centre of the forecast circle is $r_{imin}{(t)}$ and the radius is $R_i(t)$,
	\begin{equation}
    R_i(t)=\frac{1}{count(t)}\sum_{k=1}^{count(t)}{\left\|{r_{imin}(t-k)}-r_{imin}(t-k+1)\right\|}.\\
    \end{equation}
	 For count(t), 
	 \begin{equation}
	 \left\{
	 \begin{array}{lr}
	 count(t)=count(t)+1, & d_{imin}(t) \leq L\\
	 count(t)=0. & d_{imin}(t) > L
	 \end{array}
	 \right.
	 \end{equation}
	 \par
There are two tangent lines for the forecast circle $O_{t_0}$ from $c_{vehicle} {(t)}$, which denoted as $l_+(t)$ and $l_-(t)$. We set a line between the $c_{vehicle}$ and the centre of the forecast circle $r_{imin}{(t)}$, the tangent line it meets is $l_+(t)$ when the tangent line rotates around the vehicle counter-clockwise, and the other one is $l_-{(t)}$. The two intersections are named $p_+(t)$ and $p_-(t)$, respectively. An illustration is shown in Fig. 4.5. The vehicle will move to $p_+(t)$ or $p_-(t)$ with the decision made by last step.
\par
	\begin{figure}[h!]
		\centering
    	\includegraphics[width=4.2in]{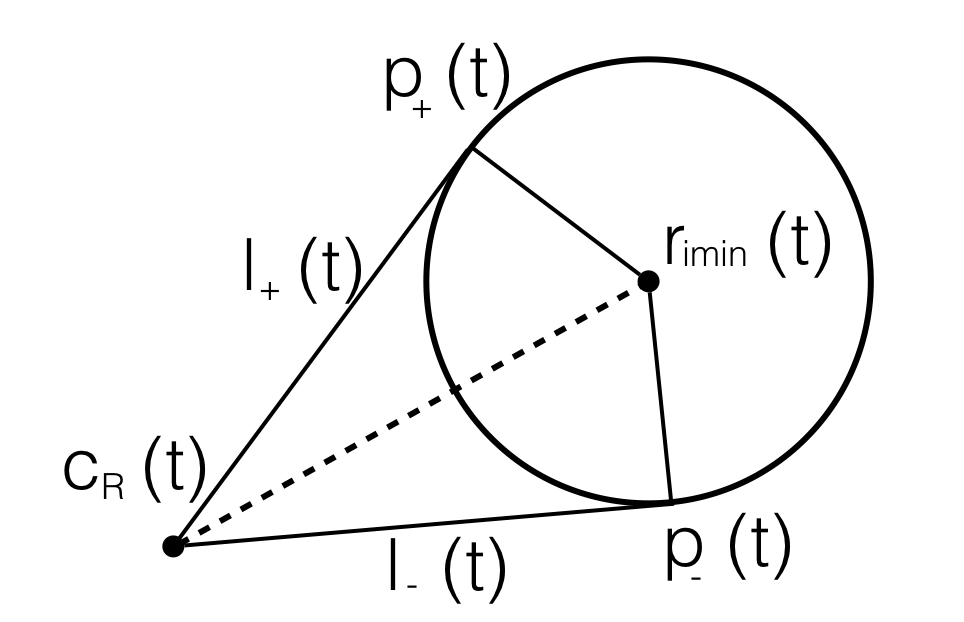}
    	\caption{The illustration of the forecast circle.}
	\end{figure}  	 
    \begin{equation}
	\left\{
	\begin{array}{lr}
	c_{vehicle}(t+1)=p_+(t), & if \quad tag=+\\
	c_{vehicle}(t+1)=p_-(t). & if \quad tag=-
	\end{array}
	\right.
	\end{equation}
We introduce a \emph{L} here, the aim of setting the \emph{L} is to make decision of switching between the two navigation strategy in order to avoid collision. As shown in Fig. 4.6, the program flow diagram, when the minimum distance between the vehicle and the obstacle $d_{imin}(t)$ is larger than \emph{L}, we suppose that there is no safety issues need to be worried about, the UUV will choose the target approaching strategy. When the minimum distance between the vehicle and the obstacle $d_{imin}(t)$ is smaller than \emph{L}, the UUV should take action to avoid the obstacle, in other word, switching to obstacle avoiding strategy. When the vehicle is away from the obstacle, switching to target approaching strategy. Switching between these two strategies until the UUV reaches its goal. So far, we have introduced our collision-free navigation algorithm for undersea unmanned vehicles.
\par
	\begin{figure}[h!]
		\centering
    	\includegraphics[width=4.2in]{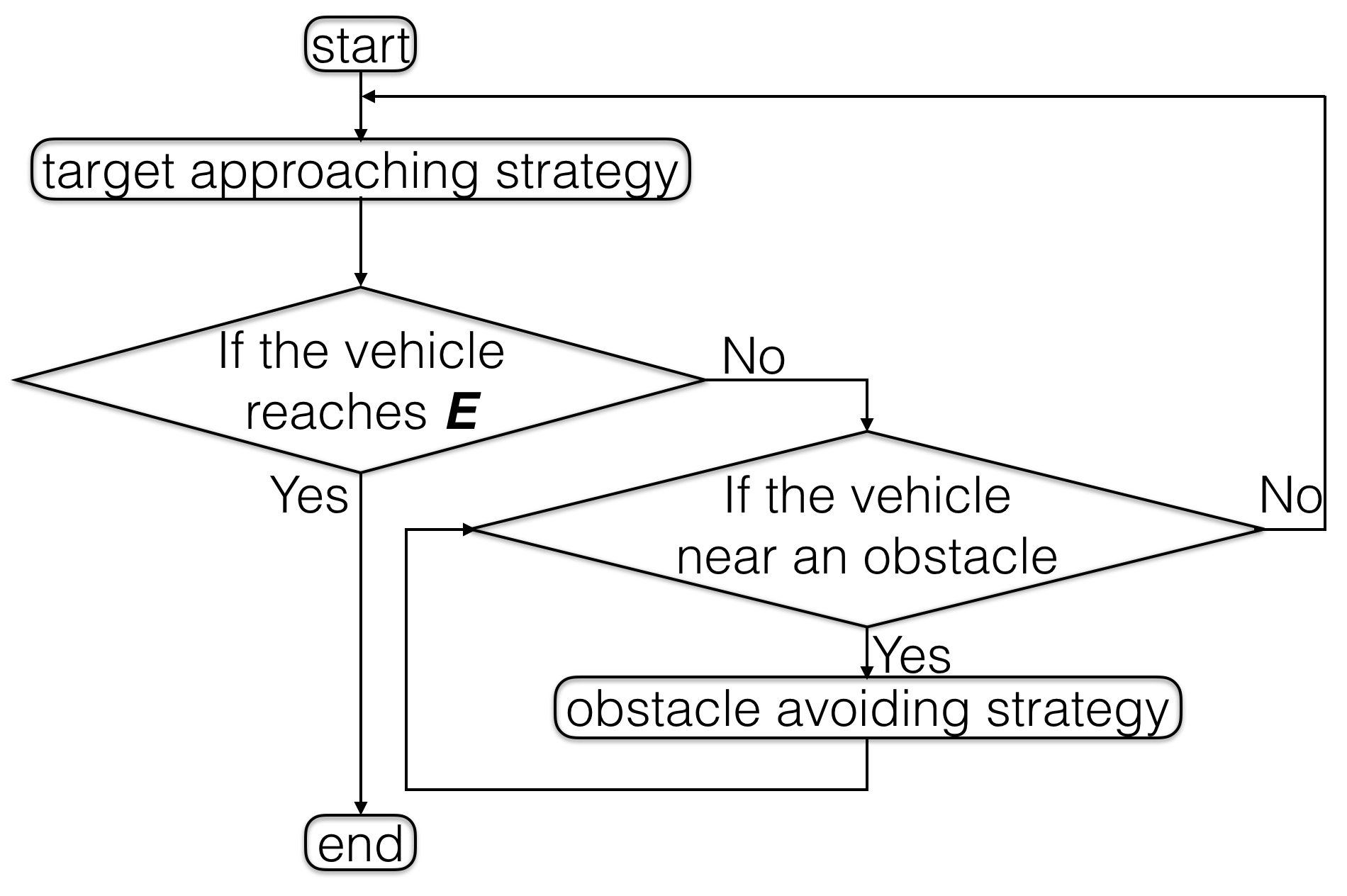}
    	\caption{Program flow diagram of proposed algorithm.}
	\end{figure}
\section{Computer Simulation}
We verify the feasibility of our proposed navigation algorithm with simulation experiments. The simulation is implemented by Matlab. We will show the simulation results of each step and a whole navigating process rigorously.  
\par
We give an example scene for testing the given strategy. There is an initial point \emph{\textbf{I}}, an end point \emph{\textbf{E}} and an obstacle in the planar, see Fig. 4.7(a). The coordinate of the initial point and the end point are [0, 0] and [10, 10], respectively. 
For the target approaching strategy, the UUV starts to move from the initial point for time $t=0$. From $t = 0$ to $t = 4$, which are shown from Fig. 4.7(a) to Fig. 4.7(d), the shape of the obstacle is changing randomly, while the UUV goes directly to its aim, and distance between the UUV and the obstacle is becoming shorter. Even so, the distance is still longer than \emph{L}. 
For the obstacle avoidance strategy, when the distance between the UUV and the obstacle is less than \emph{L}, the UUV will take action to avoid the obstacle. Fig. 4.8 shows the whole decision-making process.
\par
\par
Then, we examine a whole process of navigating an UUV to reach its aim while avoiding an obstacle on its way. The simulation result is shown in Fig. 4.9. In the planar, the initial point \emph{\textbf{I}} [0, 0], the ending point \emph{\textbf{E}} [10, 10]. As the deformation of the obstacle is randomly, it would be messy to display in the same image. So, we use points marked with green star to show $r_{imin}(t)$, the nearest point need to be avoided for the UUV. The UUV moving from the initial point \emph{\textbf{I}} to the ending point \emph{\textbf{E}}, and it avoids a moving obstacle on the way with our proposed algorithm. The trajectory of the UUV is marked with blue line. The points marked with blue star are the vehicle' s waypoint corresponding to the obstacle avoiding strategy. Thus, our proposed algorithm can navigate the UUV safely while avoiding shape changeable obstacles in 2D underwater environments.
\par
    \begin{figure}[h!]
	\centering
	\begin{subfigure}[b]{0.5\textwidth}
		\centering
		\includegraphics[width=2.9in]{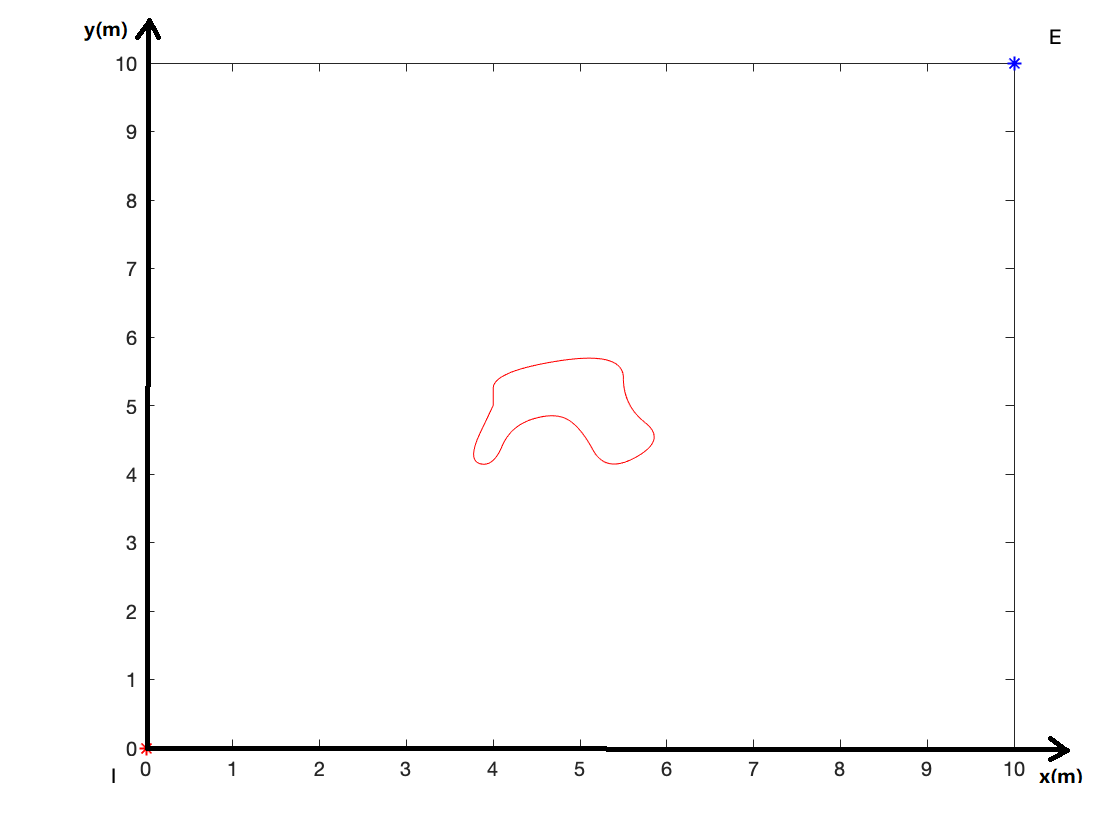}
		\caption*{(a)}
	\end{subfigure}
	\begin{subfigure}[b]{0.4\textwidth}
		\centering
		\includegraphics[width=2.9in]{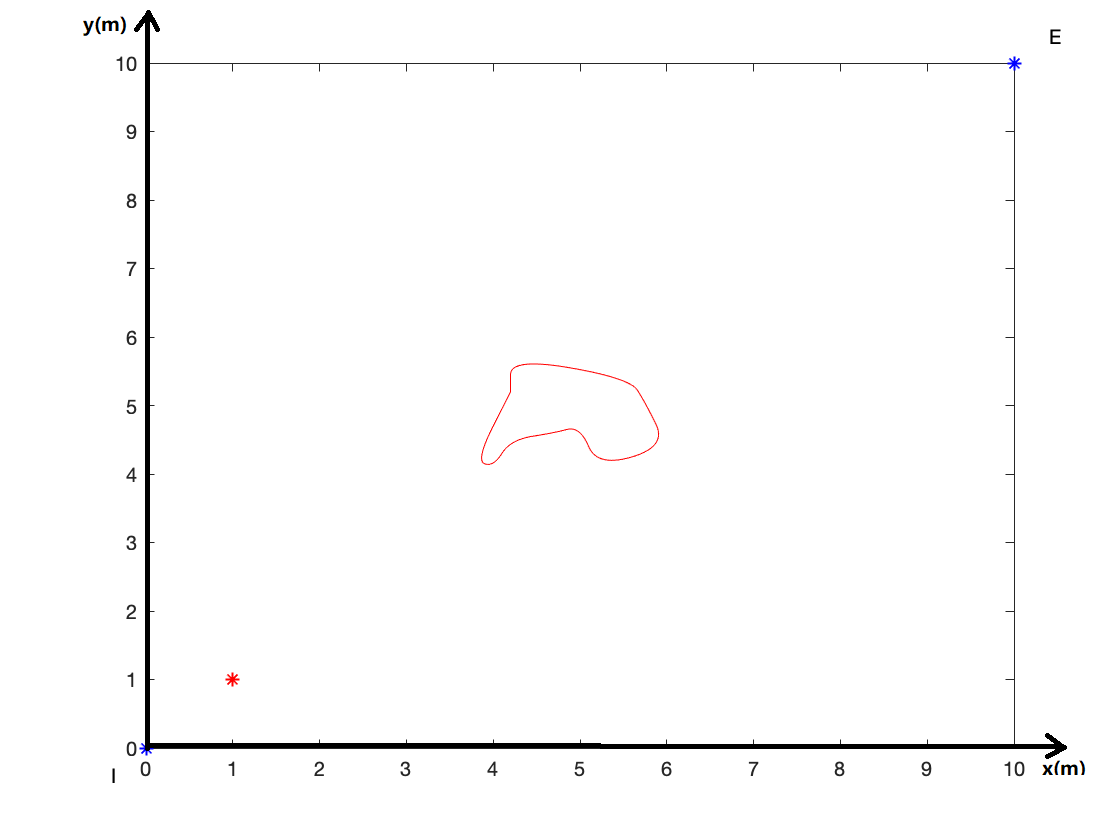} 
		\caption*{(b)}
	\end{subfigure}
	\begin{subfigure}[b]{0.5\textwidth}
		\centering
		\includegraphics[width=2.9in]{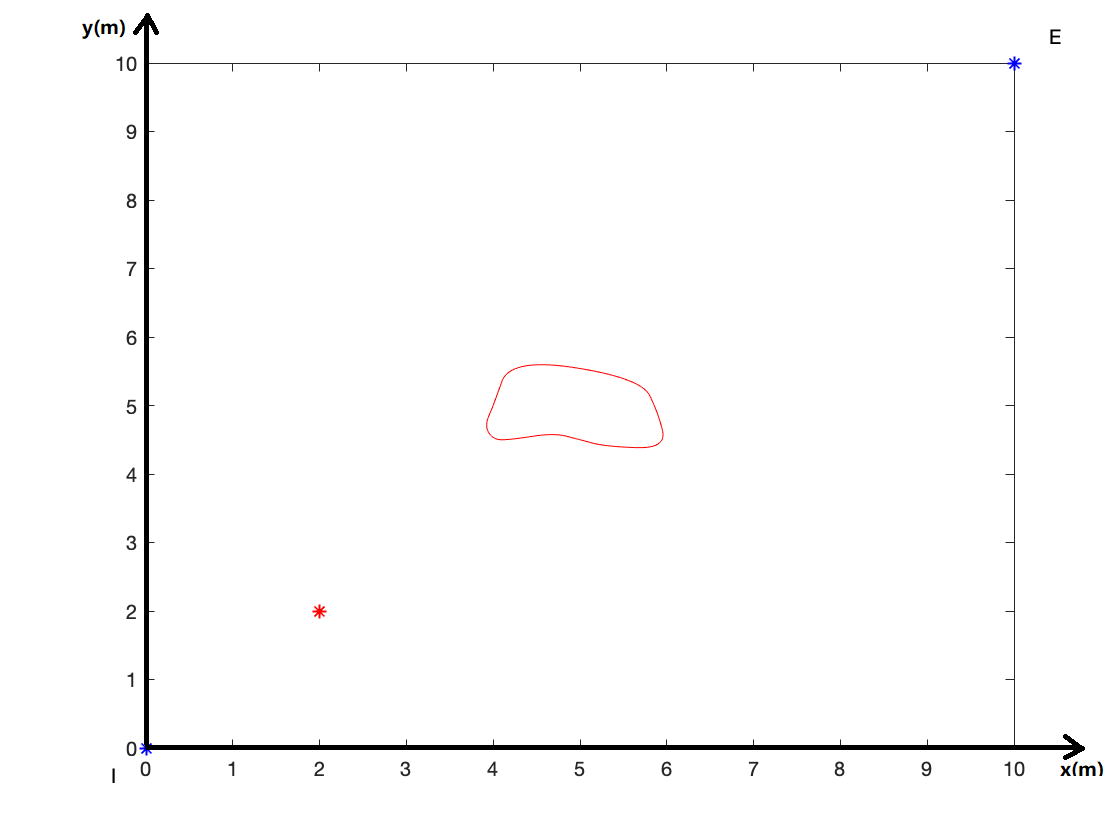}
		\caption*{(c)}
	\end{subfigure}
	\begin{subfigure}[b]{0.4\textwidth}
		\centering
		\includegraphics[width=2.9in]{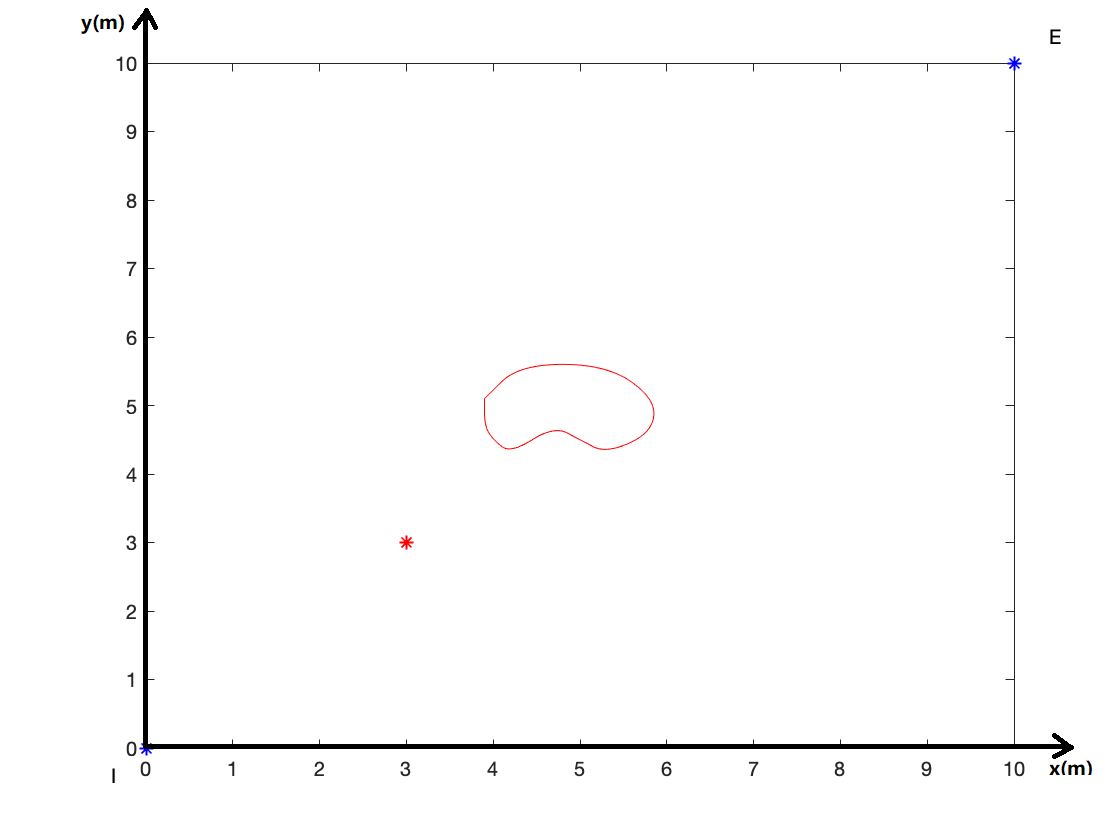} 
		\caption*{(d)}
	\end{subfigure}
	\caption{Simulation result from time $t = 0$ to $t = 4$.}
    \end{figure}
	\begin{figure}[h!]
		\centering
    	\includegraphics[width=4.2in]{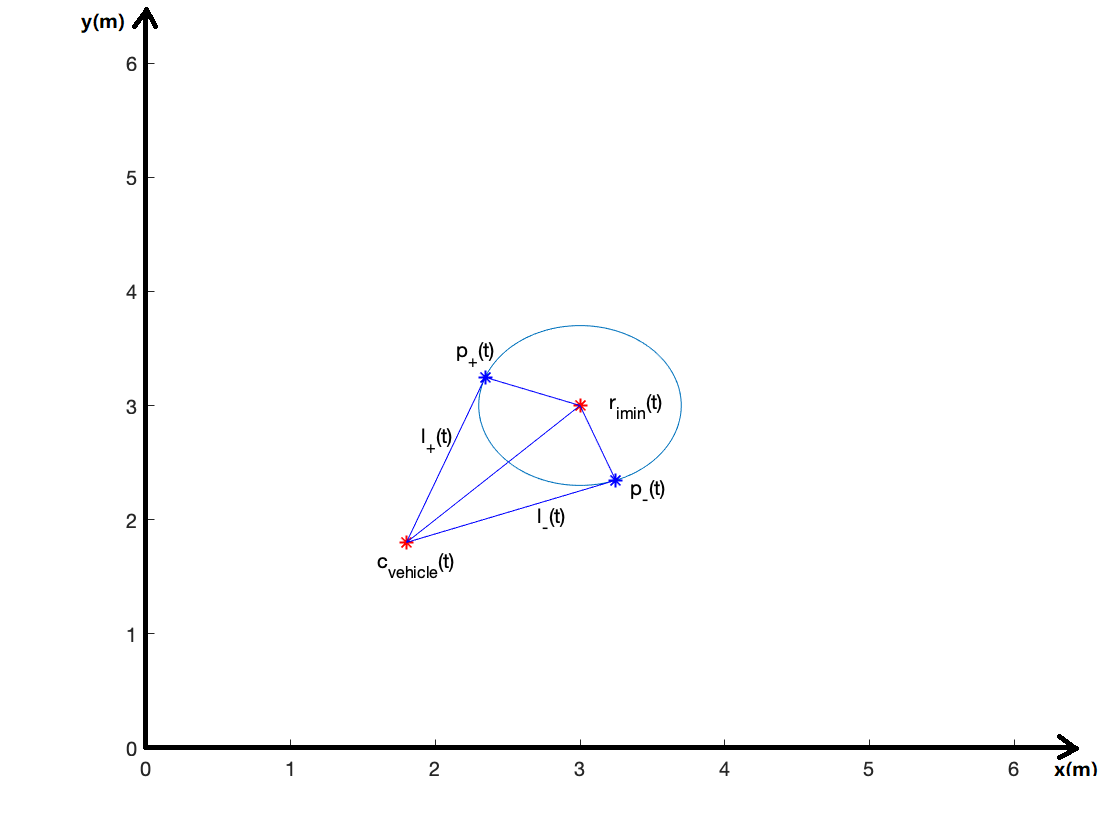}
    	\caption{The decision making process of the UUV for time $t$.}
	\end{figure}
	\begin{figure}[h!]
		\centering
    	\includegraphics[width=4.2in]{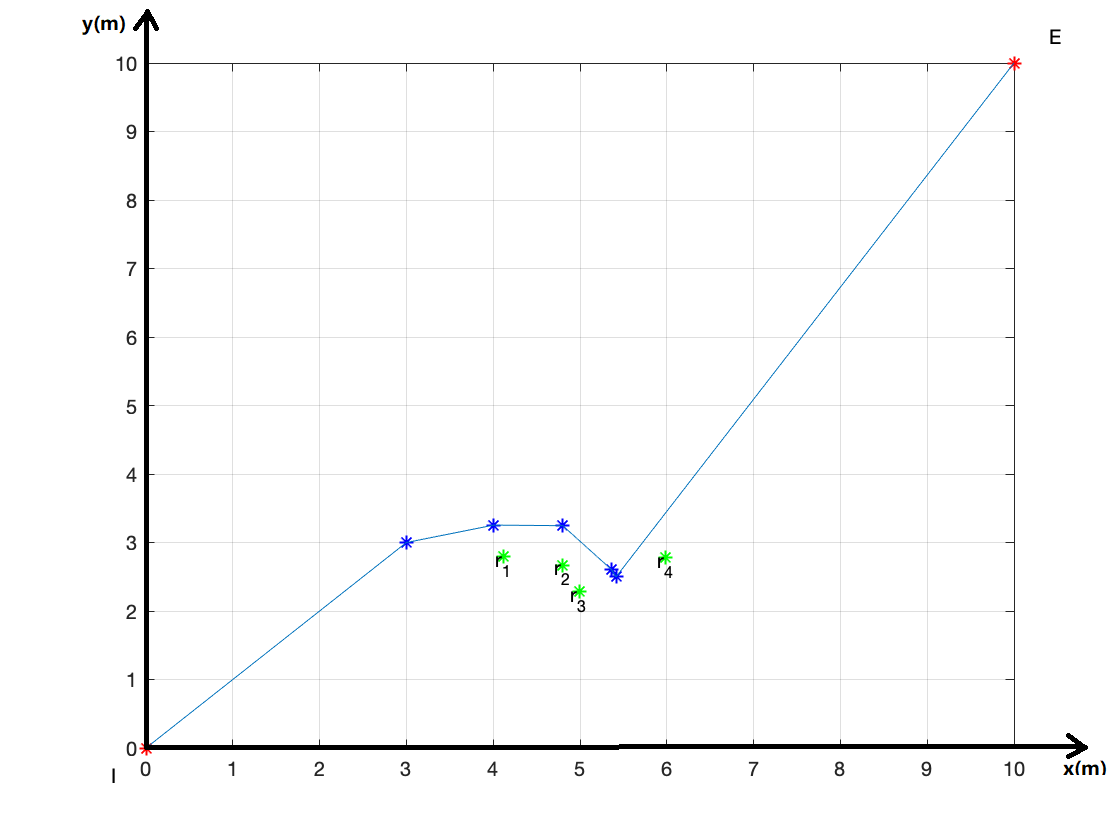}
    	\caption{The trajectory of the UUV.}
	\end{figure}
\section{Conclusion}
In this chapter, we proposed a navigation algorithm for Undersea Unmanned Vehicles (UUVs). The deformation of obstacles has been taking into consideration, and we introduced an \emph{AMAPS} to record and predict the deformation. As obtaining the information of the point to avoid, obstacle avoidance strategy could be implemented. Thus, collision free navigation can be achieved. As the proposed algorithm can be applied to 2D underwater environments, one future research direction is to extend the proposed method to 3D environments.

%% file: chapter/Chapter6.tex
\chapter{An Obstacle Avoidance Method for Sonar-based Robots Avoiding Shape Changeable Obstacles}
\label{C6:chapter6}
As we focusing on the changeable of the obstacles shape, we proposed an obstacle avoidance strategy for 2D underwater environments in the last chapter.
%
%
In this chapter, we will put efforts into the 3D environments. Compared with the planar environments, the 3D environments is a more complicated situation. It is more difficult for system modelling. Even, sensors could be used for underwater environment is limited. As we choose a group pf sonar-based sensors, the data we gain will be a mass of digital data. The Back Propagation Neural Network (BPNN) will be deployed to deal with them. To reduce computational complexity, we combine our approach 
 with the traditional method.
\par	
The results of this chapter were originally published in the conference papers: 
\textbf{Y.~{Zhang}} and J.~{Zhang}, `` An Obstacle Avoidance Method for Sonar-based Robots Avoiding Shape Changeable Obstacles in Underwater Environments,'' in {\em 2022 the 14th international Conference on Computer and Automation Engineering (ICCAE)}, Brisbane, Australia, 25-27 Mar. 2022.
\section{Introduction}
Unmanned Vehicles are widely used in various fields, from industry \cite{6.1} to agriculture \cite{6.2}, to military uses \cite{3.1}. For those applications, safe driving is always been considered as one of the most important tasks no matter where the operating location is. 
\par
At present, all the navigation strategies could be divided into two types, global path planning and local path planning, according to the environmental information they could obtain. The information of the former one is known in advance \cite{5.2}, while only part of the information is known for the latter one \cite{4.10} \cite{5.6}. Also, there are some strategies which combine the global path planning and local path planning together \cite{3.14} \cite{6.8}. In general, the requirements of the processor is not too high in order to implement the algorithms mentioned above. However, for the practical situations, the environment is always dynamic and changing from time to time. Numbers of advanced algorithms have been introduced for this purpose \cite{6.9} \cite{6.10}. In order to employ those more advanced algorithms, the requirements of the processors is higher. Here, we introduce one of those algorithms, neural network algorithm, which is also known as NN algorithm \cite{6.11} \cite{6.12}.
\par
Neural network (NN) is a model that imitates biological neural network, which consists of several artificial neurones. It is an adaptive system \cite{6.13}. For the most systems, the relationship between the input and output is complex. In general, people are always modelling with non-linear system. The traditional navigation strategies employing neural network are building a whole system for the robot and guide it from the initial point to the target point. The input is the information obtained from the sensor. The output is the position or the movement information after training with the neural network system \cite{6.14}.
\par
The back propagation neural network (BPNN) is one of the most widely used neural network (NN) models. A multi-layer feed-forward network is utilised to train the system according to the error back propagation algorithm \cite{6.15}. BPNN can learn and store a large number of input-output model mappings without revealing the mathematical equations describing these mappings in advance. Repeating learning and training with these data, the network parameters could be determined in order to achieve the minimum error of sum of square \cite{6.16}. With the development of technology and deeper research on the neural network, application of the BPNN method could become more competitive. 
\par
For the robots working on the ground, which is the problem been proposed the earliest, researchers focused on the 2D space \cite{6.17} \cite{yang.5}. When the application has been extended to the air, the height variable has been taking into consideration, as a result, the dimensionality is increasing to 3D. The modelling analysis is becoming more difficult, but the sensors employed could be similar ones compared with the ground environments. For example, the radar system and visual sensors still can provide a good performance. However, when the application scenario is the underwater environments, a critical problem arises, the sensors could be chosen is too few \cite{6.18}. As we employ sonar-based sensors, the position information of the robot and the obstacle we can get is limited compared with the visual-based one and radar-based one. In this case, the difficulty of designing algorithms is greatly increased.
\par
Moreover, for obstacles, researchers always pay more attention to the motion of the obstacles, as well as the position relationship between the robot and the obstacle \cite{6.19} \cite{yang.1}. However, based on the observation of the actual situation, it is necessary that the deformation of the obstacle should also be considered \cite{5.15}. We have proposed a strategy for 2D underwater environments in the last chapter. In this chapter, we focus on the 3D environments. A solution of the problem of avoiding shape changeable obstacles in three-dimensional underwater environments will be presented. The sensors be employed are sonar-based sensors. The method being applied is based on back propagation neural network algorithm.
\par
The remainder of the chapter is organized as follows. Section 5.2 focuses on the system construction. Section 5.3 describes the obstacle avoidance method based on the BPNN. Then, we give simulation results in section 5.4 to demonstrate the feasibility of our proposed method, which is implemented by Matlab. For section 5.5, we make a short summary.
\par
    
\section{System Identification}
As we employ Back Propagation Neural Network (BPNN) in order to solve the problem of obstacle deformation prediction, we introduce a basic BPNN structure in this part. Then we build a model for the UUV, including mathematical model and the sensors distribution. Example scenes with some different types of obstacles are also given at the end of this part.
	
\subsection{Back Propagation Neural Network}
As mentioned before, neural network is a network which simulated of human brain or other neural network of nature. It consists of numbers of processing units, which is known as neurons. There are numbers of input and only one output in one processing unit. Artificial neural network (ANN) model could be divided into two types according to the connection method of neurons, to be specific, feed forward neural network and feedback neural network \cite{6.14}. Back Propagation neural network (BPNN) is a forward multi-layer network, Fig. 5.1 is the most fundamental structure.
It contains an input layer and an output layer, also a hidden layer between these two layers. The input is named as X and the output is named as Y. The input layer receives data, the output layer outputs data, the neurons in the previous layer connect to the neurons in the next layer, collect the information from the neurons in the previous layer, and pass the value to the next layer. W and V are the weight coefficients, they are important parameters for the system to achieve the purposed goal. There is an error, which is the difference between the actual output and the expected output. With repeating learning from numbers of samples and adjusting the weight coefficients, the error will decrease along the gradient direction. The ultimate aim of repeating learning is to eliminate error. The learning will come to the end when the parameters corresponding to the minimum error are determined.
\par
    \begin{figure}[h!]
		\centering
		\includegraphics[width=4.2in]{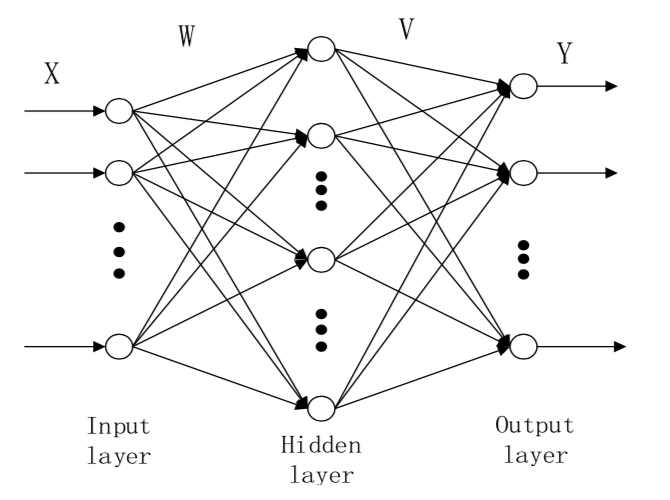}
		\caption{An example structure of BPNN.}
	\end{figure}
\subsection{Model Establishment of UUV} 
We consider a sonar-based robot working in an underwater environment. Here is a world three-dimensional coordinate, which denoted as $X-Y-Z$. 
In order to describe the motion of the robot, we give its mathematical model which shows below:
	\begin{equation}
	c_r{(t)}:=\left[x_r{(t)},y_r{(t)},z_r{(t)}\right],
	\end{equation}
	\begin{equation}
	\dot{c_r}{(t)}=V_{r}{(t)}\cdot u_{r}{(t)},
	\end{equation}
	\begin{equation}
	u_r{(t)}=\left[u_{rx}{(t)},u_{ry}{(t)},u_{rz}{(t)}\right].
	\end{equation}
Here, (5.1) is the three-dimensional vector of the robot's coordinate, (5.2) and (5.3) give the motion of the robot, it consists of linear velocity and orientation. 
Obviously,
	\begin{equation}
	{V_{r}}{(t)}^2=V_{rx}{(t)}^2+V_{ry}{(t)}^2+V_{rz}{(t)}^2,
	\end{equation}
	\begin{equation}
    u_{rx}{(t)}^2+u_{ry}{(t)}^2+u_{rz}{(t)}^2=1.
	\end{equation}
The movement range is constrained with default values, to be specific,  $V_{max}$ and $U_{max}$ as the maximum values of the linear velocity and orientation, respectively. We can get:
	\begin{equation}
	V_{r}{(t)}\in\left[0, V_{max}\right],
	\end{equation}
	\begin{equation}
	\lVert\dot{{u_{r}}}{(t)}\rVert\in\left[0, U_{max}\right].
	\end{equation}
	
\subsection{Onsite sensor distribution}
As mentioned in the previous section, the sensors we chose are sonar-based ones. We assumed that there is a smooth and convex space in the front of the robot, and it is large enough to hold the required number of sensors. We employs nine sonar-based sensors, and marked them as S1,S2,...,S9. Fig. 5.2 is an illustration of the onsite sensors distribution. Fig. 5.2(a) is from the front view and Fig. 5.2(b) is from the side view. They are arranged evenly in the dotted line area, we hold that the area is the largest area on the robot front surface where the sensors could be placed. As it is a convex surface, the robot can see the whole obstacle when there is a certain distance between the robot and the obstacle. 
\par
    \begin{figure}[h!]
	\centering
	\begin{subfigure}[b]{0.5\textwidth}
		\centering
		\includegraphics[width=2.9in]{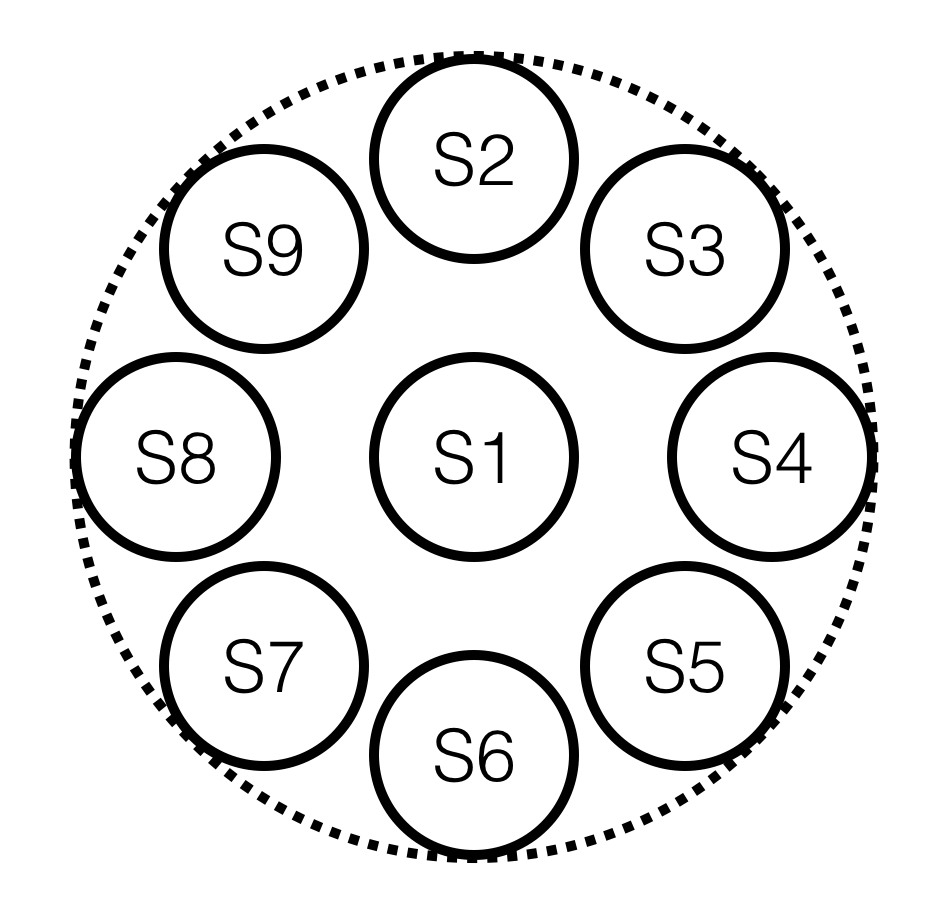}
		\caption*{(a)}
	\end{subfigure}
	\begin{subfigure}[b]{0.4\textwidth}
		\centering
		\includegraphics[width=2.9in]{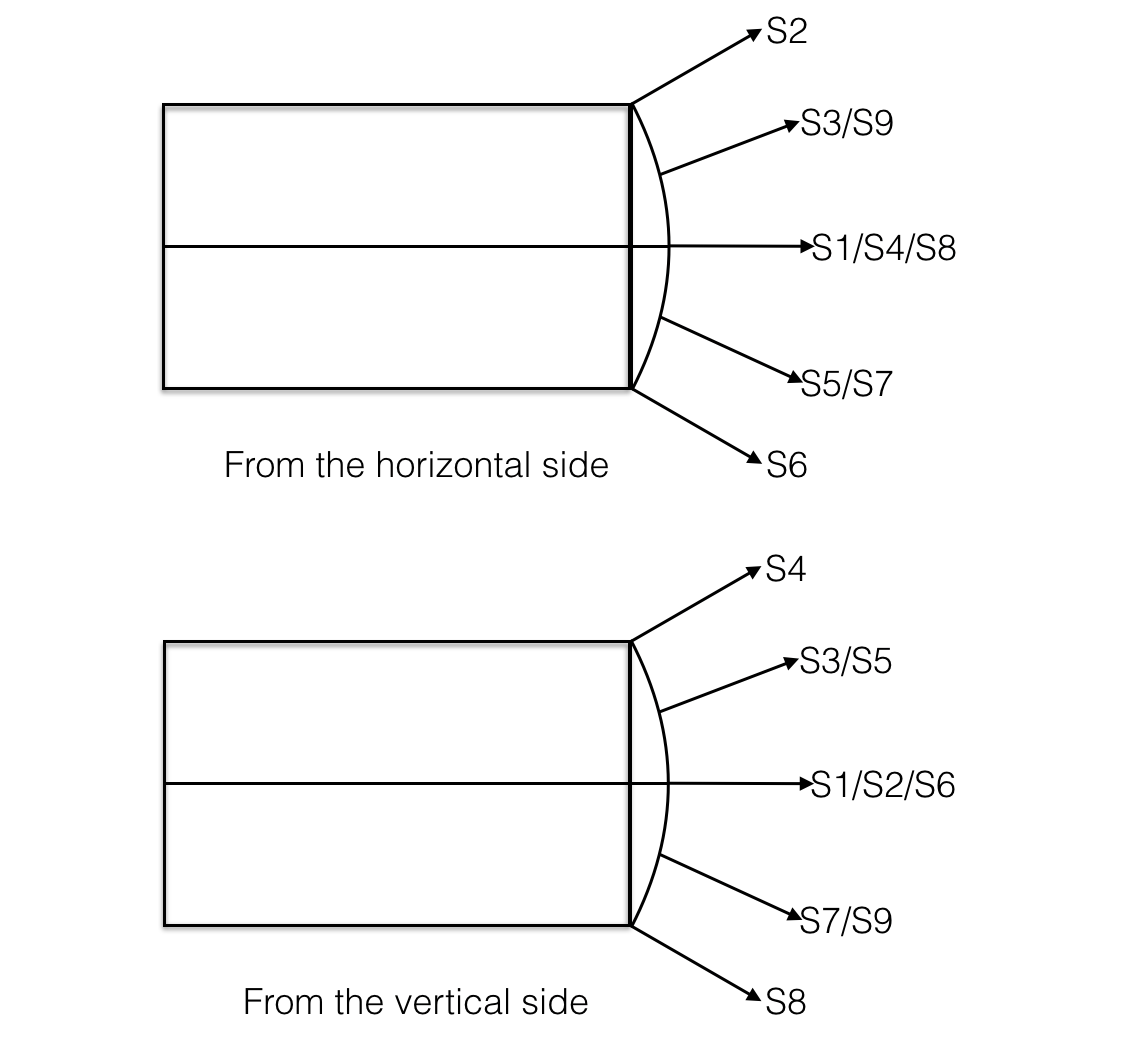} 
		\caption*{(b)}
	\end{subfigure}
	\caption{The distribution of sonar-based sensors.}
    \end{figure}
The output of each sensor is the distance between the sensor and the nearest point on the object in its vision cone. Although they are all fixed to the robot's front surface, the nine sensors are not in the plane. For the consistency of these data, we introduce a adaptive coefficient, which could help transfer the distance between the different sensors and the obstacle to the distance between S1 and the obstacle. S1 is equipped at the center of the surface, where is regarded as the leading position.
We use a matrix to express the distance sets, which denoted as $R(t)$,
	\begin{equation}
	R(t)=\left[R_{1}{(t)}, R_{2}{(t)},..., R_{9}{(t)}\right].
	\end{equation}
%
	
\subsection{Example Scenes with Obstacles}
We assumed a scene with some basic elements. Firstly, we define the initial point and the ending point, \textbf{\emph{A}} and \textbf{\emph{B}}, respectively. They are always stationary and known in advance for the robot. Then, we put some obstacles in the given scene, which are denoted as $\Diamond_1(t), \Diamond_2(t), ..., \Diamond_i(t)$. Their shapes are irregular, and we do not assume that all of them are convex. The boundaries should be smooth, the mass centre of them are fixed, and the shapes of them could be changeable in some extent. Fig. 5.3(a)(b)(c) show three scenes from bird's-eye view.
\par
\subsubsection{Scene 1}
Fig. 5.3(a) is an example scene with an ordinary obstacle. The boundary of the obstacle is smooth, the mass centre is fixed, and the deformation rate will not exceed 10\% of the length from the point to the mass centre.
\subsubsection{Scene 2}
In the example scene of Fig. 5.3(b), there is a really narrow space inside the obstacle. For the robot, it could not go into the space as the width of the robot is larger than the obstacle' s inner space, and collision could not be avoided in this case. 
\subsubsection{Scene 3}
For scene 3, which shows in Fig. 5.3(c), there are two obstacles on the robot' s way to the ending point. The robot has to avoid the obstacles when it gets access to the obstacles, or it will fall into dangerous situation.
	\begin{figure}[h!]
	\centering
	\begin{subfigure}[b]{\textwidth}
		\centering
		\includegraphics[width=3.8in]{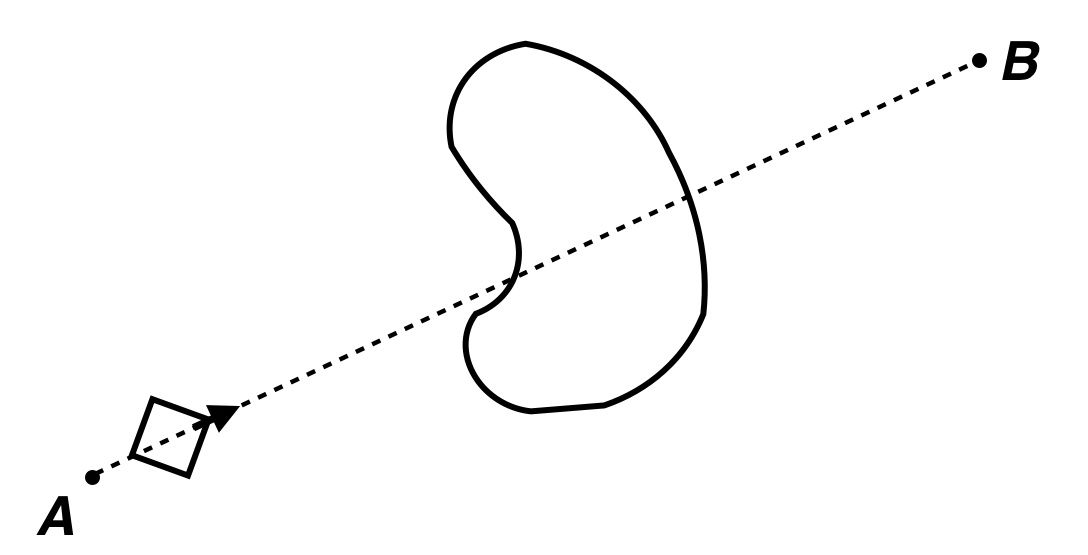}
		\caption*{(a)}
	\end{subfigure}
	\begin{subfigure}[b]{\textwidth}
		\centering
		\includegraphics[width=3.8in]{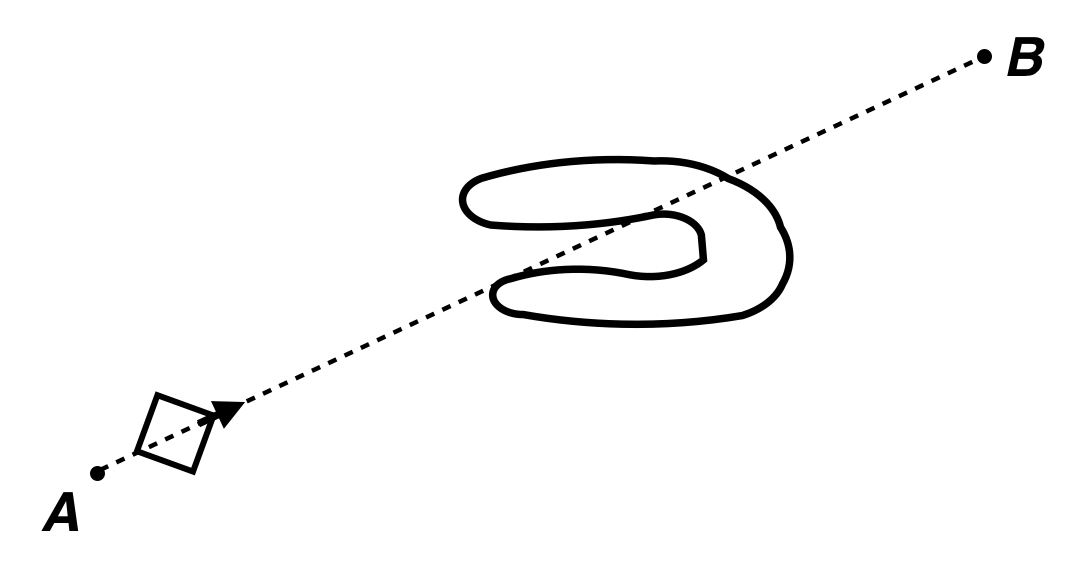} 
		\caption*{(b)}
	\end{subfigure}
	\begin{subfigure}[b]{\textwidth}
		\centering
		\includegraphics[width=3.8in]{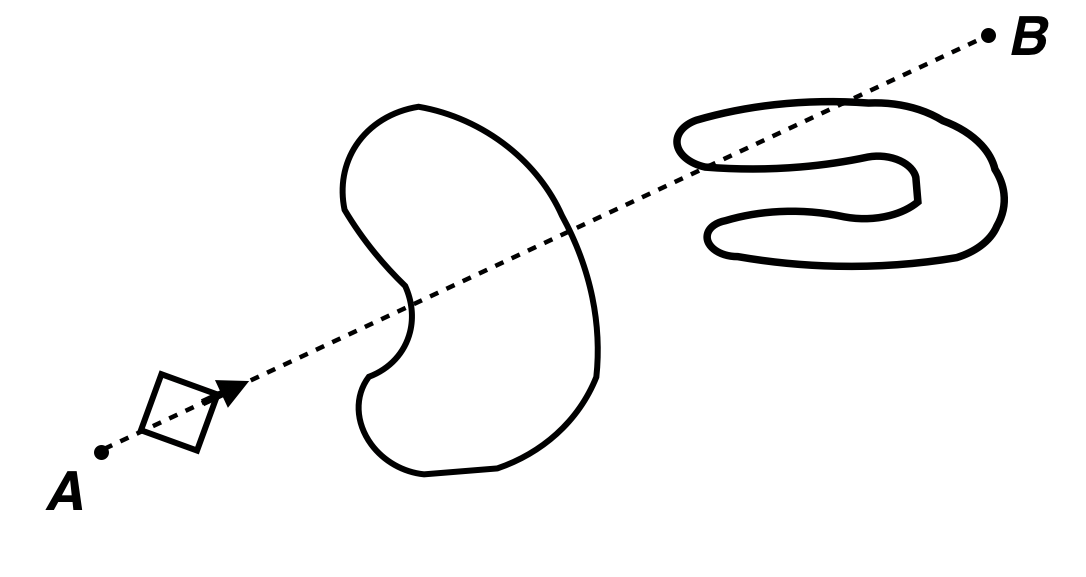} 
		\caption*{(c)}
	\end{subfigure}
	\caption{The illustration of example scenes.}
    \end{figure}

\section{Obstacle Avoidance Method}
Here, we will describe the main logic of our proposed obstacle avoidance strategy. The robot goes straight from the initial point \textbf{\emph{A}} to the ending point \textbf{\emph{B}} with the maximum speed, without any rotations. It will take actions to avoid the obstacles on its trajectory when the obstacles appear in its surroundings. For the obstacle avoidance method, we have proposed a viable solution for 2D environment in the last chapter, which is a traditional biologically inspired algorithm. In this case, the most important step is to observe and predict the deformation of the obstacle and determine the specific point for avoiding. We employ Back Propagation Neural Network (BPNN) to solve with the problem.
\par
Fig. 5.4 is an illustration of the neural network structure diagram of our proposed BPNN. The Back Propagation Neural Network consists of nine data inputs which gain from the sonar-based sensors S1, S2,...,S9. The data is consistent after dealing with the adaptive coefficient before input. The outputs are the minimum distance between the robot and the obstacle, and the marked number of sensor which achieved the minimum distance. Training with the samples, the error between the actual output and the desired output is decreasing. The learning for the neural network will come to an end when the minimum mean square error equals to zero, that means the weight coefficients have been adjusted to the appropriate ones. 
\par
So far, we obtain the position information of the specific point needs to evade. Combining with the traditional biologically inspired algorithm, obstacle avoidance method could be implemented. The trajectory for avoiding the obstacles could be generated successfully.
\par
	\begin{figure}[h!]
	\centering
	\includegraphics[width=8cm]{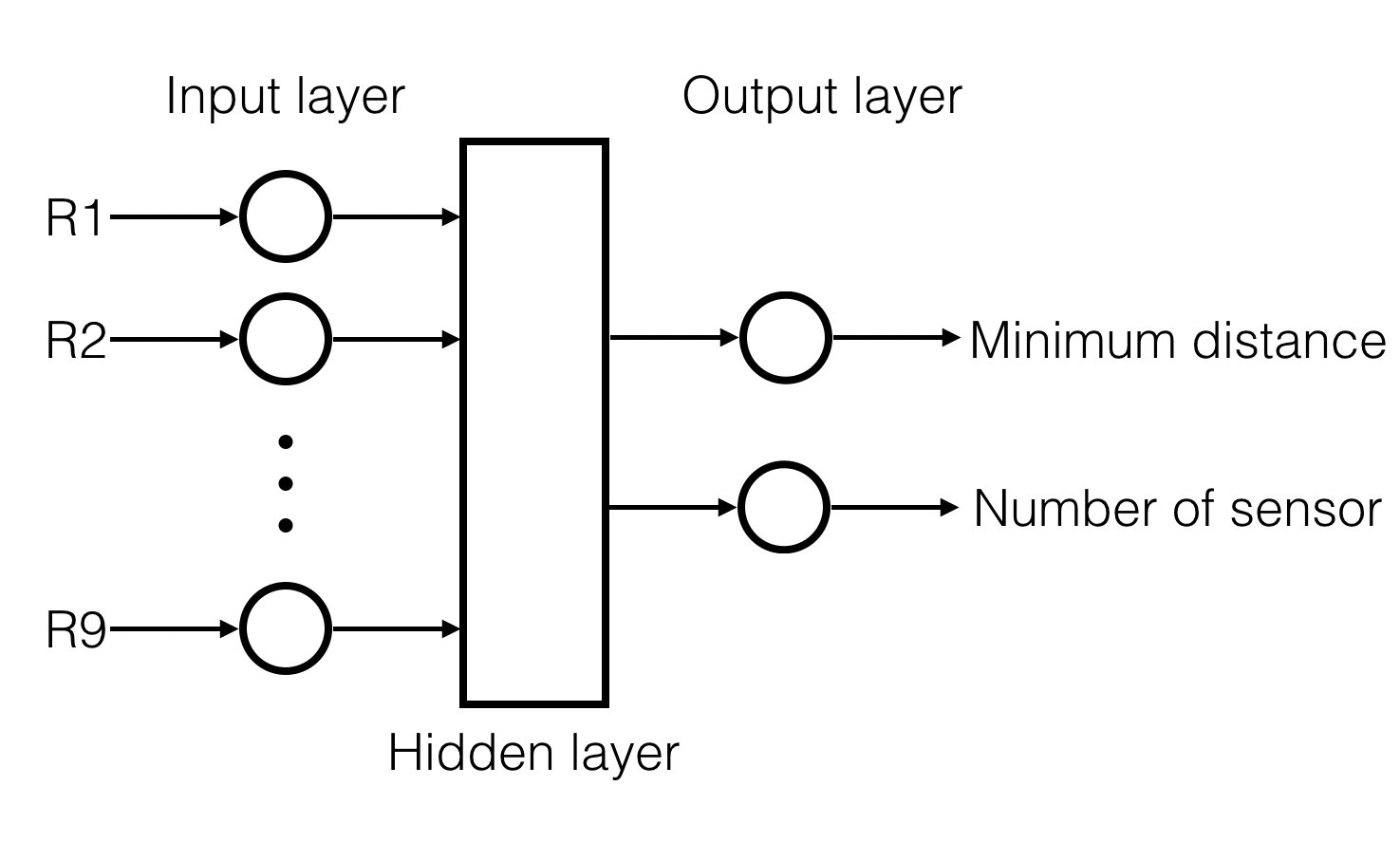}
	\caption{The neural network structure diagram of our proposed algorithm.}
	\end{figure}
\section{Simulation Results}
The experiments are implemented by Matlab Neural Network Toolbox. It is an intelligent tool for testing whether the given neural network is viable. We obtained some data in advance. They are divided randomly to three groups, training group, validation group and testing group. The training group is for the neural network to learn, the network will adjust according to its error. The validation group is used to measure network generalization, the network will stop training as the network generalization no longer improves. The testing group is an independent group to test the performance of the trained network. 
\par
As introduced in the previous section, the inputs are the distance measured from the nine sensors, and the outputs are minimum distance among the nine sensors and the sensor be marked which achieved the minimum distance. Firstly, we employed 50 groups of data. To be specific, the amount of input samples is 9*50, they present as the inputs of the Neural Network. The amount of output samples is 2*50, which show the minimum distance between the robot and the obstacle and the sensor which achieved the minimum distance. Fig. 5.5 and Fig. 5.6 show these results. For the minimum distance, the error between the targets and the actual outputs is [-1.688, 1.861], see Fig. 5.5(a). For the number of sensor, the error between the targets and the actual outputs is [-2.806, 4.454], see Fig. 5.6(a). The results are not satisfactory as they are too scattered. According to Fig. 5.5(b) and Fig. 5.6(b), the distribution of the data, we found the reason leads to the unsatisfactory results should be the amount of the samples is not enough. Thus, we increase the data to 150 groups, the results are shown in Fig. 5.7 and Fig. 5.8. This time, the errors are [-0.04884, 0.3578] and [-0.08559, 0.1147] for the minimum distance and the marked sensor, respectively. The majority of the data is around the zero error, see Fig. 5.7(a) and Fig. 5.8(a). Looking at the distribution of the data, which are shown in Fig. 5.7(b) and Fig. 5.8(b). It is clear that, training with the sufficient samples, the results are acceptable. As introduced BPNN, the aforementioned problem could be solved successfully. 
\par

    \begin{figure}[h!]
	\centering
	\begin{subfigure}[b]{\textwidth}
		\centering
		\includegraphics[width=4.2in]{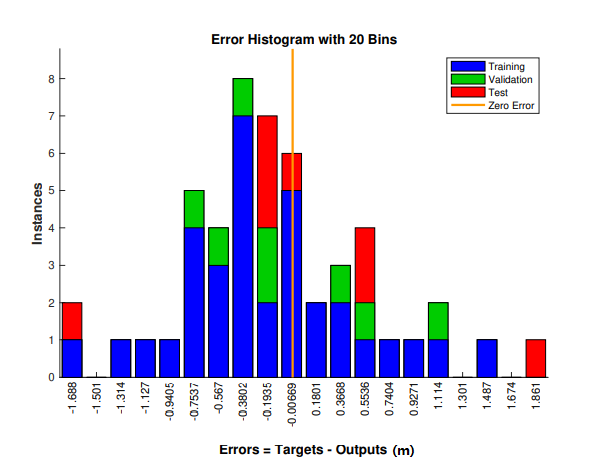}
		\caption*{(a)}
	\end{subfigure}
	\begin{subfigure}[b]{\textwidth}
		\centering
		\includegraphics[width=4.2in]{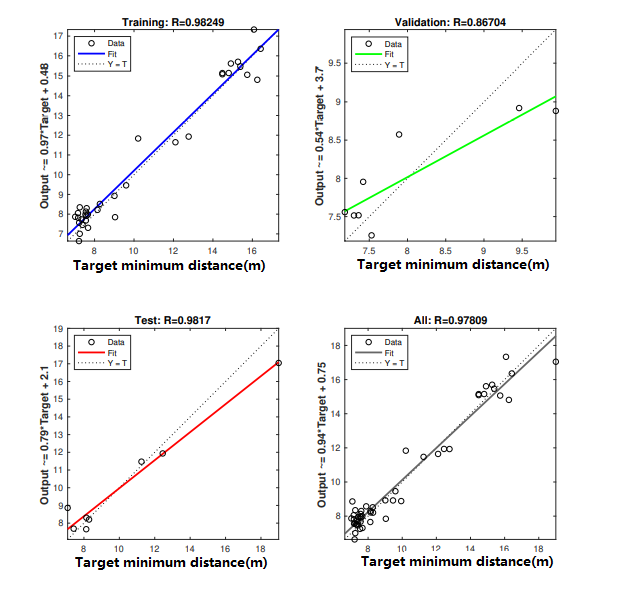} 
		\caption*{(b)}
	\end{subfigure}
	\caption{The simulation result of minimum distance with 50 groups of data.}
    \end{figure}	

    \begin{figure}[h!]
	\centering
	\begin{subfigure}[b]{\textwidth}
		\centering
		\includegraphics[width=4.2in]{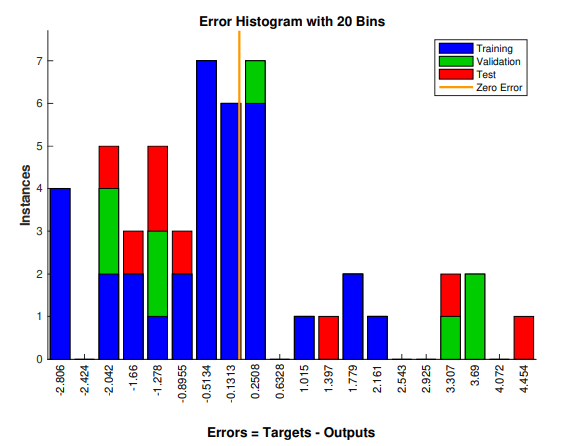}
		\caption*{(a)}
	\end{subfigure}
	\begin{subfigure}[b]{\textwidth}
		\centering
		\includegraphics[width=4.2in]{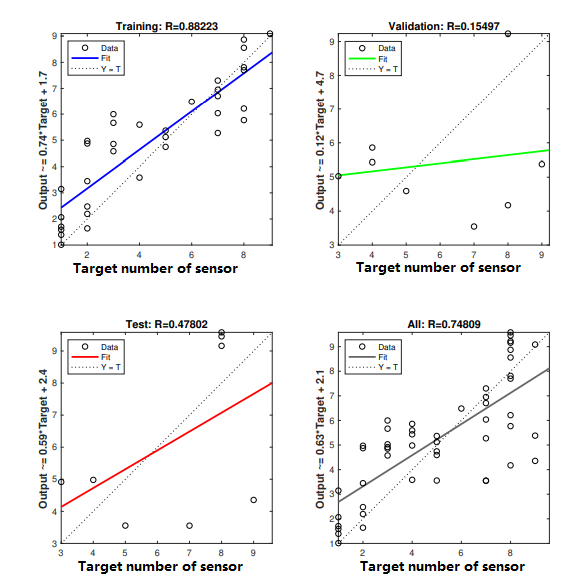} 
		\caption*{(b)}
	\end{subfigure}
	\caption{The simulation result of mark sensor with 50 groups of data.}
    \end{figure}

    \begin{figure}[h!]
	\centering
	\begin{subfigure}[b]{\textwidth}
		\centering
		\includegraphics[width=4.2in]{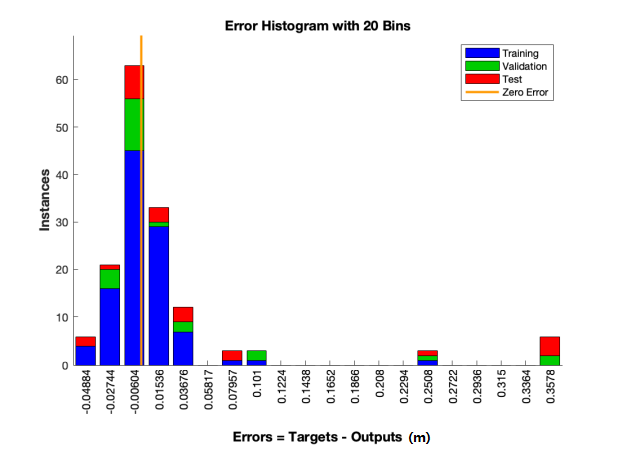}
		\caption*{(a)}
	\end{subfigure}
	\begin{subfigure}[b]{\textwidth}
		\centering
		\includegraphics[width=4.2in]{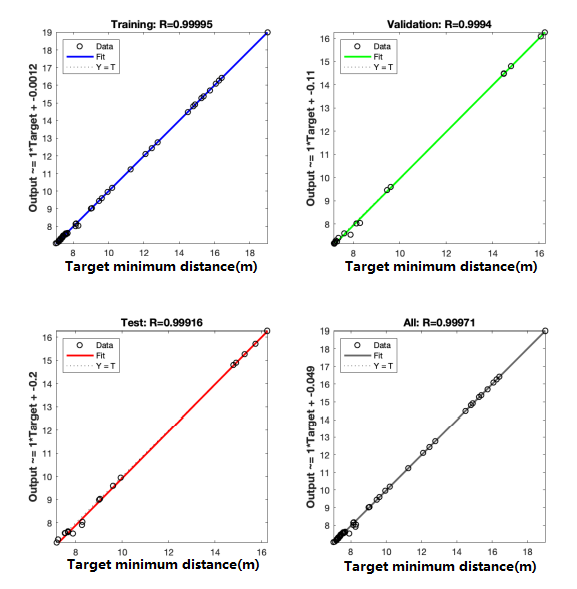} 
		\caption*{(b)}
	\end{subfigure}
	\caption{The simulation result of minimum distance with 150 groups of data.}
    \end{figure}	

    \begin{figure}[h!]
	\centering
	\begin{subfigure}[b]{\textwidth}
		\centering
		\includegraphics[width=4.2in]{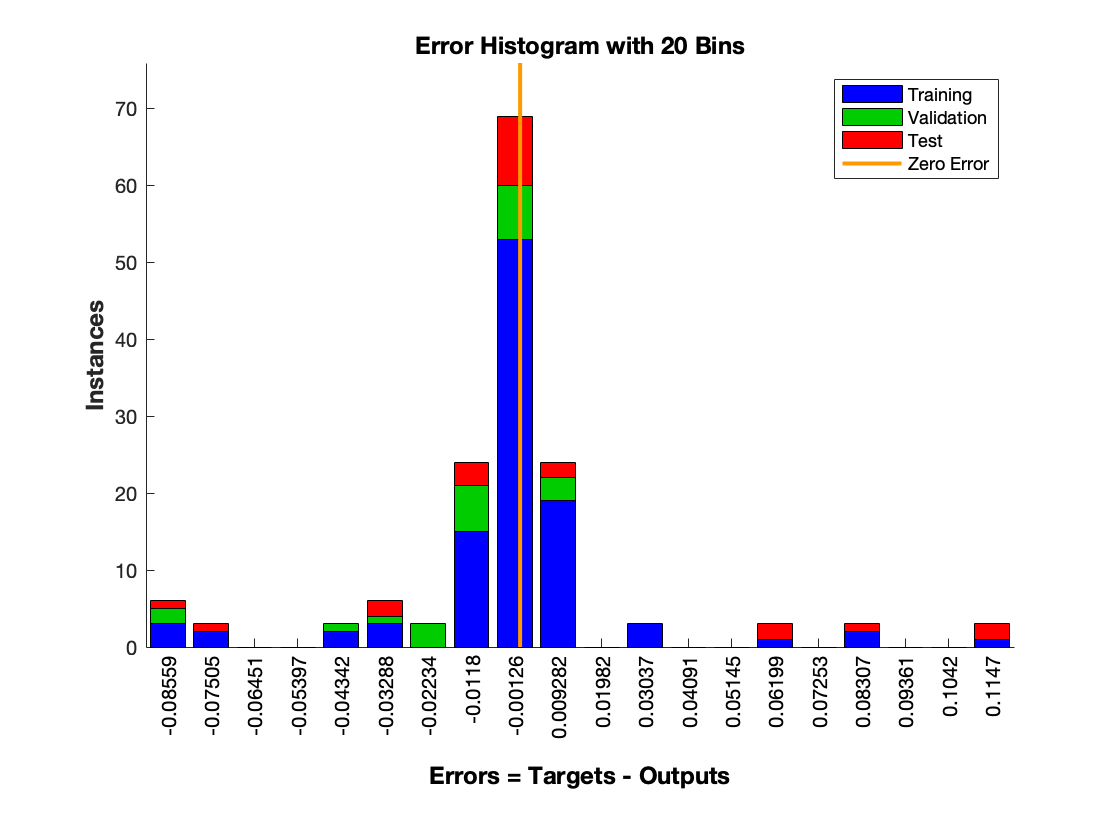}
		\caption*{(a)}
	\end{subfigure}
	\begin{subfigure}[b]{\textwidth}
		\centering
		\includegraphics[width=4.2in]{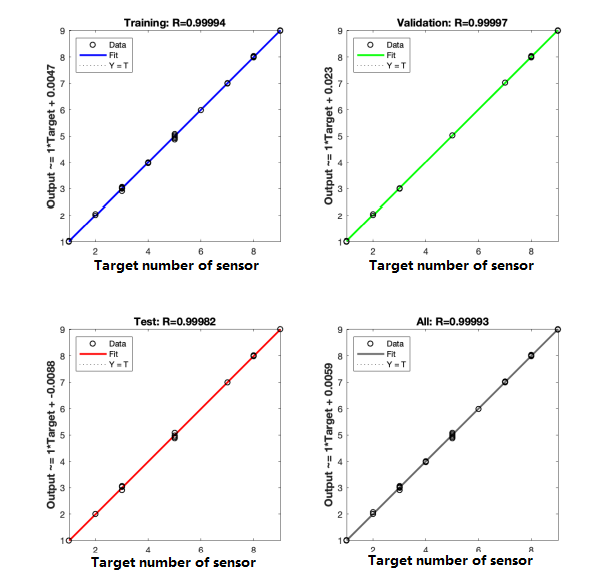} 
		\caption*{(b)}
	\end{subfigure}
	\caption{The simulation result of mark sensor with 150 groups of data.}
    \end{figure}

\section{Conclusion}
With a large number of recently proposed algorithms, the field of robot navigation develops rapidly. It is obvious that both the traditional algorithms and the algorithms with high-tech have their own advantages. Combining them properly could overcome some difficulties and achieve a better result. As we trying to propose an obstacle avoidance method for shape changeable obstacles in three-dimensional underwater environment, it is a tough task to predict the obstacles deformation due to the difficulty of modeling. In this chapter, we present a solution based on Back Propagation Neural Network (BPNN). After successful training with enough number of samples, the network could output accurately. Then, the traditional biologically inspired algorithm could be implemented for robots' navigation.
\par

%% file: chapter/Chapter71.tex
\chapter{A Convolutional Neural Network Method for Self-Driving Cars}
\label{C71:chapter71}
%
%
%
In Chapters 2-5, we focused on the task of obstacle avoidance. The strategies we proposed could solve the problems of avoiding collisions with motion changeable and shape changeable obstacles. In the next two chapters, we address path planning problems. There are many applications requiring path planning. In particular, self-driving cars and reconnaissance drones. In this chapter, we develop a trajectory planning strategy for self-driving cars. The specific problem we want to solve is predicting steering angle and speed for the cars in order to keep the vehicle going along the determined track. As we propose a CNN-based method, we will also deal with the problem that the amount of data for training the neural network is not large enough.

\par
%
%
The results of this chapter were originally published in the conference papers: 
%
J.~{Zhang}, H.~{Huang} and \textbf{Y.~{Zhang}}, `` A Convolutional Neural Network Method for Self-Driving Cars,'' in {\em 2020 Australian and New Zealand Control Conference (ANZCC)}, Gold Coast, Australia, 26-27 Nov. 2020.
\section{Introduction}
The conception of self-driving was first proposed in the 1920s. During the last hundred years, the technology of self-driving cars developed rapidly, and it has changed people' s lives and has brought huge economic benefits. In order to evaluate the ability of a driverless vehicle, the Society of Automotive Engineers proposed a classification system, which consists of six levels. The six levels system could identify the degree of automation. From Level 0 to Level 5, represent ‘\emph{No Automation}' to ‘\emph{Fully Automation}'. The larger the number, the higher the degree of automation. The ultimate aim of developing driverless vehicles is to replace humans to accomplish driving tasks completely. Although researchers all over the world out their efforts into this topic, there are still several problems that need to be solved\cite{7.21}.
\par
Here, we introduce Udacity, which is an open source for driverless vehicle simulation. The inputs are the images obtained from the three video cameras. The three video cameras are fixed on the left, centre, right of the front windshield of the simulating vehicle. Fig. 6.1 is a group of sample inputs. The values of steering angle and throttle are also recorded. In this platform, there are two modes that could be chosen, training mode and autonomous mode. The maximum speed and the maximum steering angle of the simulating vehicle are 30 miles per hour and 25 degrees, respectively \cite{3.1}. 
\par 
	\begin{figure}[h]
		\centering
		\includegraphics[width=4.2in]{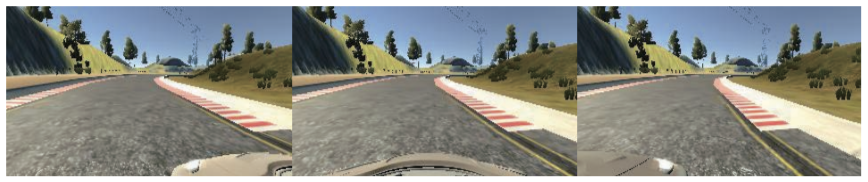}
		\caption{Images from left, center and right camera}
	\end{figure}
A typical architecture of a driverless vehicle system could be divided into two parts, the perception system and the decision-making system. The perception task includes obstacle detection, localization, mapping, and decision-making task includes path planning, behavior selection and control \cite{7.22} \cite{huang1} \cite{huang2}. Moreover, the equipped sensors are an important segment \cite{3.1}. There are a number of types of sensors, for example, lidar-based ones \cite{7.24}, radar-based ones \cite{7.25}, optical cameras \cite{7.26} \cite{7.27} and even ultrasound-based ones \cite{7.28}. The characteristics of sensors are different, and the method of extracting and analysing the information obtained by sensors are difficult. Even, extensive works need to be carried out to build model for the traditional path planning strategies. Neural Network (NN) is an adaptable approach, as it could learn from human behavior, and train itself until it can conduct itself drive safely. 
\par
Pomerleau was the first researcher in the world who employed the Neural Network for self-driving cars. He built an Autonomous Land Vehicle on a Neural Network (ALVINN) system, which is equipped on a semi-autonomous vehicle \cite{7.26}. In simple driving scenarios with obstacles, the steering angle could be directly predicted from the pixel inputs. The ALVINN model shows the potential of implementing a neural network in the field of driverless vehicles. It could be regarded as pioneer of Neural Network based driverless cars navigation strategy. In 2016, NVIDIA appeared, which was inspired by ALVINN, and employed deep neural network to extract features from the inputs effectively and robustly \cite{7.27}.
\par
In this chapter, we employ a Convolutional Neural Network (CNN) for driverless vehicles. Convolutional Neural Network (CNN) is one of the most popular Neural Network architectures. It consists of convolutional layers, fully connected layers, pooling layers, and dense layers. It is a feasible technology for autonomous vehicles to learn self-driving. Compared with the other Neural Networks, the advantages of CNN are that the accuracy of extracting features is higher and the training time it needs is less. It could learn from the training samples automatically. 
A scheme is given in Fig. 6.2 to illustrate the training and self-driving process. The platform we implement our proposed method is Udacity. For the training process, original images are taken by the three video cameras, which are equipped on the front windshield. The collected images with preprocessing will be the inputs of CNN. After training, the steering angle and speed values will be calculated. Then, the values will be compared with the expected values for the specific image, and the weight coefficients will be adjusted to improve the accuracy of the Neural Network. Thus, the trained network could be employed to generate suitable values of steering angle and speed by the real-time captured image to fulfill the task of self-driving for humans.
\par
	\begin{figure}[h]
		\centering
		\includegraphics[width=4.2in]{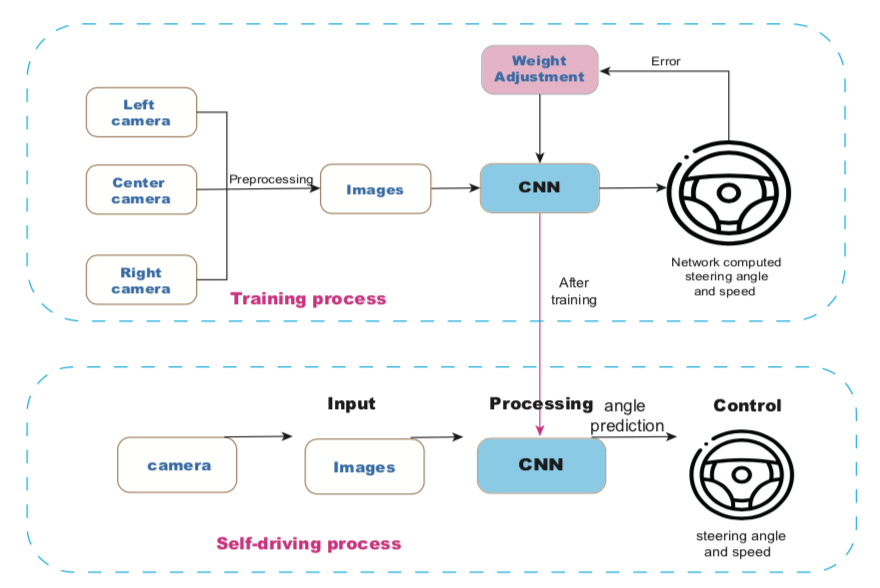}
		\caption{The training and self-driving scheme of the proposed vehicle}
	\end{figure}	
However, there is a fatal disadvantage of CNN, that the amount of training samples it requires is huge. The learning effect may be non-ideal when the amount of training samples are not enough. Fortunately, this problem could be solved by introducing data augmentation. For the Neural Network, one could be regarded as a completely new data when mall changes have been made to the existing data if the form of data is picture. Commonly used data enhancement methods include flip, rotation, scale, crop, translation, and other advanced enhancement techniques. In addition to increasing the amount of training data and improving model generalization ability, data augmentation can also avoid over-fitting, improve model robustness, and avoid sample disequilibrium.
\par

\section{Network Architecture}
We proposed a Convolutional Neural Network, the architecture is shown in Fig. 6.3. It consists of a normalization layer, four convolution layers, three dropout layers, a flattening layer, and three dense layers. The four convolutional layers are placed after the normalization layer. The neurons of the four convolutional layers are 24, 36, 48, 64, respectively. The ELU activation function is deployed between each convolutional layer as the convergence speed of the ELU is faster than ReLU \cite{7.214}. A dropout layer, with a 50 percent dropout rate, is placed after the convolutional layers in order to prevent overfitting. Then a flattening layer is followed, the aim of it is to convert the multi dimensional output of the convolutional layers to a one dimensional. The amount of neurons is 3840, the kernel is 3*3 size and the stride is 0. For the three dense layers, the number of neurons are 50, 10, 2, respectively, and the kernel is 5*5 size and the stride is 2. In addition, after the first two dense layers, there is a dropout layer for preventing overfitting.
\par
	\begin{figure}[h]
		\centering
		\includegraphics[width=4.2in]{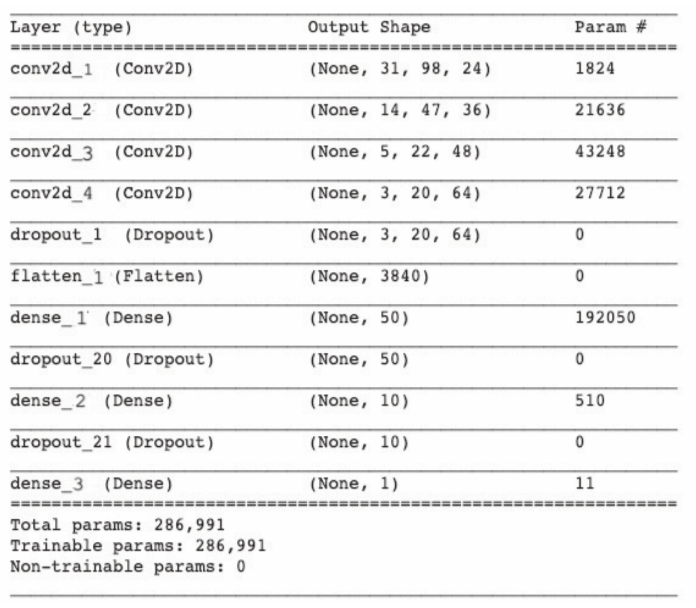}
		\caption{The network architecture}
	\end{figure}	
	\begin{equation}
	MSE = \frac{1}{n}\sum{(y_i-\hat{y_i})}^2
	\end{equation}
Here, we introduce a Mean Square Error (MSE). It is used to indicate the minimum error of steering command between our network and the actual one. Then, training our network with the result of the proper weight.
\par
\section{Experiments}
We employed the open-source simulation platform Udacity to collect our dataset \cite{7.215}. The images are collected from the three video cameras on the front windshield, the total amount of the data is 17592 160 * 320 pixels frames, also the values of steering angle and throttle are collected. Fig. 6.4 shows the distribution of the steering angle. The maximum value and minimum value are 1 and -1, respectively. The mean is -0.009 and the standard deviation is 0.204. We can get from the figure that the majority of the steering angle values are 0 degree. Then, we set the threshold value at 200 in order to balance the data, which is shown in Fig. 6.5. 
\par
	\begin{figure}[h!]
		\centering
		\includegraphics[width=4.2in]{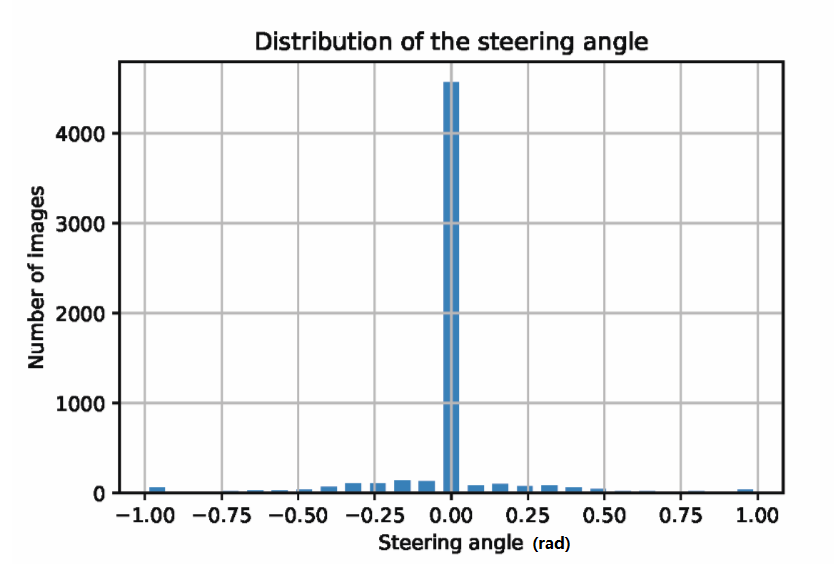}
		\caption{Steering Angle Distribution Histogram}
	\end{figure}	
	\begin{figure}[h!]
		\centering
		\includegraphics[width=4.2in]{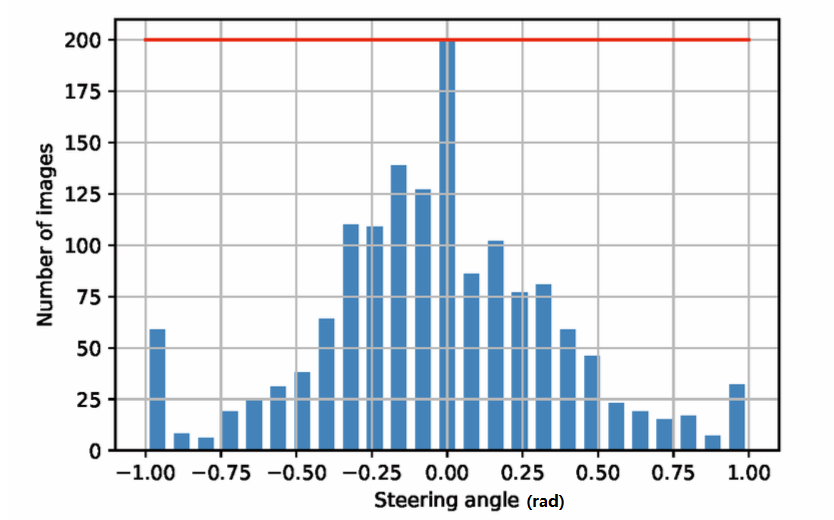}
		\caption{Steering Angle Distribution Histogram after Balancing}
	\end{figure}
Furthermore, the data is split randomly into a training set and a validation set with portion 8:2, Fig. 6.6 shows the exact distribution of the two sets, which are 1199 images for the training set and 300 images for the validation set, respectively. 
\par
Fig. 6.7 is the illustration of apply the augmentation techniques to improve the accuracy of our proposed network, to be specific, flipping, brightness changing and zooming.
\par
	\begin{figure}[h!]
		\centering
		\includegraphics[width=4.2in]{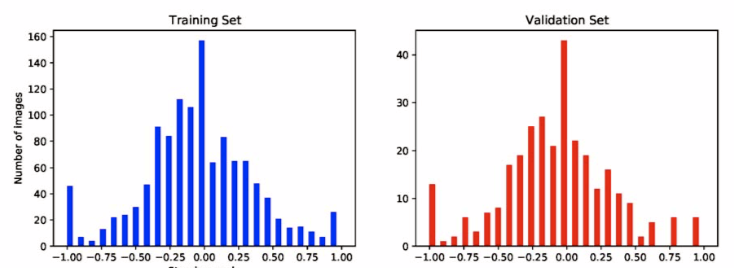}
		\caption{Steering angle distribution of two sets}
	\end{figure}
	\begin{figure}[h!]
	\centering
	\begin{subfigure}[b]{\textwidth}
		\centering
		\includegraphics[width=4.2in]{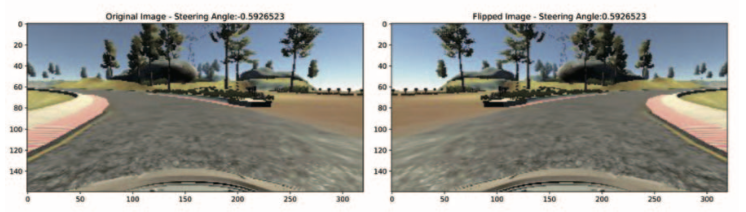}
		\captionsetup{}
		\caption*{(a)}
	\end{subfigure}
	\begin{subfigure}[b]{\textwidth}
		\centering
		\includegraphics[width=4.2in]{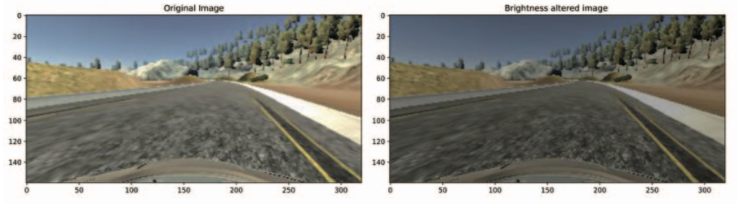}
		\caption*{(b)}
	\end{subfigure}
	\begin{subfigure}[b]{\textwidth}
		\centering
		\includegraphics[width=4.2in]{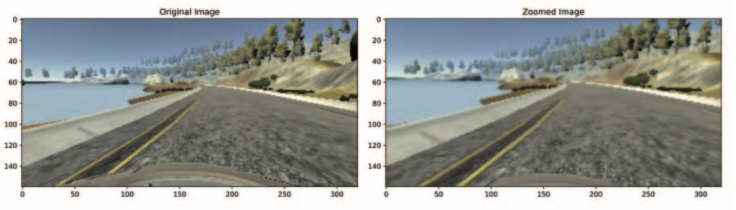}
		\caption*{(c)}
	\end{subfigure}
	\caption{Data Augmentation Methods}
    \end{figure}
	\begin{figure}[h!]
		\centering
		\includegraphics[width=4.2in]{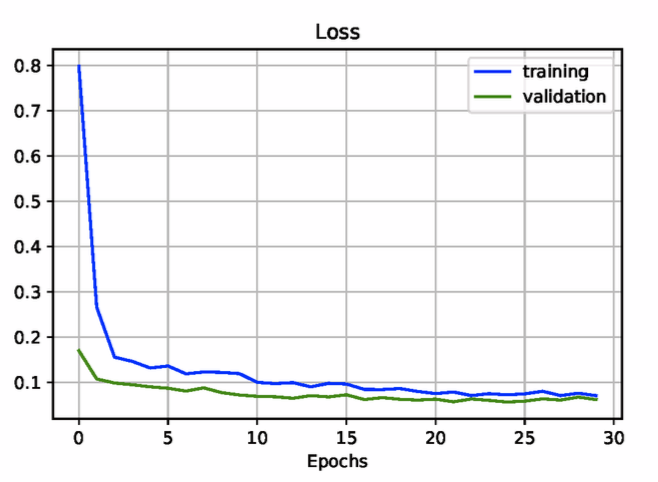}
		\caption{MSE Loss}
	\end{figure}
The Adam Optimizer is employed for our method, the learning rate is 0.001, the batch size is 100 and the epoch value is 30 \cite{7.216}. 
\par
Fig. 6.8 is the training result for our proposed method. Obviously, the MSE Loss could be reduced to around 0.05. As implemented our trained network, the accuracy of predicting steering angles and throttle values is viable, the self-driving vehicle could drive safely without any collisions, which is tested on the Udacity platform.
\par
\section{Conclusion}
In this chapter, we present a path planning method based on convolutional neural network (CNN), which could predict the steering angle and throttle values for self-driving cars. Data augmentation is introduced to solve the problem that the amount of data for the CNN training is not enough. The driving simulation platform we adopted is Udacity, a open source platform for autonomous vehicles. The images for inputs are collected by three video cameras equipped on the front windshield. With implementing our proposed method, the features could be extracted automatically, and the driving commands could be predicted successfully. Satisfactory performance and training parameters of the proposed CNN could show that our proposed method is rational. However, there is still a long way to go for self-driving cars to completely replace human drivers.
\par



%% file: chapter/Chapter72.tex
\chapter{A Method for UAV Reconnaissance and Surveillance in Complex Environments}
\label{C72:chapter72}
%
%
%
We are focusing on the path planning problem for some popular scenarios. 
In the previous chapter, a CNN-based strategy to predict steering angle and throttle values for driverless cars has been addressed. 
In this chapter, we will put effort into the other scenario, UAVs for reconnaissance and surveillance. 
A technical challenge need to be overcome is reaching a compromise between coverage of the drones and resolution of the video camera equipped on the drones. 
We will propose a path planning strategy for the reconnaissance and surveillance UAVs flying at a given altitude, which is the lowest altitude it could be, while ground targets are fully covered. 
%
%
%
\par	
The results of this chapter were originally published in the conference papers: 
%
J.~{Zhang} and \textbf{Y.~{Zhang}}, `` A Method for UAV Reconnaissance and Surveillance in Complex Environments,'' in {\em 2020 6th International Conference on Control, Automation and Robotics (ICCAR)}, Singapore, 20-23 Apr. 2020.
\section{Introduction}
Unmanned Aerial Vehicles (UAVs), which are also known as aerial drones, could be applied in different areas \cite{7.1} \cite{huang3}. For example, delivering goods and merchandise, research and rescues, infrastructure inspection, maintaining surveillance and security, border patrolling \cite{7.8} \cite{tuft2}\cite{7.2} \cite{7.3} \cite{7.4} \cite{7.5}. For the reconnaissance and surveillance UAVs, finding an optimal trajectory to execute the coverage task well is challenging. Generally, the coverage problem could be divided by the status of the drones, static coverage and dynamic coverage, respectively. The static coverage, which always employed drones who hovering in the air to monitor the specific terrains \cite{7.2} \cite{7.7} \cite{7.8}. For the dynamic coverage, the UAVs be employed are moving ones for reconnaissance and surveillance duties \cite{7.3} \cite{7.5} \cite{7.9}.
\par
The UAVs for reconnaissance and surveillance are always equipped with a video camera on the bottom surface. The targets they monitor are on the ground, from vehicles, to humans, to animals \cite{7.10}. Their performance is evaluated by the quality of the coverage and resolution of the equipped camera \cite{7.11}. Lower attitude is preferred for the better resolution, while the drones are also required to fully cover the given target on the ground with the minimum number of the waypoints. That means all the points of the given target should be seen at least once by the equipped video camera each surveillance circle. Apparently, these two evaluation criteria are inconsistent. A compromise needs to be reached for an impressive performance when we propose a strategy.
\par
Some solutions have been proposed previously. The strategy proposed by Semsch et al. ignored one of the evaluation criteria, resolution \cite{7.12}. Geng et al. gave a better solution which is based on a generic algorithm, in which the waypoints are generated and ant colony approach is deployed for visiting those waypoints \cite{7.11}. In addition, Zhang proposed an occlusion-aware UAV reconnaissance and surveillance strategy \cite{7.3}. As the applied environment is complex with varying altitudes and different occlusion situations, it is decomposed into a drone version 3D Art Gallery Problem and a variant of the combinatorial traveling salesman problem when solving the problem.
\par
Inspired by \cite{7.3} \cite{7.11} and \cite{7.12}, we proposed a two-phase strategy, which is more practical and efficient. We solved the two problems separately, generating the waypoint and planning the trajectory. 
In the first stage, generate the waypoints by constructing a triangulation of the area on the ground, which consists of numbers of congruent equilateral triangles, every point in the given area on the ground could be seen from at least one equilateral triangle. 
In the second stage, a short and smooth trajectory will be planned by the trajectory planning strategy. It consists of straight segments and Dubins curves, and each point will be covered at least once.
\par
The remainder of the chapter is organized as follows. The detailed problem statement is given, then describe the proposed method for UAV reconnaissance and surveillance thoroughly. We also perform the computer simulations in order to show the viability of our proposed solution. Finally, we present a short conclusion.
\par
\section{Problem Statement}
For each point $(x_g, y_g)$, the elevation on the ground could be expressed as $z_g=F(x_g, y_g)$. Let \emph{D} be a given subset of $z_g=0$. $Z_{max}$ and $Z_{min}$ are the maximum and minimum altitude for the UAVs. The following constraints for the generated waypoint $(x, y, z)$ should be satisfied. 
\par
	\begin{equation}
	(x,y)\in \emph{D}, z\in [Z_{min}, Z_{max}], 0\leq Z_{min}\leq Z_{max}.
	\end{equation}
The visible angle of the UAV is $0 < \theta < \pi$, the targets of interest $(x_g, y_g, z_g)$ on the ground could be seen. If it is out of the space, it is invisible. An illustration is shown in Fig. 7.1. The field of view (FOV) is a cone-shaped, and the radius of it is:
	\begin{equation}
	R := z \cdot tan{(\frac{\theta}{2}})
	\end{equation}
	\begin{figure}[h]
		\centering
		\includegraphics[width=4.2in]{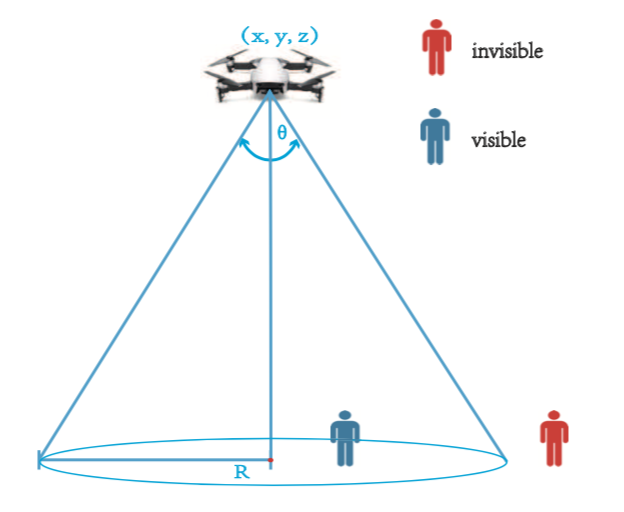}
		\caption{Occlusion effects on camera sensing. \cite{huang.4}}
	\end{figure}
\par
As mentioned in the previous section, we have to make a compromise between coverage and resolution. The higher the drone flies, the worse the onboard camera resolution. In order to get better resolution, we place the UAV at the lowest altitude as possible to obtain the best resolution. In this case, the specific problem we need to dealing with is fully coverage in the given altitude. 
\par
The problem could be considered as a similar problem to a traveling salesman problem, finding the trajectory of visiting $\delta$-neighborhood of all initial waypoint locations. We suppose that there is an initial waypoint location $p_i \in R^3$ which belongs to a vantage waypoint set $P = \{p_1, ..., p_n\}$, the UAV should visit and record certain part of the target area from $p_i$ within distance $\delta$. The waypoint location $p_{i}^{*}$ should be included in the vantage waypoint set, $ \left\|( p_i^*, p_i)\right\| \leq \delta $. The trajectory $\tau$ consists of a number of $Be^`zier$ curves, in which the curves are uninterrupted and shared in the same direction. For the complete trajectory, it is a closed and smooth curve.
	\begin{equation}
	\tau(t) = \mathbf{B_0}{(1-t)}^3+3\mathbf{B_1}t{(1-t)}^2+3\mathbf{B_2}t^2(1-t)+\mathbf{B_3}t^3,
	\end{equation}
where \textbf{$B_f$} is the k-th control point, and $0 \leq t \leq 1$.
\par

Moreover, in order to avoid collisions, the following constraints should be hold: 
	\begin{equation}
	q_{ij} \geq c_1, q_{i} \geq c_2, 
	\end{equation}
$q_{ij}$ is the minimum distance between waypoints i and j, and $q_{i}$ is the minimum distance  between waypoint i and the terrain. $c_1 > 2\delta >0$ and $c_2 > \delta >0$ are given safety margins. 
\par
\section{Reconnaissance Algorithm}
In this section, we describe our proposed reconnaissance algorithm in detail. The strategy could be divided into two stages, waypoints generation stage and the UAV path planning for surveillance stage, respectively.
\subsection{Stage One: Waypoints Generation}
First, we consider a parallel to the ground plane $z_p = z$, and estimate the minimal number of waypoint locations on the plane. Then, seek z, which is the minimum altitude of sensing with the achieved 2D coordinate $(x_i, y_i)$. $P = \{p_1, ..., p_n\}$ is the vantage waypoint set, where $p_i \in R^3$.
\par
	\begin{figure}[h!]
		\centering
		\includegraphics[width=4.2in]{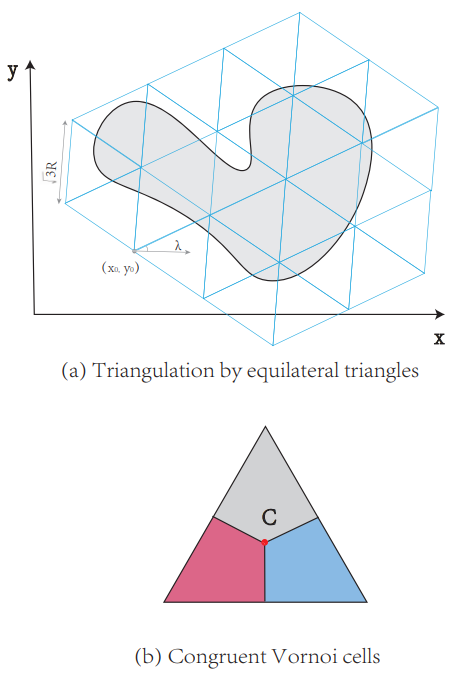}
		\caption{Waypoint Generation Method}
	\end{figure}
We assumed that the triangulation consists of equilateral triangles with the side $v = \sqrt{3}$. For any of those triangulation, $\Delta (\lambda, x_0, y_0)$, where $\lambda \in [0, \frac{\pi}{3})$ is the angle between one direction of $\Delta(\lambda, x_0, y_0)$ and x-axis, and $(x_0, y_0 )$ is the coordinate of one of the vertices, see Fig. 7.2(a). The center is C and the triangle is divided equally into three congruent Voronoi cells, which are marked in different colors, see Fig. 7.2(b). An illustration of the locations of waypoints is shown in Fig. 7 .3. The target area is 300 x 300 $m^2$ terrain, and the waypoint set is generated by the regular triangulation method.
\par

	\begin{figure}[h]
		\centering
		\includegraphics[width=4.2in]{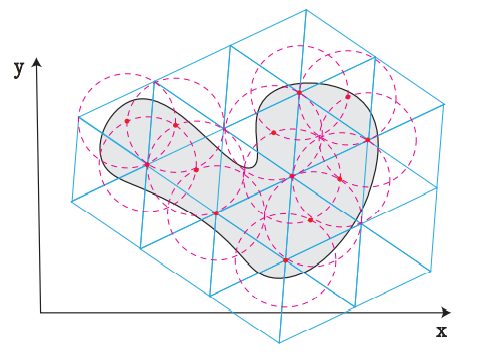}
		\caption{Locations of waypoints. The coverage areas are shown in pink circles}
	\end{figure}
\subsection{Stage Two: UAV Path Planning for Surveillance}
In order to save traveling time, the neighborhoods of each individual vantage waypoint location are allowed to visit. As a certain part of the target area from $p_i$ within distance $\delta$ can be recorded, we introduce clustered spiral-alternating and $Be^`zier$ curves here, and the combinatorial traveling salesman problem will be readily solved. In addition, the trajectory follows the UAV' s kinematic constraints.
\par
It has been proved that, for the situation that the distance between waypoints is sufficiently large for the turning radius, the spiral algorithm \cite{7.12} performs better when the waypoints are dense, and also for the low altitudes surveillance tasks. However, after observing the result of the complete trajectory, the total path length is less when employing the clustered spiral-alternating algorithm \cite{7.13}. The methods of the waypoints being clustered into different clusters are manually or using a cluster analysis algorithm.
\par
We suppose that the UAV will return to the initial locations $p_1$ when it finishes the reconnaissance tour, no matter where it ends. The initial trajectory could be planned by the clustered spiral-alternating algorithm \cite{7.13}, with the given vantage waypoint set $P = \{p_1, ..., p_n\}$,   where $p_i \in R^3$, distance $\delta$, the initial waypoint location $p_1$. The final smooth trajectory $\tau$ is $\tau = (\mathbf{T_{\sigma_1},...,T_{\sigma_n}})$, $1 \leq \sigma_i \leq n$, which is a sequence $\sum = (\sigma_1,...,\sigma_n) $ of $Be^`zier$ curves $\mathbf{T_i}$, $1 \leq i \leq n$.
\par
\section{Results}
We give computer simulation results to demonstrate the potential of our proposed strategy. In the experientment, the target area is 600 x 600 $m^2$ terrain. In stage one, the minimum number and the exact location of the waypoints are shown in Fig. 7.4(a), and the $\delta$-neighborhood of the vantage waypoints are shown in Fig. 7.4(b). 
Then, we set two example scenarios to verify the practicability of our proposed strategy. The first scenario is a single-area with multiple UAVs. The result is shown in Fig. 7.4(c). Two UAVs monitor two clusters, respectively. The target area is covered completely, as both of them finish their own tasks. A quick parallel surveillance is achieved, and the following task allocations are simple. The second scenario is a single-area with a single UAV. The result is shown in Fig. 7.4(d). In this scenario, only one drone can be employed to cover the whole area. We reserve one trajectory and regenerate the other. Finally, the integrated trajectory with the $Be^`zier$ curves is generated successfully by our proposed strategy. So far, the validation of the solution is demonstrated carefully with the simulation results.
\par
	\begin{figure}[h]
		\centering
		\includegraphics[width=4.2in]{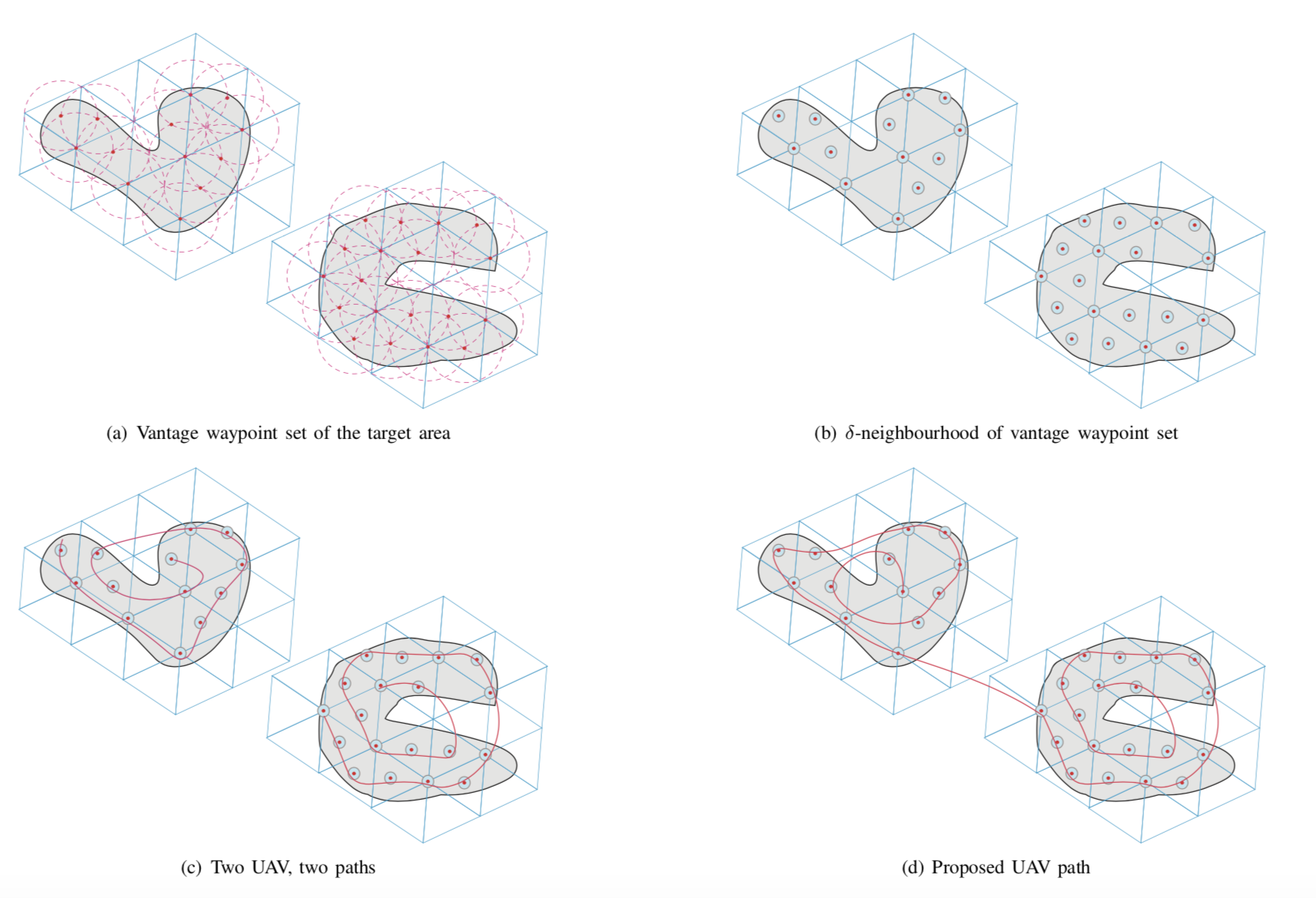}
		\caption{Proposed UAV surveillance strategy}
	\end{figure}
\section{Conclusion}
In this chapter, we propose a two-phase strategy for surveillance and surveillance UAVs. For the two-phase strategy, deciding the waypoints to be visited in the first phase and plan the path for the drones in the second phase. Our proposed strategy generates an optimal trajectory for the drones, which achieves a compromise between the completeness of the coverage and the resolution. The given altitude for the drones is the lowest possible one, while the target area on the ground could be completely covered. The number of waypoints to be visited is the minimum one. 
Simulation results have been given to show the performance of our proposed strategy. 
\par

%% file: chapter/Chapter8.tex
\chapter{Conclusions and Future Works}
\label{C9:chapter9}
In this chapter, we make a conclusion for the whole report and then give some future research directions.
\section{Conclusions}
In this report, we focus on the problem of navigating and controlling mobile robots. For our research topic, it is been divided into two main directions, path planning and obstacle avoidance.
In general, robot navigation is calculating a feasible path before the robot moves according to the static map, which has always been known as global path planning, and employing local obstacle avoidance strategy to avoid the obstacles on the road. The combination of global path planning and local obstacle avoidance enables the robot not only to reach the target quickly from the starting position, but also to avoid dynamic obstacles in the environment. 
\par
Based on the studies we conducted on the previous relevant literature and the observation of the practical situations, we determined to design algorithms for those more complex application scenarios, while the computational efficiency is acceptable, compared with the existing ones. 
The specific points of our research are reducing the restrictions of the dynamic obstacles in the environment when designing obstacle avoidance methods and providing novel path planning solutions for two popular scenarios. We gave each viable obstacle avoidance strategy for four different situations from Chapter 2 to Chapter 5. The path planning methods for the two popular scenarios are shown in Chapter 6 and 7. For each strategy we proposed, we gave a detailed description of the robot system and example experimental environments the method could be applied to. Algorithms with necessary analysis and simulation results are also given to illustrate the viability of the proposed algorithms.  
The conclusions of each chapter are drawn as follows.
\par
%
%
In Chapter 1, we presented a brief introduction of the research topic. At the first beginning, we introduced the origin of the robot and the applications of robots in order to show the necessity of navigating and controlling robots. Then, we reviewed the previous navigation strategies to figure out the aim of our research. We also gave the organization of the whole report and presented the main contribution of our research. 
\par
%
In Chapter 2, the algorithm we proposed is for two-dimensional environments, where the motion of the obstacles is changeable. Both the speed and angular velocity of the obstacles are limited by given constants. They could move in the given planar with the restrictions. The robot moves from the initial point with the maximum speed without any rotations, which is named as target approaching mode, shifts to obstacle avoidance mode when an obstacle is detected. The obstacle avoidance mode consists of three steps, deciding the direction to avoid, getting the nearest point on the boundary of the obstacle, and choosing the tangent point, respectively. Shifting between the two modes until successfully reaches the target point. The restriction of the obstacles motion is less compared with the previous works. Even the whole path length is shorter for the same scenario, which has been shown with simulation results. 
\par
In Chapter 3, we put efforts into three-dimensional environments with motion changeable obstacles. The navigation strategy could be regarded as an extended application of the algorithm proposed in Chapter 2. Also, it is composed of two modes, target approaching mode and obstacle avoidance mode. Target approaching mode is to find the shortest path from the position of the robot to the ending point, which is a straight line. For the obstacle avoidance mode, predicting the motion of the obstacle, finding the obstacle avoidance point, and transforming between the on-line coordinate and the geometry coordinate are needed. The proposed method is a user-friendly one to fulfill the mentioned obstacle avoidance problem as it is computationally efficient compared with those high-level decision-making algorithms. 
\par
In Chapter 4, we tried to solve the problem of navigating the robot to avoid shape changeable obstacles in underwater planar environments. The sensors we employed are sonar-based ones as the others are not applicable for underwater environment. The distance between the robot and the nearest obstacle is the input for the given algorithm. We introduced an \emph{AMAPS} for recording and predicting the deformation of the obstacles. Recording the times covered by the obstacle in each square, and the exact point for avoidance will be indicated. Then the robot will take action to avoid collisions with the obstacles. The core of the proposed algorithm is to avoid the obstacles safely while the trajectory chosen is as short as possible.
\par
In Chapter 5, the three-dimensional underwater environments with deformation obstacles are taken into consideration. We arranged a sonar-based sensor group for robot to detect obstacles in its line of sight. Because the task of predicting the deformation of obstacles for 3D environment is more challenging, we deployed Back Propagation Neural Network (BPNN), the popular high-level decision-making technique which could solve the problems difficult to modelling. Observing and learning from a certain amount of the samples, the shape change of obstacles could be predicted. The elements required by viable obstacle avoidance strategy could be obtained. Combining with the traditional path planning algorithm, successful navigation is realized.
\par
In Chapter 6, we focused on the problem of path planning for self-driving cars. The method we adopted to predict the steering angle and throttle values for an autonomous vehicle is Convolutional Neural Network (CNN). The performance of CNN-based self-driving algorithm is good, however it requires a huge amount of the training samples. We introduced data augmentation to solve the problem. The approaches we used for data enhancements are flipping, brightness changing and zooming. The network is trained and tested with the simulation platform, Udacity. Safe driving without any collisions is achieved. 
\par
In Chapter 7, we put efforts into the problem of planning path for reconnaissance and surveillance drones. We tried to find an optimal trajectory for the UAVs in a given altitude, which is the lowest altitude it could be, while all the points of the given target on the ground could be seen at least once. Our proposed approach is a two-stage approach, using triangles to make all points be involved in the first stage, and planning trajectory in the second stage. For a single area, the strategy is always effective for both a single drone and multiple drones.
\par

\section{Future prospects}
Although we have proposed solutions to some specific problems that are not often addressed by other researchers, there is still a long way to go for navigation and control of mobile robots so that robots can fulfill a number of tasks in all walks of life on behalf of humans.
\par
During our research period, we kept thinking and some original ideas, which seemed feasible were come up with. We made some of them come true, but there were still many things that we have not tried. Only simple research and analysis, even preliminary ideas popped up. Here, we would like to introduce some of them.
\par
Firstly, combining traditional algorithms with high-tech algorithms together. Although we have tried to combine the biologically inspired algorithm with the neural network method when solving the problem of avoiding shape changeable obstacles in 3D underwater environments, a further combination of algorithms is still a valuable research direction, so as to improve navigation ability while saving the cost of computing.
\par
Secondly, extending the obstacle avoidance algorithms developed in the report for ground robots to autonomous UAVs. The applicable scenarios of UAVs are more than ground robots \cite{8.8} \cite{8.9} \cite{8.10} \cite{8.11} \cite{8.12}. We take three-dimensional tunnel-like environments as an example scenario, they proposed methods of map building and autonomous navigation \cite{8.1} \cite{8.2} \cite{8.3}. Combing with our proposed obstacle avoidance methods should be a feasible idea.
\par
Moreover, cooperative working among multi-robots. No matter where they work, on the ground, in the air, or under the water, it is inevitable that more than one robot is needed when completing some tasks. An important direction of future research is to extend the results of this report for single ground robots to the case of systems of several communication ground mobile robots. Such systems of communicating autonomous vehicles can be viewed as an example of Networked Control System (NCS) \cite{8.4} \cite{8.5} \cite{8.6} \cite{8.7} \cite{8.13} \cite{8.14}. Under the situations of multi-robots participating, how to avoid collisions between each other and make each of their routes as optimal as possible is a worthy topic to study \cite{8.15} \cite{8.16} \cite{8.17}.
\par
With the rapid development of technology and the huge investment in navigation and control of robots,  we believe that solutions will be proposed in the future for those inspirations arising from our work. It can be expected that making robots do work what people are not willing to do. 
\par